\documentclass[fleqn,usenatbib]{mnras}

\usepackage[T1]{fontenc}

\DeclareRobustCommand{\VAN}[3]{#2}
\let\VANthebibliography\thebibliography
\def\thebibliography{\DeclareRobustCommand{\VAN}[3]{##3}\VANthebibliography}

\usepackage{graphicx}	% Including figure files
\usepackage{amsmath}	% Advanced maths commands
\usepackage{amssymb}	% Extra maths symbols

\usepackage{newtxtext,newtxmath}

\usepackage[usenames,dvipsnames]{xcolor}

\usepackage{hyperref}
\usepackage{float,svg} 
\usepackage[skip=7pt]{caption}         % pie de foto o tabla
\usepackage[skip=3pt]{subcaption}      % pie de subfigura o subtabla 
\usepackage[rightcaption]{sidecap}       % pie de foto/tabla lateral
\usepackage{floatflt,wrapfig}           % figura o tabla rodeada de texto
\usepackage{import} 
\usepackage{duckuments}
\usepackage{pgfplots}
\usepackage{stackengine}

\title[EAGLE meets \texttt{gsf}]{Milky Way-like galaxies: stellar population properties of dynamically defined disks, bulges and stellar halos}

\author[S. Ortega-Martinez et al.]{Sara Ortega-Martinez$^{1, 2}$\thanks{E-mail: sara.ortega@dipc.org}, Aura Obreja$^{3}$\thanks{E-mail: obreja@usm.lmu.de}, Rosa Dominguez-Tenreiro$^{2,4}$, Susana E. Pedrosa$^{2,5}$, 
\newauthor Yetli Rosas-Guevara$^{1}$, Patricia B. Tissera$^{6,7}$
\\
$^{1}$ Donostia International Physics Centre (DIPC), Paseo Manuel de Lardizabal 4, 20018 Donostia-San Sebastian, Spain\\
$^{2}$ Departamento de F{\'i}sica Te{\'o}rica, Universidad Aut{\'o}noma de Madrid, E-28049, Cantoblanco, Madrid, Spain\\
$^{3}$ Universitäts-Sternwarte München, Scheinerstraße 1, 81679 München, Germany\\
$^{4}$Centro de Investigaci{\'o}n Avanzada en F{\'i}sica Fundamental, Universidad
Aut{\'o}noma de Madrid, E-28049 Cantoblanco, Madrid, Spain\\
$^{5}$ Instituto de Astronom{\'i}a y F{\'i}sica del Espacio, CONICET, C1428ZAA Buenos Aires, Argentina\\
$^{6}$ Instituto de Astrof{\'i}sica, Pontificia Universidad Cat{\'o}lica de Chile, 8970117, Santiago, Chile\\
$^{7}$ Centro de Astro-Ingenier{\'i}a, Pontificia Universidad Cat{\'o}lica de Chile, 8970117, Santiago, Chile
}
\date{Accepted XXX. Received YYY; in original form ZZZ}

\pubyear{2022}

\begin{document}
\label{firstpage}
\pagerange{\pageref{firstpage}--\pageref{lastpage}}
\maketitle

\begin{abstract}
The formation of galaxies can be understood in terms of the assembly patterns of each type of galactic component. To perform this kind of analysis, is necessary to define some criteria to separate those components. Decomposition methods based on dynamical properties are more physically motivated than photometry-based ones. We use the unsupervised Gaussian Mixture model of \texttt{galactic structure finder} to extract the components of a sub-sample of galaxies with Milky Way-like masses from the EAGLE simulations. A clustering in the space of first and second order dynamical moments of all identified substructures reveals five types of galaxy components: thin and thick disks, stellar halos, bulges and spheroids. We analyse the dynamical, morphological and stellar population properties of these five component types, exploring to what extent these properties correlate with each other, and how much they depend on the total galaxy stellar and dark matter halo masses. All galaxies contain a bulge, a stellar halo and a disk. 60\% of objects host two disks (thin and thick), and 68\% host also a spheroid. The dynamical disk-to-total ratio does not depend on stellar mass, but the median rotational velocities of the two disks do. 
Thin disks are well separated in stellar ages, [Fe/H] and $\alpha$-enhancement from the three dispersion-dominated components, while thick disks are in between.
Except for thin disks, all components show correlations among their stellar population properties: older ages mean lower metallicities and larger $\alpha$-enhancement. Finally, we quantify the weak dependence of stellar population properties on each component's dynamics.
\end{abstract}


%%%%%%%%%%%%%%%%%%%%%%%%%%%%%%%%%%%%%%%%%%%%%%%%%%

%%%%%%%%%%%%%%%%% BODY OF PAPER %%%%%%%%%%%%%%%%%%

\section{Introduction}

Observed galaxies display a wide variety of morphologies, ranging from round ellipticals to disk-shaped spirals and irregular objects. Within galaxies of different types, various components with common features (e.g. shapes, ages, rotational support) can be distinguished. 

Bulges, occupying the very inner regions of galaxies, are thought to host old and metal poor stellar populations on hot orbits. The so-called classical bulges are assumed to be the more spherically symmetric, more concentrated (small effective radii), older, less metal enriched, and fully velocity dispersion supported components. 
This kind of galaxy substructure is thought to form from gas that suffered significant merger-induced dissipation and from accreted stars \citep{Katz:1992}, mostly in the early stages of galaxy formation \citep[e.g.][]{Obreja:2013}. Pseudo or disky bulges, on the other hand, are made of younger and more metal enriched stellar populations, are less concentrated, and can have significant coherent rotation \citep[e.g.][]{Fisher:2011}. In observational studies, various formation scenarios have been proposed for pseudo-bulges, most of them falling under the umbrella of secular evolution \citep[internal dynamical processes of galaxies, e.g.][]{Samland:2003,Kormendy:2004,Okamoto:2013,Laurikainen:2014}. Simulations revealed that boxy-peanut bulges are yet another class of bulges, present in galaxies that host bars \citep[][]{Athanassoula2002,Athanassoula:2005b,Fragkoudi:2017}. 
A different type of dispersion supported components are the stellar halos, covering large areas around galaxies with very low surface brightness \citep[e.g.][]{MartinezDelgado:2010,Iodice:2016,Trujillo:2016}. Stellar halos encode precious information on the build-up of galaxies, as they preserve fossil tracers of past mergers \citep[][]{Font2012, Tissera2013, Tissera2014, Monachesi2019}.       

The other major type of galaxy component are the disks. Some galaxies host one, while others host multiple disks. Their unifying property is the larger extent and lower surface brightness, $\Sigma$, when compared to bulges. The $\Sigma$ radial profiles in most individual disks of observed galaxies can be described by declining exponential profiles \citep{Freeman1970} or broken exponentials \citep{Erwin:2008, Breda2020, Varela2021}. Among disks, two separate classes stand out: thin and thick ones \citep[e.g.][]{Gilmore,Dalcanton,Seth,Comeron:2011}.
Apart from their thickness, the two classes of disks are differentiated by their age, metallicities, and rotational support: thin disks are younger, more rich in heavy elements, have higher rotational velocities and smaller vertical velocity dispersions than thick disks. Also, for the Milky Way it is known that the thick disk has enhanced $[\alpha/\rm Fe]$ abundance compared to thin disk populations of similar metallicities \citep[][]{Edvardsson1993, Prochaska2000, Reddy2003, Reddy2006}, suggesting shorter star formation time-scales. Many of the observations supporting the split in properties between thin and thick disks are from within our own Galaxy \citep[e.g.][]{Gilmore:1989, Fuhrmann:1998, Bensby:2003, Soubiran:2003, Reddy:2006, Ivezic:2012}.

While thin stellar disks form within the very thin cold gas disks of galaxies, there is still considerable debate on how thick disks originate. Three main scenarios have been proposed for thick disk formation. One possibility is that this type of disks form  thick from cold gas disks with high vertical velocity dispersions in the early universe \citep[e.g.][]{Brook2004}. This scenario is supported by observations of high redshift galaxies, which have larger gas vertical velocity dispersions than their local universe counterparts \citep[e.g.][]{ForsterSchreiber:2006}.  
However, not all high redshift disks are thick, as shown by recent studies that found some very highly rotational dominated galaxies \citep{Rizzo2020, Neeleman2020, Fraternali2021, Lelli2021}.
Another possibility is that thick disk formation is entirely driven by secular evolution: stars of a thin disk can be heated up by internal dynamical processes of galaxies like radial migration, caused by the interactions of stars with the time-dependent gravitational potential of transient features like bars and spiral arms \citep[e.g.][]{Dehnen:2000,Sellwood:2002,Schonrich:2012,Roskar:2008,Roskar:2013}. In this second scenario, all disks form thin, but some parts of a thin disk can evolve into a thick one. A third possibility is that stellar disks increase their vertical heating by interactions with satellites \citep[e.g.][]{Kazantzidis:2008,Tissera2012,Gomez2017, Bignone2019,Grand2020}. 

Though the fraction of mass in each of these types of observed galaxy components varies along the Hubble sequence, there are many objects that can simultaneously host all of them \citep[e.g.][]{Gadotti:2009,Erwin:2015,MendezAbreu:2017}. Also, even if there is a fair degree of agreement on the formation paths for some of these components (classical bulges), many open questions remain for others (pseudo-bulges, thick disks). Robust identification of stellar substructure is, therefore, an essential step towards understanding how galaxies have been assembled in terms of the formation of each of their separate components.     
As first proposed by \cite{Sersic1968}, the traditional photometric decomposition separates galaxy into bulges and disks, using combinations of S{\`e}rsic and exponential profiles.
Another option, which became widely feasible only recently, is to characterise galaxy components by means of stellar kinematic maps obtained from Integral Field Spectroscopy (IFS) data. A motivation to work with kinematic decomposition is the connection between the orbits of stars (and thus, their dynamics) and their history: stars in different orbit types are usually linked to different formation processes \citep[e.g.][]{Brook2004,Bird2013,Stinson2013}. 

One example of kinematically-motivated decomposition in observations was applied by \cite{Zhu2017},  \cite{Zhu2018a}, and  \cite{Zhu2018b} to 260 isolated (and not biased by dust lanes)  galaxies from the CALIFA survey \citep{IFUCalifa}. In these works four galaxy components have been distinguished by the circularity\footnote{Ratio between the azimuthal angular momentum of a star (particle) and the angular momentum associated with a circular orbit having the same binding energy as the star (particle).} of their stellar orbits: a cold one (with strong rotation), a hot component (dominated by random motion), a warm component (in between the cold and hot components) and a counter-rotating component. Additionally, \cite{Zhu2018b} also studied the mass fraction of each component in terms of morphological types, and found that the cold component fraction decreases from Sb to Sa, to S0 to elliptical galaxies (while the hot component increases). 
\cite{Zhu2018b} also found that the properties (including mass fractions) of disks (bulges) and the cold (hot) components are positively correlated, No clear correlation was found between the warm component and any "labelled" structure (disk, bulge or bar), but it can contribute to any of them depending on the morphology.
A similar analysis has been  performed in \cite{Zhu3}, but taking into account  the age and metallicity as well, obtaining a cold component  identifiable with young, metal-rich disks and a hot component composed of older stars.

Galaxy simulations are the most comprehensive tool to discern among various formation paths proposed for galaxy components.
In the last decade, simulations have been able to reproduce the stellar masses, sizes and shapes of observed galaxies, as well as to obtain realistic bulge-to-disk ratios and angular momentum contents \citep[][]{Agertz2011, Rosa1998, NIHAO, Marinacci2014, Pedrosa2015, Buck_2019}, thanks to the inclusion of supernovae and active galactic nuclei feedback processes, more realistic interstellar medium models, and ever increasing resolutions \citep[e.g.][]{Stinson:2006,Stinson:2013b,DallaVecchia:2008,Richings:2016,AnglesAlcazar:2017,Hopkins:2018}. 

Dynamical decomposition methods can be easily applied to simulations, where the complete phase space of the stellar particles is known (masses, positions, velocities, stellar population properties). 
Once the dynamical features on which to perform the decomposition are chosen, there are two possible paths to classify stellar particles into different components. In the first case, the values of the features that characterise each type component are chosen, and then stellar particles are classified in terms of that criteria. This means that component labels are imposed a priory, and the decomposition has to be validated on other characteristics like stellar population properties. This separation criteria can be motivated by the distribution shape of different parameters \citep[e.g. circularity, as introduced by][]{Abadi2003} or by assuming certain physically motivated thresholds for the binding energy and the angular momentum content \citep[][]{Tissera2012,Pedrosa2015}. 
The second option is to first find coherent components in an input feature space (for example, via clustering algorithms), and label them subsequently based on their characteristics. This approach was followed in \cite{Domenech2012, Obreja2016, Obreja2018, Obreja2019,Du2019}. Both kinds of decomposition methods have been applied to large volume hydrodynamical and zoom-in cosmological simulations, mostly using (specific) angular momentum and binding energy to distinguish between the different components. 

In this work, we use the new version of \texttt{galactic structure finder} \citep{Obreja2018} to analyse a large sample of galaxies in the MW-mass range identified in the EAGLE simulation \citep{Crain, Schaye}. This code uses Gaussian Mixture Models in a three dimensional dynamical space of normalised angular momentum -- normalised binding energy to define stellar galaxy components in simulated galaxies, and it is based on the previous work of \cite{Domenech2012} who used k-means as clustering algorithm. The updated version of the code \texttt{gsf2}\footnote{available upon request from the second author.}, which we use in this work, implements an automatic method to decide on the optimal number of components in a galaxy. The application of \texttt{gsf} to simulations from the NIHAO Project \citep[][]{NIHAO}, revealed eight types of galactic components from thin and thick disks to classical and pseudo bulges, and stellar halos with properties in agreement with observational expectations \citep[][]{Obreja2019}. Our aim in this work is to quantify to what extent the dynamical properties of components of MW-mass galaxies at $z=0$ are a proxy for their stellar ages, metallicities and $\alpha$-enhancement, and to look for relations between observables and properties that can only be inferred through modelling. 
This work is organised as follows. Section~\ref{sec:sample} introduces the simulated galaxy sample. Section~\ref{sec:deco} describes the dynamical decomposition method and shows the example of one galaxy. Section~\ref{sec:results} presents our results for the full sample, starting from the classification into generic galaxy components (Section~\ref{generic_components}), to their observational properties (Sections~\ref{sec:formas}, to \ref{pop_properties}), the correlations between their different properties (Section~\ref{sec:correlations}), and their relations to the stellar and the dark matter halo mass (Section~\ref{sec:massdep}). Finally, our summary and discussion of future work is presented in Section~\ref{sec:summary}.

%%%%%%%%%%%%%%%%%%%%%%%%%%%%%%%%%%%%%%%%%%%%%%%%%%%%%%%%%%%%%%%%%%%%%%%%%%%%%%%%%%%%%%%%%%%%%%%%%%%%%%%%%%%%%%%%%%%%%%%%%%%%%%%%%%%%%%%%%%%%%%%%%%%%%%%%%%%%%%%%%%%%%%%%%%%%%%%%%%%%
\section{Sample of simulated galaxies}

We extracted our galaxy sample from the 100 Mpc sized box L100N1504 of the EAGLE Project \citep[][details in \autoref{tablaeagle}]{Crain, Schaye}. This experiment is part of a suite of hydrodynamical cosmological simulations that follow the formation of structure of a representative volume of the universe. They are consistent with a $\Lambda$ Cold Dark Matter ($\Lambda$CDM) cosmology. The different physical processes included in these simulations are: radiative heating and cooling \citep{Wiersma}, stochastic star formation \citep{Schaye2008}, stochastic stellar feedback \citep{DallaVecchia2012} and AGN feedback implemented in a similar way to the feedback from star formation \citep{Yetli2015}. The energy injected into the gas by supernovae explosions depends on the local gas metallicity and density. The initial mass function is Chabrier-type \citep[][]{Chabrier:2003}.
EAGLE adopts the cosmological parameters of \cite{Planck2016}, with: $ \Omega_m = 0.307$, $\Omega_\lambda = 0.693$, $\Omega_b = 0.04825$, $H_0 = 67.77 \text{h km s}^{-1}  \rm Mpc^{-1}$. %, with $h = 0.6777$.
The L100N1504 box is resolved by $1504^3$ dark matter particles and the same initial number of gas particles, and the maximum gravitational softening is of $0.7$ kpc.
The halo catalogue was constructed using a Friends-of-Friends algorithm, while the subhalos were identified with \texttt{SUBFIND} \citep{Springel2001}.

\begin{table}
\centering
\begin{tabular}{ll}
\hline
Feature & Value \\ \hline
Box size & 100 cMpc\\
Number of particles (per type) & 1504$^3$ \\
Gas particle mass & $1.81 \times 10^6M_\odot$ \\
DM particle mass & $9.70 \times 10^6M_\odot$ \\
Maximum gravitational softening & 0.7 kpc\footnote{When we use kpc, we refer to proper kpc (pkpc).}\\
\hline
\end{tabular}
\caption{Main parameters of the L100N1504 box of the \textsc{EAGLE} simulation.}
\label{tablaeagle}
\end{table}

\label{sec:sample}

The subsample of this study was chosen imposing two criteria on central galaxies (no satellites were considered):

i) Stellar mass range: 10.4 $<$ log($M_\text{*}$(in 30kpc) [$M_\odot$]) $<$ 11.2

ii) Rotation dominated galaxies: $K_{\text{rot}}>0.4$ \citep{Sales2010}, with:
\begin{equation}
K_{\text{rot}}=\frac{\sum_i m_i (v_{i,\phi}^2/v_i^2)}{M_*},
\end{equation}
where $m_i$ is the mass of each stellar particle, M$_*$ is the total stellar mass of the galaxy, and $v_{i\phi}$ and $v_i$ are the tangential velocity in the plane of the disk, and the modulus of the velocity of the $i$-th stellar particle, respectively.

From this sample of galaxies, we further excluded the barred ones following \cite{Yetli2020, Yetli2022}, as it is not yet clear to what extent \texttt{gsf} can cleanly separate bars. The threshold for the presence of a bar was set at $A_2 >$ 0.3, where $A_2$ is the ratio of the second and zeroth terms of the Fourier expansion of the galaxy stellar surface mass density \citep{Athanassoula2002}. With these three criteria, our final sample contains 464 galaxies.

\begin{figure}
\centering
\includegraphics[width=0.45\textwidth]{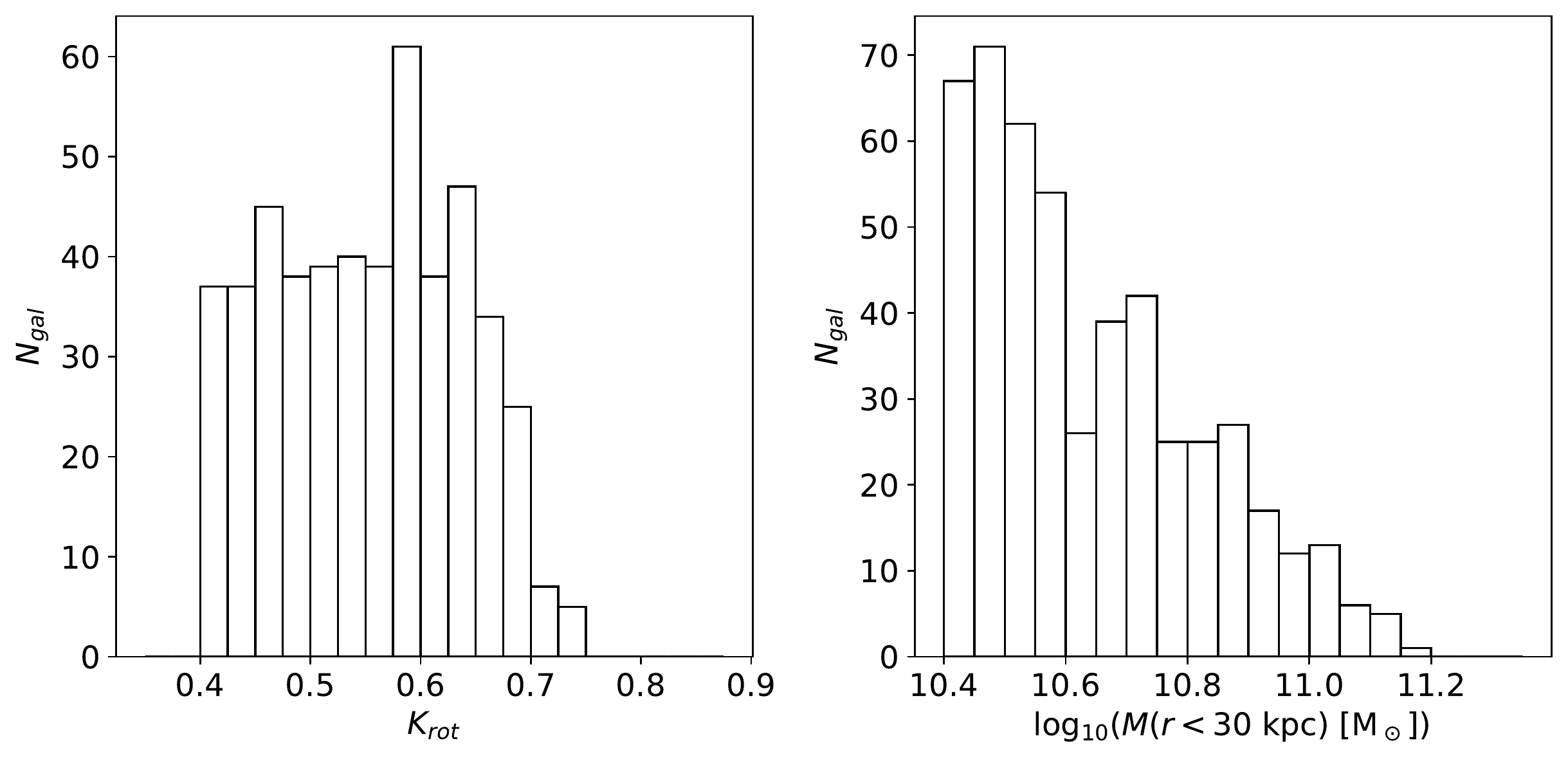}
\caption{The distribution of galaxies in the sample as a function of $K_{\rm rot}$ (left panel) and total stellar mass within 30~kpc (right panel). }
\label{fig:sample_hist}
\end{figure}

\section{DYNAMICAL DECOMPOSITION OF GALAXIES}
\label{sec:deco}
\subsection{Overview}
\label{sec:methods}

Simulations provide the whole phase space information of their stellar particles.
Therefore, various selection criteria (physically motivated thresholds) or clustering methods (e.g. agglomerative clustering, spectral clustering, Gaussian Mixtures, DBSCAN) can be used on various stellar dynamical properties (e.g. angular momentum, energy, action-angle variables) to divide a simulated galaxy into various components.

One of the first methods to do such a decomposition was proposed by \cite{Abadi2003}. Practically, for each stellar particle ``$i$" in a given galaxy, these authors used the binding energy ($E_i$) and the angular momentum in the direction $z$ of the galaxy spin ($J_{i,z}$) to define the $i$-th particle \textit{circularity}:
\begin{equation}
\varepsilon_i = \frac{J_{i,z}}{J_{c}(E_i)},
\label{eq:epsilonmal}
\end{equation}
where $J_c(E_i)$ is the magnitude of the angular momentum of circular orbit (maximum) with the same energy as the $i$-th particle. Thus, stars in circular orbits would have $\varepsilon_i \sim$ 1. 
\cite{Abadi2003} then assumed that the circularity distribution of spheroids, comprising both the central bulge and the stellar halo, is symmetric and centred on $\varepsilon_i = 0$, and that the sharp peak at $\varepsilon_i \sim 1$ is the thin disk. The particles left unassigned were assigned to a thick disk component.   

A third parameter was added by \cite{Domenech2012} to the kinematic space formed by $\epsilon_i$ and $e_i$, where $e_i$ is the specific binding energy, the $i$-th particle \textit{planarity}:
\begin{equation}
(j_p/j_c)_i = \frac{j_{i,p}}{j_{c}(e_i)},
\end{equation}
where $j_{i,p}$ is the projected specific angular momentum of each stellar/gas particle on the plane of the disk (perpendicular to $z$). A particle moving on a planar (vs polar) orbit will have $j_p/j_c = 0$ (vs $j_p/j_c = 1$). 
The circularity $\varepsilon_i$ can also be written in terms of specific angular momentum and specific binding energy:
\begin{equation}
\varepsilon_i = (j_z/j_c)_i = \frac{j_{i,z}}{j_{c}(e_i)},
\label{eq:epsilon}
\end{equation}
Aside of using specific angular momenta and energies, $e_i$ can also be normalised to the modulus of its maximum possible value in the system, $|e|_{max}$ \citep[see also][]{Tissera2012}. 

Other possible choices for the decomposition parameter space include the rotational kinetic energy, the ratio between rotation or dispersion velocities, or the binding energy. The application of some of these methods to EAGLE galaxies, including fixed cuts in the ($\varepsilon_i, e_i$) parameter space, is discussed in \cite{Thob2019}.

The classification methods range from assuming the shapes of the distributions on the parameter space ($\epsilon_i$), as was already mentioned for \cite{Abadi2003}, to choosing intervals on the kinematic space ($\varepsilon_i, e_i$) \citep{Tissera2012} or, more recently, the application of machine learning tools such as clustering algorithms.
Within the latter, it is worth mentioning the "kk-means" algorithm \citep{Schol,Karat,Dhillon}, used in \cite{Domenech2012}, substituted by Gaussian Mixtures in more recent works \citep{Obreja2016,Obreja2018,Obreja2019,Du2019}. 

\subsection{Previous decompositions of EAGLE galaxies}

\cite{Thob2019} performed a comparison between the most used kinematic diagnostics using galaxies from the EAGLE simulation. They characterise the galaxy morphology using the shape parameters, the flattening and the triaxiality. For the kinematic properties, they calculate and compare the disk to total fraction $D/T$, the rotational kinetic energy \citep[][]{Correa2017}, the spin parameter \citep[][]{Lagos2018}, the circularity parameter \citep[][]{Abadi2003} and the ratio of rotation to dispersion velocities \citep[][]{vandesande2017}. \citeauthor{Thob2019} found the different estimators to be consistent, but with relations between some estimators having significant scatter.

From a more observational kind of approach, in \cite{Irodotou2021} the orbital plane of the stellar particles is defined through their angular separation in the angular momentum projected plane. Using the HEALPix packet \citep{Gorski:2005}, a pixelisation of the angular momentum map is generated resulting in a Mollweide type distribution. In this projection particles with an angular separation lower/greater than 30 degrees are then assigned to the disk/spheroid component. In this manner, they are able also to defined a counter-rotating component. 
Through this kind of angular separation, \cite{Irodotou2021} distinguish rotation supported structures from dispersion supported ones, and find the properties of these components comparable to the ones obtained with more traditional decomposition methods, such as those considered by \citeauthor{Thob2019}.

\subsection{\texttt{GSF} method}
\label{sec:GSFbasics}
\texttt{GSF} is based on a descriptive method. It first finds coherent mathematically-identified components \citep[i.e. looking for clusters in a three dimensional kinematic space, as introduced by][]{Obreja2016, Obreja2018}, and subsequently checks if they correspond with classical definitions of bulges, disks, etc. based on their kinematic or observational-like properties (instead of defining them a priori).

\texttt{GSF} uses \textit{Gaussian Mixture Models} (GMM), as implemented in the Python package for Machine Learning \texttt{scikit-learn} \citep{scikit} to find the clusters within the input data space ($j_z/j_c$, $j_p/j_c$, $e/|e_{\rm max}|$). GMM is an iterative probabilistic method (i.e. assigns to each particle a probability of belonging to a cluster), which assumes that the distribution of data points in an $n$-dimensional space can be modelled as the sum of $n_k$ $n$-dimensional Gaussian distributions, where $n_k$ is the number of clusters. These Gaussians can have different covariance structures, and the ``distance" between each point and the centre of the Gaussians (their means) is given by a metric (Mahalanobis distance). 
Further information and a complete description of the algorithm is found in \cite{Obreja2018} and \cite{Obreja2019}.

%%%%%%%%%%%%%%%%%%%%%%%%%%%%%%%%%%%%%%%%
\begin{figure}
\centering
    \centering

        \includegraphics[width=0.45\textwidth]{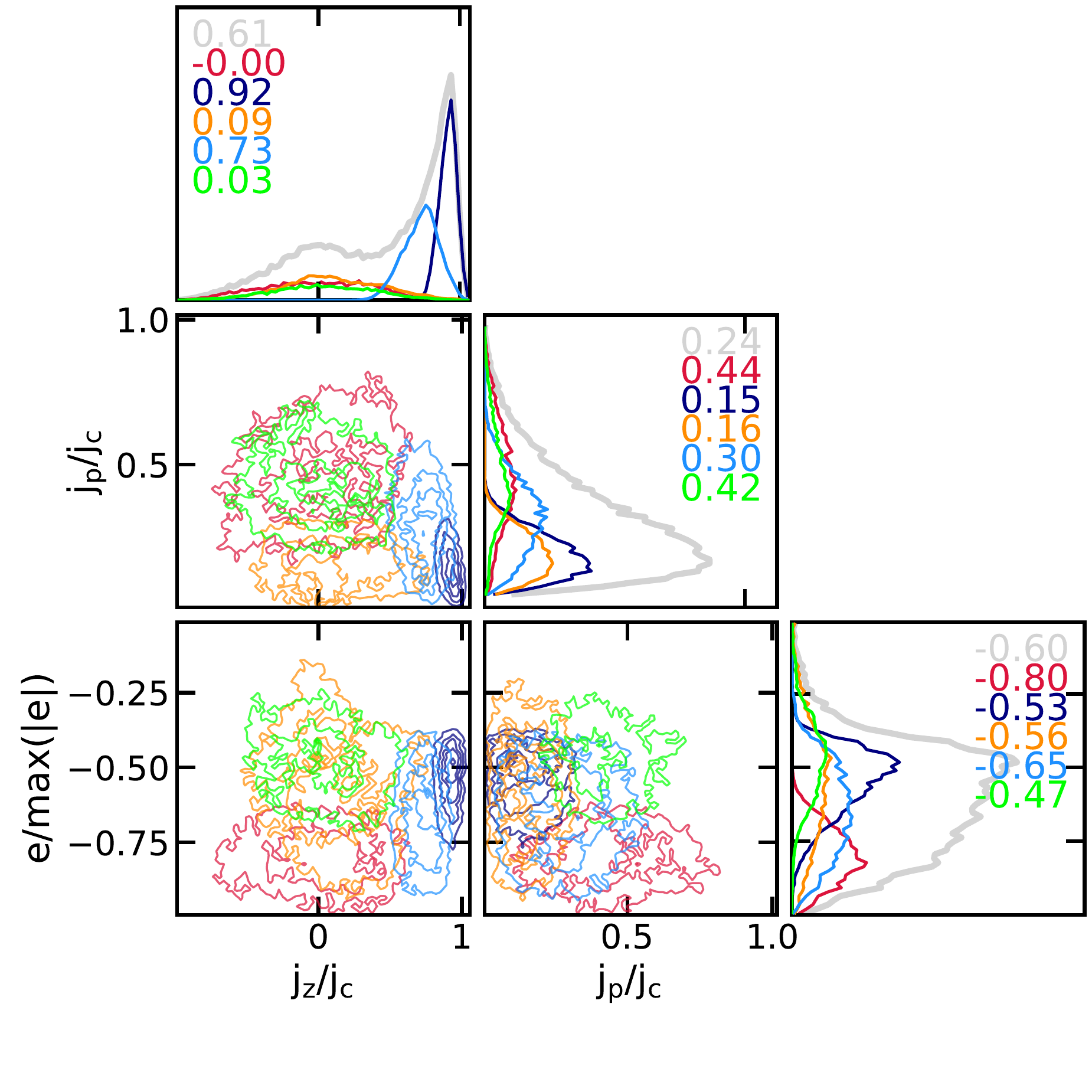}

\caption{Example of \texttt{gsf}-clustering in the ($j_z/j_c$, $j_p/j_c$, $e/|e_\text{max}|$) space for one of the galaxies for which the best model has $n_k=5$. The three diagrams in the bottom-left corner show the projected 2D histograms for each component, ordered by their mass-weighted mean $j_z/j_c$. Numbers on the top of the histograms represent the median values of each parameter for the different components (red, lime, orange, light blue and indigo) and the original distribution (grey).}
\label{fig:complete-2nk}
\end{figure}
%%%%%%%%%%%%%%%%%%%%%%%%%%%%%%%%%%%%

%%%%%%%%%%%%%%%%%%%%%%%%%%%%%%%%%%%%%%%%
\begin{figure*}
\centering
\includegraphics[width=1\textwidth]{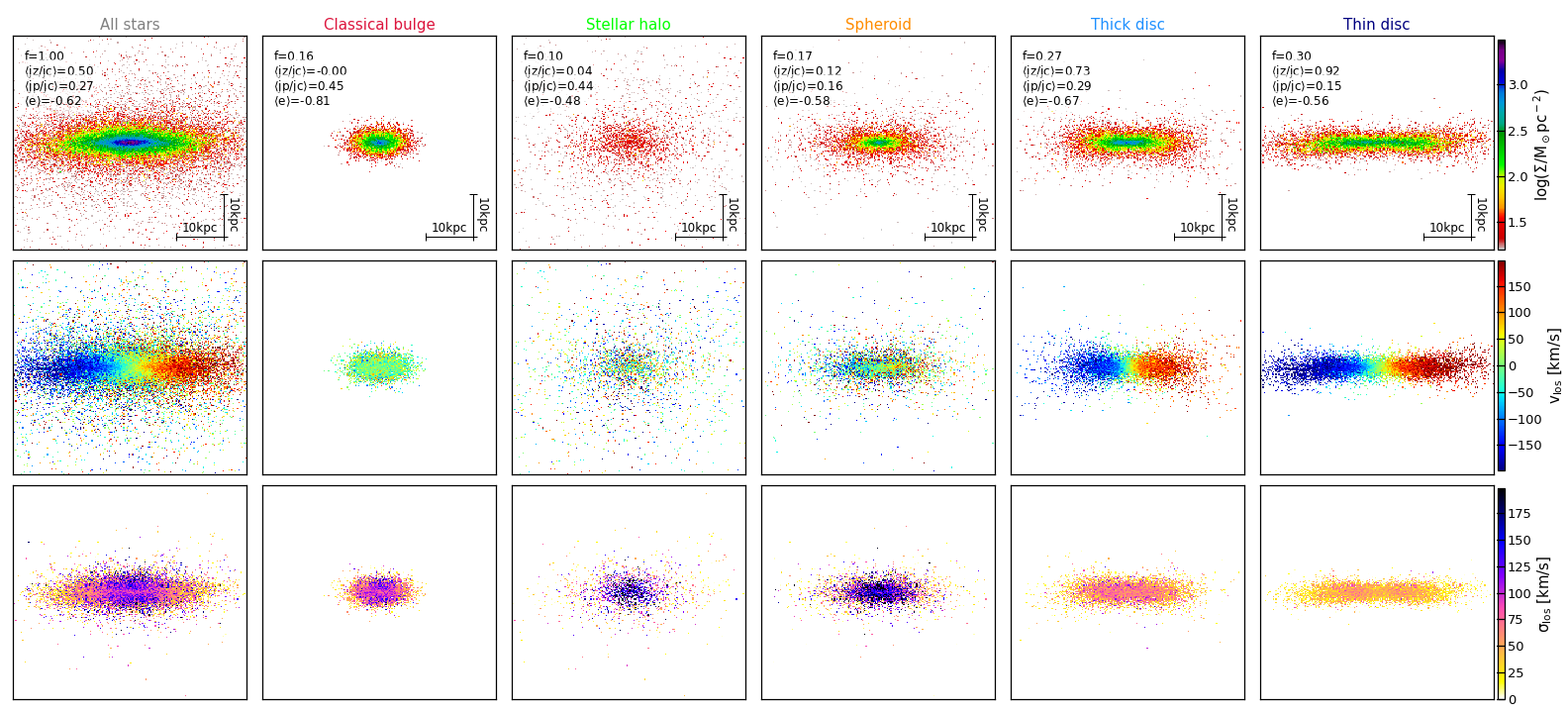}
\caption{Edge on projections for the components of the galaxy in Figure~\ref{fig:complete-2nk}. The left-most column shows the complete galaxy. 
The text in the upper left corner of each panels give the corresponding mass fraction ($f$), and mass-weighted means of $j_z/j_c$, $j_p/j_c$ and $e/|e|_{\rm max}$. The top row shows the surface mass densities, while the first and second order velocity moment maps are given in the central and bottom rows, respectively. The component labels (column titles), assigned after inspecting visually these maps, are in agreement with the results of the clustering.}
\label{fig:complete-2nk-moments}
\end{figure*}
%%%%%%%%%%%%%%%%%%%%%%%%%%%%%%%%%%%%

In its first version\footnote{available online at \url{https://github.com/aobr/gsf}}, the code only provides the decomposition for a required number of components $n_k$, and it is up to the user to decide which $n_k$ model makes more physical sense. Here, we use a new version of the code (\texttt{gsf2}, Obreja et al. in prep), which besides allowing for the freedom of choosing from a wide variety of input spaces, ranks the models according to the CHull statistics \citep[][]{Ceulemans2006}.
This statistics is agnostic -- it can be used even when it is not possible to define a likelihood function --, and for the particular case of mixture models it outperforms the more widely used Akaike Information Criteria \citep[AIC][]{Akaike:1974} and Bayes Information Criteria \citep[BIC][]{Schwarz:1978}, especially in complex problems \citep[e.g. many clusters, significant cluster overlap, noise, as discussed by][]{Bulteel2013}. 
The \texttt{gsf} method complemented with a heuristic BIC was used by \cite{Du2019} on their analysis of TNG100's galaxies \citep[from the IllustrisTNG simulations suite,][]{Pillepich2018, Nelson2018, Springel2018, Naiman2018, Marinacci2018}. 

To use the CHull statistics, one needs to chose a measure of the goodness of the fit and a measure for the model complexity. \texttt{gsf2} uses the log-likelihood ln($L$) to measure the goodness of fit, and the number of free parameters $n_{\rm param}$ for the model complexity, the latter depending on the type of covariance assumed for the GMM. In this paper we use a 'full' or unconstrained covariance. The name of the statistics comes from the step in which the convex hull in the plane goodness of fit vs model complexity ($n_{\rm param}$--ln($L$) in our case) is constructed, and all points not belonging to the convex hull are discarded. On the remaining points/models, one can compute the scree test statistics ($st$) \citep[][]{Cattell1966}, which quantifies whether
the increase in complexity between adjacent models is justified by the increase in the goodness of fit, and rank the models in decreasing order of $st$. Using the CHull statistics, \texttt{gsf2} now is able to automatically decompose large samples of galaxies, choosing for each object the most likely model\footnote{Each galaxy will have its own optimal $n_k$, but the code saves all the ranked models such that the user can also easily check the e.g. second or third best ones.}.

We applied \texttt{gsf2} to our sample of EAGLE galaxies, for each object varying $n_k$ from 1 to 15. For most of the sample (65\% of the galaxies), the CHull returned the $n_k=2$ model as the best model. However, from previous analysis we know that most galaxies host more than two components. The reason for which the code tends to rank the $n_k=2$ model as best is likely to be related to the resolution of these simulations.
The code is capable of separating even merger remnants (Obreja et al. in prep), but how well the fainter components of galaxies are identified naturally depends on the resolution.

An example of this type of decomposition is shown in Figure~\ref{fig:complete-2nk} for a galaxy with $n_k=5$ selected as the most likely model. The off-diagonal panels give the 2D iso-probability densities for each of the $n_k$ clusters in the $j_p/j_c$ vs $j_z/j_c$ (top left), $e/|e_{\text{max}}|$ vs $j_z/j_c$ (bottom left), and $e/|e_{\text{max}}|$ vs $j_p/j_c$ (bottom right) variables. The diagonal panels give the 1D distributions for the different parameters for the whole galaxy (grey) and for each of the components. 
The two components shown in indigo and light blue are clearly the dynamically cold (thin) and warm (thick) disks, respectively: they have large $j_z/j_c$ and low $j_p/j_c$. The thin disk (indigo) in particular is the component with the largest $j_z/j_c$ and lowest $j_p/j_c$, and is more loosely bound than the thick one. The other three components are dynamically hot (circularities centred closed to 0), and can be distinguished among themselves by planarity and binding energy. The red component is the most bound ($e/|e_\text{max}|$ close to -1) and has the largest $j_p/j_c$, so we nickname it the (classical) bulge. The orange one has the smallest planarity among the three, hence is the least spherically symmetric, and has a very extended distribution in binding energies, and therefore we can call it a spheroid. The last component (lime colour) has large planarity and low binding energy, so we can consider it to be the stellar halo.  

Figure~\ref{fig:complete-2nk-moments}  shows the edge-on projections of the surface mass densities (top), line-of-sight velocities (center) and velocity dispersions (bottom), for all five components (second to sixth columns) and the complete galaxy (left most column). Text boxes on the upper-left corner indicate the mass fraction, ellipticity\footnote{Ellipticity $= \sqrt{1-(\text{Minor axis}/\text{Major axis})^2}$, modelling each component as a 2D ellipse.} and mass-weighted mean circularity of each component (or of the whole galaxy). Between Figures~\ref{fig:complete-2nk-moments} and \ref{fig:complete-2nk}, we can identify the dispersion dominated components as a stellar halo (disperse, radial extended) a classical spherically symmetric bulge, a more flattened but still round spheroid, and the rotation dominated components as the thin and the thick disks.

%\section{RESULTS. GALAXY SAMPLE STATISTICS}
\section{Properties of galaxy components}
\label{sec:results}

We have applied the method of Section~\ref{sec:GSFbasics} to all the galaxies in our sample.
As it was mentioned above, \texttt{gsf2} returns the ranking given by the scree test (for each galaxy) taking all the values of $n_k$ which are not discarded by the CHull statistics. In Table~\ref{tab:CHull} we show the ranking of the models according to the $st$ for our sample. Even if higher values of $n_k$ were taken into account to build the table, we only show the results for $2 \leq n_k \leq 6$ (thus, the sum of the firs-ranked component percentages does not add to 100\%). When considering up to 14 components per galaxy, we found that $\sim$6\% of the galaxies have the optimal $n_k>6$. For these galaxies, we choose instead the highest ranked model with $3 \leq n_k \leq 6$. We chose $n_k=6$ as the maximum accepted based on the following argument.  
A large optimal $n_k$ can arise when some of the mergers suffered by the galaxy did not mix completely with the host. In such cases, (part of) the merger remnants would be identified as separate components. Since here  we are just  interested in the major substructures in galaxies,
we do not consider the small, multiple components.

\begin{table}
\centering
\begin{tabular}{cc}
\hline
Model & Ranking\\
\hline
\hline 
& 65\% ranked 1st\\
$n_k=2$ & 74\% ranked 2nd or best\\
 & 81\% ranked 3rd or best\\
 \hline
 & 19\% ranked 1st\\
$n_k=3$ & 33\% ranked 2nd or best\\
 & 48\% ranked 3rd or best\\
\hline
 & 6\% ranked 1st\\
$n_k=4$ & 26\% ranked 2nd or best\\
 & 36\% ranked 3rd or best\\
\hline
 & 4\% ranked 1st\\
$n_k=5$ & 22\% ranked 2nd or best\\
 & 33\% ranked 3rd or best\\
\hline
 & 1\% ranked 1st\\
$n_k=6$ & 17\% ranked 2nd or best\\
 & 23\% ranked 3rd or best\\
\hline
\end{tabular}
\caption{Ranking of the models according to the CHull criteria for the number of components per galaxy considered in this work. There were some galaxies whose 1st-ranked model had $n_k > 6$ ($\sim6$\%). They are included in the analysis, but the chosen model for those galaxies would be the highest ranking one with $3 \leq n_k \leq 6$.}
\label{tab:CHull}
\end{table}

\begin{table}
\centering
\begin{tabular}{ccccc}
\hline
Model & $n_k=3$ & $n_k=4$ & $n_k=5$ & $n_k=6$\\
\hline 
Percentage & 38\% & 32\% & 20\% & 10\%\\
%central & 40\% & 29\% & 19\% & 12\%\\
\hline
\end{tabular}
\caption{Statistics (percentage of galaxies for each kind of model) of the final models considered for all galaxies.}
\label{tab:statistics}
\end{table}

Ideally, the best method to choose the optimal number of components would combine a critical evaluation of the result of the scree test scores and visually-inspecting the components.  
However, when dealing with large samples, going through all the models for all the galaxies one-by-one is not feasible.
In this work, to select the optimal $n_k$ we use the $st$ ranking. If the difference between the two highest scores is larger than a chosen value, we take the highest as the optimal number of components. When the scree test does not return a clear diagnostic, we get back to visual inspection. This criteria reduced the number of galaxies that needed to be inspected to $\sim$15\% of the sample, which is a number that can be managed. 
For the visual inspection we use the same criteria as \cite{Obreja2019}: the components make sense, have a non-negligible mass fraction and are not repeated (in this case, we would be dealing with overfitting). The summary of the selected models is given in Table~\ref{tab:statistics}.

\def\bigimageb{\includegraphics[width=1\textwidth]{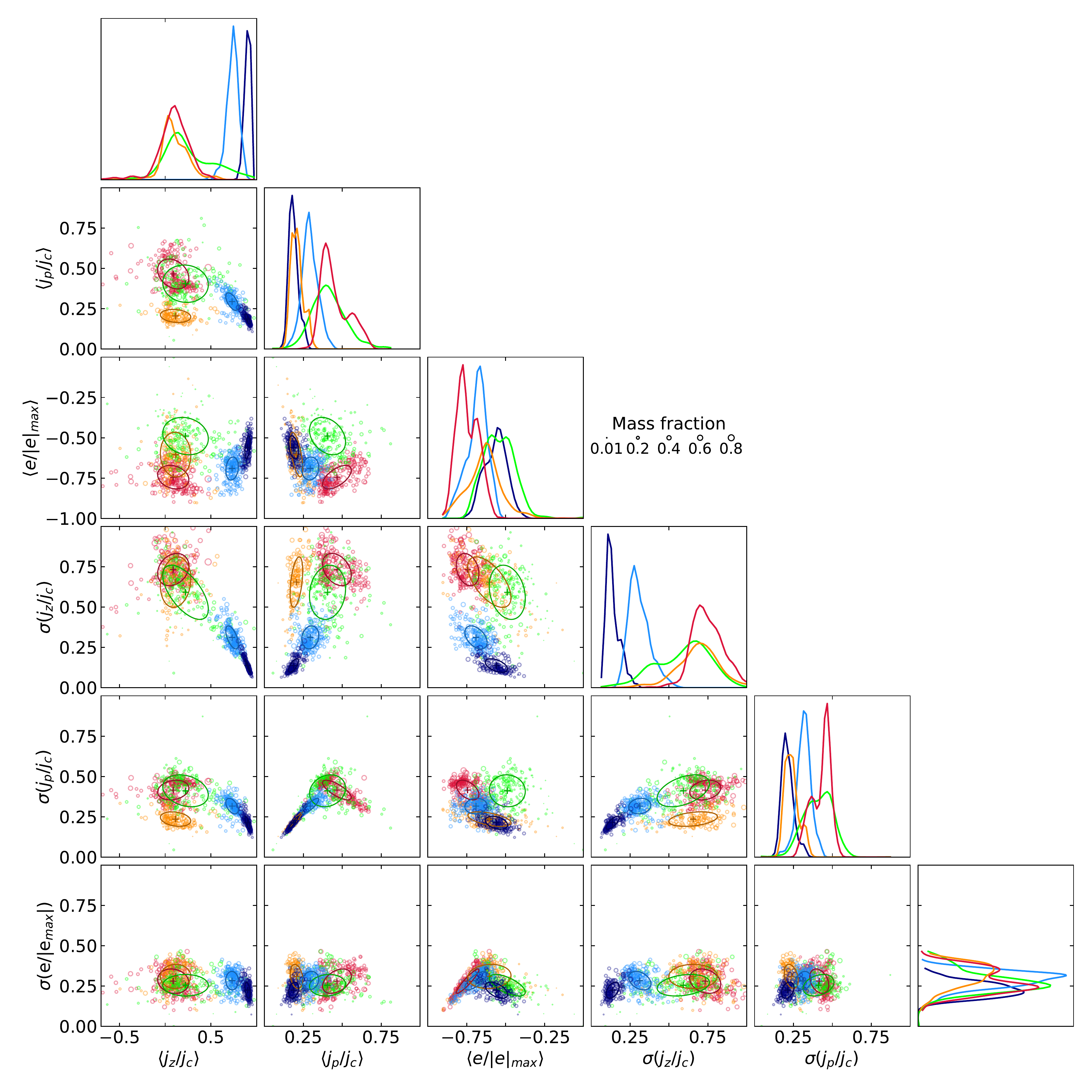}}
\def\littleimageb{\includegraphics[width=0.45\textwidth]{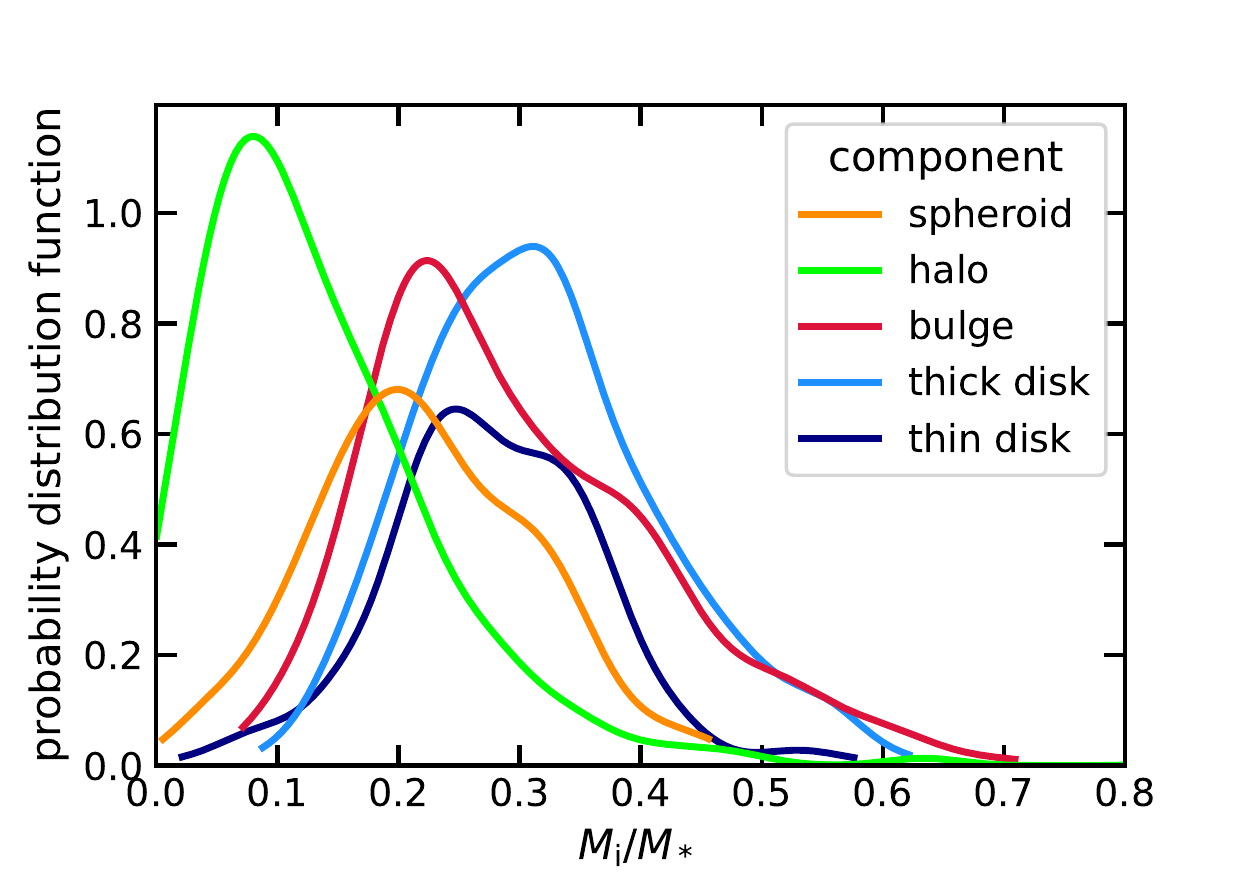}}

\setlength\fboxrule{0pt}
\newcommand{\figurebox}[5]{%
\begin{figure*}%
\fbox{%
\centering
\begin{minipage}[c]{0.98\textwidth}
\centering #1\par
    \stackinset{r}{3pt}{t}{3pt}{#2}{#3}
    \caption{#4}
    \label{#5}
\end{minipage}%
}%
\end{figure*}%
}%

\figurebox{}{\littleimageb}{\bigimageb}{Corner diagram showing the result of applying GMM on the 6-dimensional space 
($\langle j_z/j_c\rangle$, $\langle j_p/j_c\rangle$,$\langle e/|e|_{\text{max}}\rangle$, $\sigma(j_z/j_c)$, $\sigma(j_p/j_c)$, $\sigma(e/|e|_{\text{max}})$) 
representing all the components of all the galaxies selected by \texttt{gsf2} in the complete sample. Ellipses represent the 1$\sigma$ confidence regions (calculated using the 2D covariance matrix). The inset in the top right corner shows the PDFs of the mass fractions for the five types of `generic' components, the areas under the five curves summing to $1$.  }{corner-SP-galaxy-fit}

\subsection{How many types of components are in the sample?}
\label{generic_components}
Even though each galaxy in the sample has its own particular decomposition, 
individual components across the whole sample have some common features, e.g.: the thinner (dynamically colder) disks have strongly peaked distributions in $j_z/j_c$ close to $1$, or the most extended dynamically hot components have the binding energy peak at the lowest values ($e/|e|_{\rm max}$ closest to 0) with wide distributions. The presence of these common features means that we can in principle define some sort of `generic' galaxy substructures. We do such an exercise with the same method used to separate components in individual galaxies, namely GMM. In this case we use as input space for the clustering the mass-weighted means and dispersions of the three dynamical properties comprising the input space of \texttt{gsf}. Therefore, each component of each galaxy is represented by a point in the 6D space ($\langle j_z/j_c\rangle$, $\langle j_p/j_c\rangle$,$\langle e/|e|_{\rm max}\rangle$, $\sigma(j_z/j_c)$, $\sigma(j_p/j_c)$, $\sigma(e/|e|_{\rm max})$). Since it is not generally true that the distributions of the circularity, planarity and normalised binding energy for the various components are perfect 1D Gaussians, we define the dispersion $\sigma(x)$ of a variable $x$ as half the difference between the mass-weighted 84$^{\rm th}$ and 16$^{\rm th}$ percentiles of $x$. 

We run GMM on this 6D space with no constrain on the covariance for fixed number of clusters $n_k$ between 1 and 15. Similar to what happens in individual galaxies, the ln($L$), $BIC$ and $AIC$ as functions of the number of free parameters do not have clear extrema, but smooth knee-like behaviours. Therefore, we use the same CHull statistics as in Section~\ref{sec:GSFbasics}, which unambiguously ranks first the model with 5 components. Figure~\ref{corner-SP-galaxy-fit} shows the results of this optimal GMM model, each colour representing one `generic' galaxy component. The diagonal panels contain the 1D PDFs of the different clustering parameters (where the areas under the curves sum to 1), while the 2D projections appear on the lower corner. In this case, we do not represent the density curves, but the distributions are characterised using the projected points and the 1$\sigma$ confidence ellipses (calculated using the 2D covariance matrix) for each of the 5 components. These ellipses provide information about how clustered each `generic' component is (smaller size corresponds to more compact component definition) and how tight the various features correlate with each other within one type of substructure (narrow ellipses indicate that the properties correlate tightly). 
Each point in these panels represents one component of one galaxy, with a size proportional to its mass fraction within its host. 
Based on the values of the parameters assigned to each of the distributions, we identify the following components:
\begin{itemize}
    \item Thin disk (indigo): cold component, planar, low $\sigma(j_z/j_c)$.
    \item Thick disk (light blue): cold component, but with a lower $\langle j_z/j_c\rangle$ compared to the thin disk. Even if the differences on the binding energies of these components are not significant, the binding energies of thick disks are slightly larger than of thin ones.
    \item Halo (lime): dynamically hot component ($\langle j_z/j_c\rangle \sim 0$), loosely bound. 
    \item Spheroid (orange): hot component, but with planarities similar to the ones of the thick disk.
    \item Bulge (red): tightly bound hot component.
\end{itemize}
The mass fractions of these five types of components
are shown in the inset panel of Figure~\ref{corner-SP-galaxy-fit} using the same colour code for the different components as in the main figure. The areas under the five curves add to $1$. In general, the mass fractions of all the most bounded components are ``relevant" (i.e., $>$ 10\% of the total mass of the galaxy), which is consistent with the fact that we are not decomposing the galaxy into an excessive number of components.
This minimum mass constraint does not apply to halos due to their morphology: halos' stellar particles are clustered by their low binding energies, and their negligible (even counter-rotating) average rotational support. Since we are working with a sample dominated by disk galaxies, there are not many stellar particles far from the centre of the galaxy (i.e., with low binding energies) which do not ``live" on the plane of the disks or do not belong to the bulge or the spheroid.

In the following subsections we look in more detailed at the potentially observable properties of the five types of `generic' galaxy components. Our aim is two-fold, first to check if the choice of names makes physical sense and is in agreement with the literature on the topic, and second to quantify the observable properties of these components.

Most of these properties for a given component of a given galaxy (e.g. age, metallicity, $\alpha$-enhancement, rotational velocities) are characterised by distribution functions, while masses, mass fractions, axes ratios and half mass radii are single values.   
Therefore, for the first type of properties we quantify each component of each galaxy by the mass-weighted mean or median of their respective distributions. In general, both mass-weighted means and medians returned similar results, except in the case of [Fe/H] and [$\alpha$/Fe], where the medians are more robust given that these two properties are in a logarithmic scale. To quantify the spread of these distributions we use the differences between the mass-weighted 84$^{\rm th}$ and 50$^{\rm th}$, and between the 50$^{\rm th}$ and 16$^{\rm th}$ percentiles, and refer to them as inter-percentile ranges.

\begin{table*}
\begin{tabular}{cccccc}
\hline

Component & bulge& spheroid& halo& thick disk& thin disk\\ 

\hline

$\langle j_z/j_c\rangle$ & 
0.11$^{+0.11}_{-0.12}$ & 0.10$^{+0.13}_{-0.09}$ & 0.16$^{+0.36}_{-0.17}$ & 0.74$^{+0.07}_{-0.08}$ & 0.91$^{+0.02}_{-0.03}$\\ 

\rule{0pt}{3ex}$\sigma(j_z/j_c)$ & 
0.73$^{+0.09}_{-0.07}$ & 0.70$^{+0.12}_{-0.16}$ & 0.61$^{+0.14}_{-0.24}$ & 0.29$^{+0.08}_{-0.05}$ & 0.12$^{+0.04}_{-0.03}$\\ 

\rule{0pt}{3ex}$\langle j_p/j_c\rangle$ & 
0.44$^{+0.14}_{-0.05}$ & 0.20$^{+0.05}_{-0.03}$ & 0.39$^{+0.13}_{-0.09}$ & 0.29$^{+0.05}_{-0.05}$ & 0.18$^{+0.03}_{-0.03}$\\ 

\rule{0pt}{3ex}$\sigma(j_p/j_c)$ & 
0.45$^{+0.03}_{-0.11}$ & 0.23$^{+0.06}_{-0.03}$ & 0.40$^{+0.09}_{-0.10}$ & 0.32$^{+0.04}_{-0.05}$ & 0.20$^{+0.04}_{-0.03}$\\ 

\rule{0pt}{3ex}$\langle e/|e|_{\text{max}}\rangle$ & 
-0.77$^{+0.09}_{-0.05}$ & -0.63$^{+0.10}_{-0.11}$ & -0.52$^{+0.11}_{-0.11}$ & -0.68$^{+0.06}_{-0.08}$ & -0.55$^{+0.07}_{-0.08}$\\ 

\rule{0pt}{3ex}$\sigma(e/|e|_{\text{max}})$ & 
0.26$^{+0.10}_{-0.06}$ & 0.33$^{+0.07}_{-0.08}$ & 0.26$^{+0.06}_{-0.05}$ & 0.29$^{+0.05}_{-0.06}$ & 0.22$^{+0.05}_{-0.05}$\\ 

\hline

\rule{0pt}{3ex}$M_i/M$ & 
0.27$^{+0.14}_{-0.08}$ & 0.21$^{+0.10}_{-0.08}$ & 0.11$^{+0.10}_{-0.07}$ & 0.31$^{+0.11}_{-0.09}$ & 0.27$^{+0.08}_{-0.07}$\\ 

\rule{0pt}{3ex}log($M_i/M_{\rm\odot}$) & 
10.07$^{+0.29}_{-0.25}$ & 9.92$^{+0.39}_{-0.24}$ & 9.68$^{+0.40}_{-0.40}$ & 10.10$^{+0.26}_{-0.21}$ & 10.03$^{+0.27}_{-0.20}$\\ 

\rule{0pt}{3ex}$b/a$
& 0.94$^{+0.03}_{-0.09}$ & 0.93$^{+0.05}_{-0.15}$ & 0.93$^{+0.04}_{-0.11}$ & 0.96$^{+0.02}_{-0.04}$ & 0.96$^{+0.02}_{-0.04}$\\ 

\rule{0pt}{3ex}$c/a$ & 
0.70$^{+0.13}_{-0.11}$ & 0.58$^{+0.10}_{-0.10}$ & 0.69$^{+0.12}_{-0.19}$ & 0.36$^{+0.09}_{-0.08}$ & 0.20$^{+0.05}_{-0.04}$\\ 

\rule{0pt}{3ex}$r_{\rm 50}$ [kpc] & 
3.54$^{+1.76}_{-0.84}$ & 6.07$^{+3.85}_{-2.78}$ & 12.29$^{+6.62}_{-5.77}$ & 5.44$^{+2.76}_{-1.95}$ & 11.62$^{+4.39}_{-3.74}$\\ 

\rule{0pt}{3ex}$\sigma(z)$ [kpc] & 
3.56$^{+2.91}_{-1.08}$ & 5.04$^{+3.78}_{-2.29}$ & 11.59$^{+8.17}_{-5.99}$ & 2.62$^{+1.23}_{-0.87}$ & 3.00$^{+0.83}_{-0.69}$\\ 

\rule{0pt}{3ex}$v_{\rm\phi,50}$ [km~s$^{\rm -1}$] & 
27$^{+21}_{-28}$ & 
23$^{+35}_{-20}$ & 
53$^{+89}_{-55}$ & 
149$^{+35}_{-23}$ & 
188$^{+33}_{-22}$\\ 

\rule{0pt}{3ex}$\sigma(v_z)$ [km~s$^{\rm -1}$] &  
175$^{+64}_{-29}$ & 
133$^{+65}_{-33}$ & 
224$^{+82}_{-64}$ & 
104$^{+27}_{-24}$ & 
66$^{+14}_{-11}$\\ 

\rule{0pt}{3ex}$[\rm Fe/H]_{\rm 50\%}$ & 
-0.46$^{+0.09}_{-0.11}$ & -0.45$^{+0.14}_{-0.19}$ & -0.68$^{+0.30}_{-0.25}$ & -0.19$^{+0.12}_{-0.12}$ & -0.10$^{+0.07}_{-0.08}$\\ 

%\rule{0pt}{3ex}$\sigma([Fe/H]_{\rm 50\%})$ & 1.19$^{+0.13}_{-0.09}$ & 1.17$^{+0.11}_{-0.11}$ & 1.27$^{+0.19}_{-0.14}$ & 1.08$^{+0.11}_{-0.15}$ & 0.92$^{+0.06}_{-0.06}$\\ 

\rule{0pt}{3ex}$[\rm O/Fe]_{\rm 50\%}$ & 
0.25$^{+0.06}_{-0.05}$ & 0.23$^{+0.06}_{-0.06}$ & 0.26$^{+0.07}_{-0.07}$ & 0.15$^{+0.06}_{-0.04}$ & 0.10$^{+0.03}_{-0.03}$\\ 

%\rule{0pt}{3ex}$\sigma([O/Fe]_{\rm 50\%})$ & 0.49$^{+0.06}_{-0.05}$ & 0.49$^{+0.05}_{-0.06}$ & 0.54$^{+0.08}_{-0.08}$ & 0.47$^{+0.07}_{-0.05}$ & 0.42$^{+0.05}_{-0.03}$\\ 

%\rule{0pt}{3ex}$\textlangle age\textrangle$ [Gyr] & 0.66$^{+0.07}_{-0.09}$ & 0.62$^{+0.09}_{-0.08}$ & 0.65$^{+0.08}_{-0.11}$ & 0.52$^{+0.08}_{-0.11}$ & 0.36$^{+0.07}_{-0.07}$\\ 
\rule{0pt}{3ex}$\langle age\rangle$ [Gyr] & 
8.99$^{+1.01}_{-1.19}$ & 8.62$^{+0.96}_{-1.37}$ & 8.69$^{+1.29}_{-1.58}$ & 6.88$^{+1.28}_{-1.60}$ & 4.90$^{+0.99}_{-1.08}$\\ 

\rule{0pt}{3ex}$\sigma(age)$ [Gyr] & 
4.05$^{+1.47}_{-1.36}$ & 4.50$^{+1.54}_{-1.38}$ & 4.43$^{+1.70}_{-1.23}$ & 5.61$^{+1.35}_{-1.34}$ & 5.92$^{+0.98}_{-1.42}$\\ 

\hline\\
\end{tabular}
\caption{Summary of properties for the five types of `generic' galaxy components. %The values in the table are the percentiles of the distributions of properties in column 1.  
The central values are the 50$^{\rm th}$ percentiles, the values in the subscript give the range between the 50$^{\rm th}$ and 16$^{\rm th}$ percentiles, while the ones in the superscript the range between the 84$^{\rm th}$ and 50$^{\rm th}$ percentiles.}
\label{tab:bigtable}
\end{table*}

\subsection{Shapes, sizes and rotational velocities}
\label{sec:formas}

\begin{figure}
\includegraphics[width=0.45\textwidth]{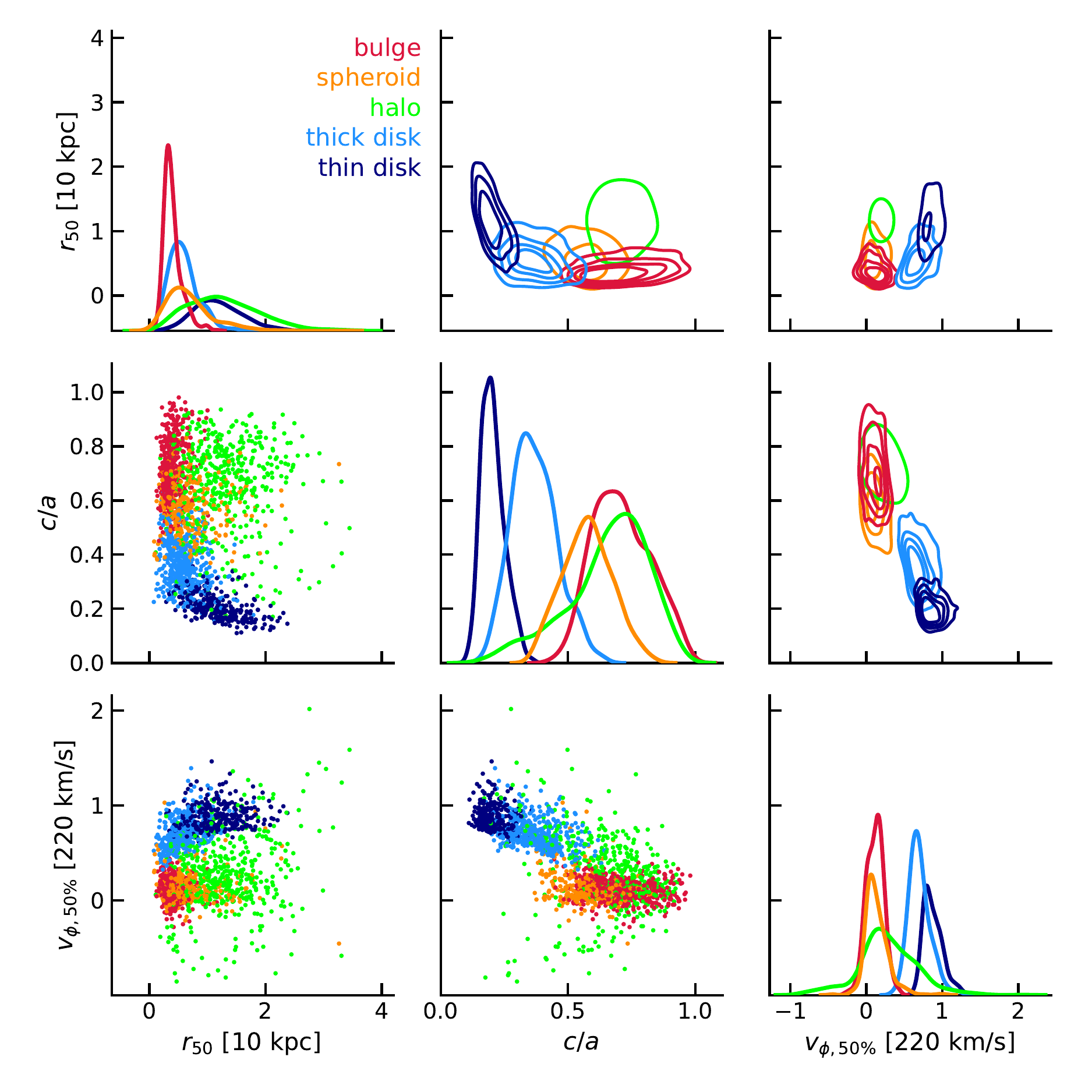}
\caption{Shapes (minor-to-major axis ratio from the inertia tensor $c/a$), sizes (3D half mass radii $r_{\rm 50}$) and median rotational velocities ($v_{\rm\phi,50\%}$) for the five types of components. Radii are in units of 10 kpc, while velocities are in units of 220~km~ s$^{\rm -1}$.}
\label{fig:shapessizes}
\end{figure}

\begin{figure}
\includegraphics[width=0.45\textwidth]{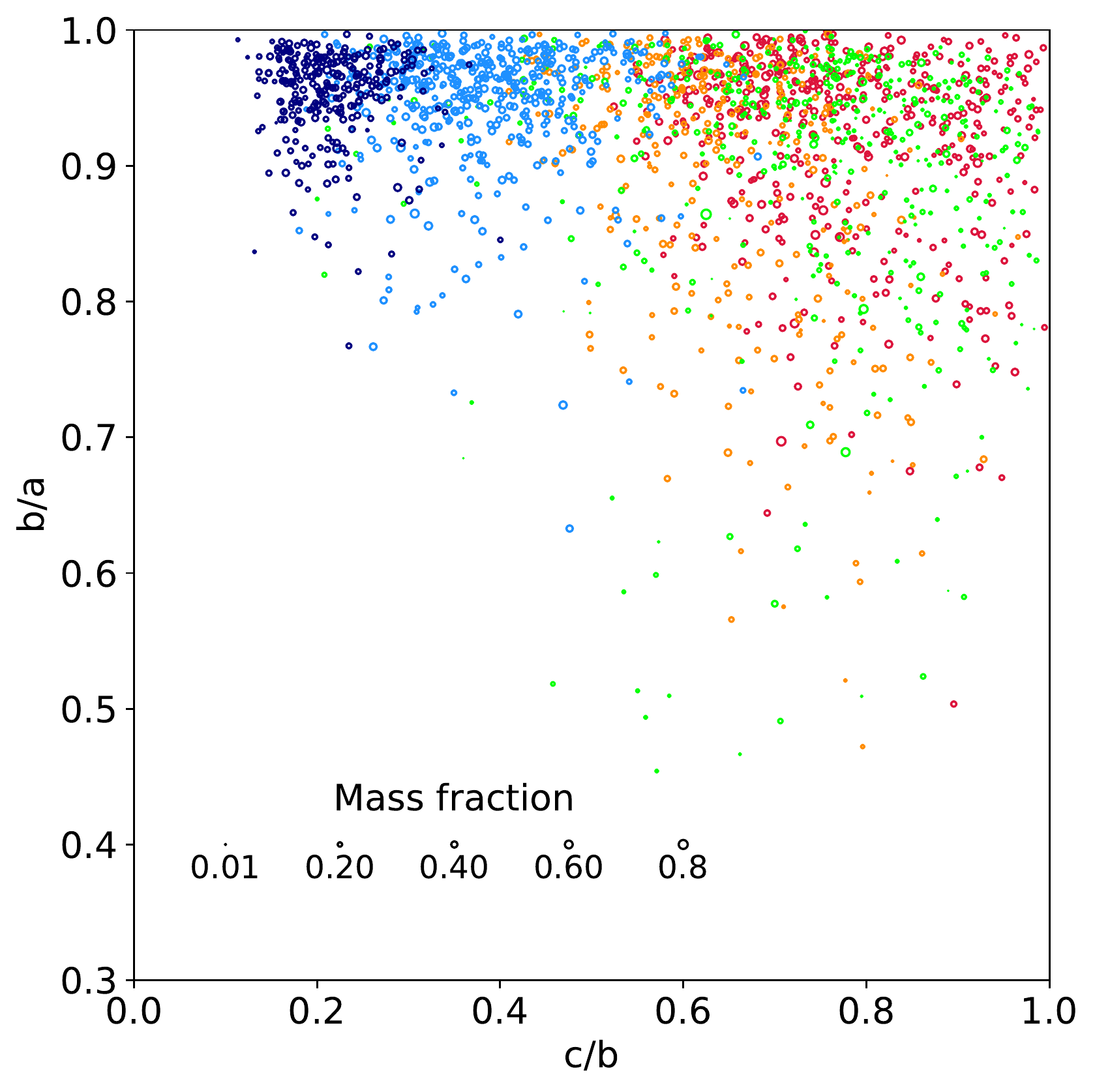}
\caption{Axis ratios diagram for the five types of components in the sample. The different colours correspond to the five types of generic components in Figure~\ref{corner-SP-galaxy-fit}. Disks are located on the upper-left corner ($c/a < b/a$), while spheroids are in the upper-right corner ($a = b = c$). The size of the points scales with the mass fraction in each component, as given in the legend.}
\label{fig:formas}
\end{figure}

Once we have checked that components are coherent in the input feature space for clustering, we can look at kinematic and morphological variables that have observational counterparts. Quantities such as the size, ellipticity and rotational velocity will be related to our clustering variables, but not fully determined by them. For example, one would expect that components with higher mean energies (i.e. least bound) would be more radially-extended, since stellar particles further away from the galaxy centre tend to have lower binding energies. 

Figure~\ref{fig:shapessizes} shows the 3D half mass radius $r_{\rm 50}$, the shape as quantified by the minor-to-major axis ratio of the inertia tensor $c/a$, and the median rotational velocity $v_{\phi\rm ,50}$. The 1D distributions are on the diagonal, while in the lower triangle we give the scatter plots, and in the upper triangle the Gaussian kernel density estimates (kde).  

Bulges are the most compact components (as measured by $r_{\rm 50}$), followed by spheroids and thick disks. The components with the largest $r_{50}$ are the disks and the stellar halos. If we look at the rotational velocities, there is a clear division between rotation-dominated (thin and thick disks) and dispersion-dominated ones (bulges, spheroids and stellar halos). 
We measure the shapes of individual galaxy components using the minor-to-major axes ratio $c/a$. Dispersion dominated components have higher values of $c/a$, meaning that they are closer to spherical symmetry. Disks, on the other hand, have $c/a$ much closer to $0$, particularly the thin disks. Based on these three properties, our `generic' structures match the expected features of bulges, disks, and stellar halos. The medians and the inter-percentile ranges for these three quantities of each component are given in Table~\ref{tab:bigtable}. 

There are a few other interesting features in Figure~\ref{fig:shapessizes}. The disks and the stellar halos that co-rotate with the galaxy ($v_{\phi\rm ,50}>0$) form together a sequence from low $c/a$ -- high rotational velocities to high $c/a$ -- low rotational velocity. 
There are also the stellar halos that counter-rotate with the galaxy which fall along a different branch in the $v_{\phi\rm ,50}$ vs $c/a$ diagram. In this case the more flatten a counter-rotating halo is, the faster it counter-rotates. In Figure~\ref{fig:appendix_spearman} we give the Spearman's rank correlation coefficients $C_S$ between all the properties we discuss, and for all five types of components. Finally, the only components that show a relation between shape and extent are the thin disks, for which a larger $r_{\rm 50}$ implies a lower $c/a$ ($C_S=0.59$).            

One can further investigate the shapes of the components using the diagram $(b/a)$ vs $(c/a)$ shown in Figure~\ref{fig:formas}, where $b$ is the intermediate axis of the inertia tensor. %How the various components occupy this diagram is shown in \autoref{fig:formas}. 
Disks are found in the upper-left corner ($b/a$ close to $1$ and $c/a$ close to $0$). As we get closer to the upper-right corner, components start to get more spherically symmetric. The upper-right corner (both $b/a$ and $c/a$ close to $1$) is occupied by the dispersion dominated components. When we analysed the components in Figure~\ref{corner-SP-galaxy-fit}, we mentioned that the difference between the spheroids (orange) and the bulges (red) was their planarity. This can be confirmed looking at their shapes, since they occupy different regions in Figure~\ref{fig:formas} ($b/a_{\rm spheroid}<b/a_{\rm bulge}$, and $c/a_{\rm spheroid}\leq c/a_{\rm bulge}$).

%%%%%%%%% FIGURE Densidades %%%%%%
\begin{figure*}
\centering
\includegraphics[width=1\textwidth]{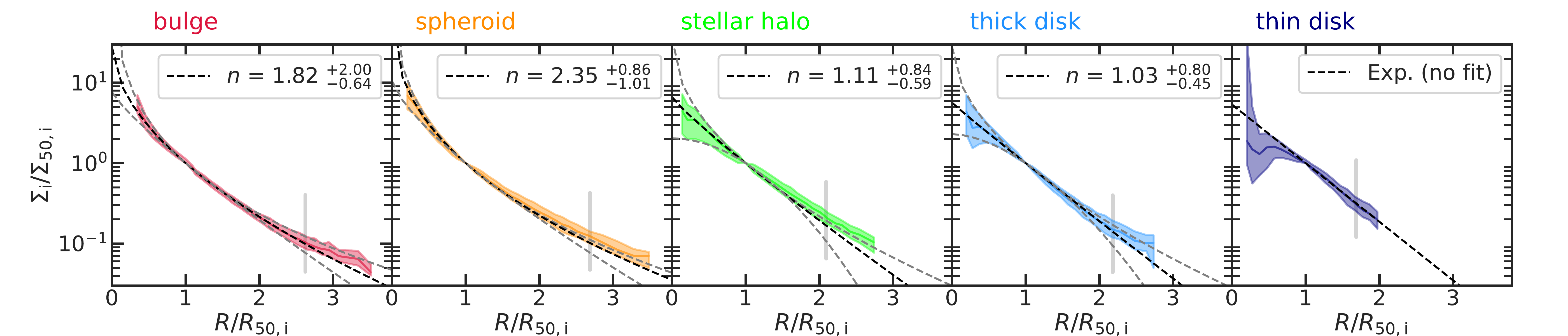}\\
\includegraphics[width=0.98\textwidth]{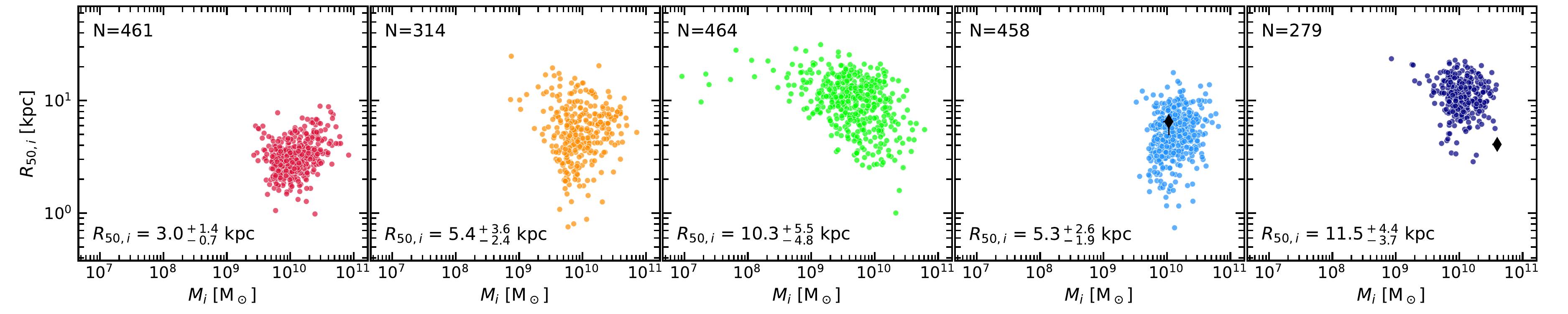}\\
\caption{\textbf{Top:} face-on density profiles for the five types of components. The coloured curves and shaded regions show the stacked profiles. The dashed black curves show the result of the median of the S{\`e}rsic fits for each component using Equation~\ref{eq:Sersic} in all but the right most panel, where it gives the theoretical expected exponential function. Gray dashed lines show the profiles corresponding to the 16th and 84th percentiles of the corresponding individual fits for $n$.
The vertical grey line in each panel marks the region with contains 95\% of the points of the different profiles (without stacking).
\textbf{Bottom:} 2D half mass radius (face-on perspective) as a function of mass for the five types of components. The labels in the bottom left corner of each panel give the corresponding median and spread for $R_{\rm 50,i}$, while the labels in the top left give the total number for each type of component in the sample $N$. The black diamond symbols in the two right most panels are the MW estimates of \citet{Cautun:2020}.}
\label{fig:densityprofile}
\end{figure*}
%%%%%%%%%%%%%%%%%%%%%%%%%%%%%%%%%%

\begin{figure*}
\includegraphics[width=0.95\textwidth]{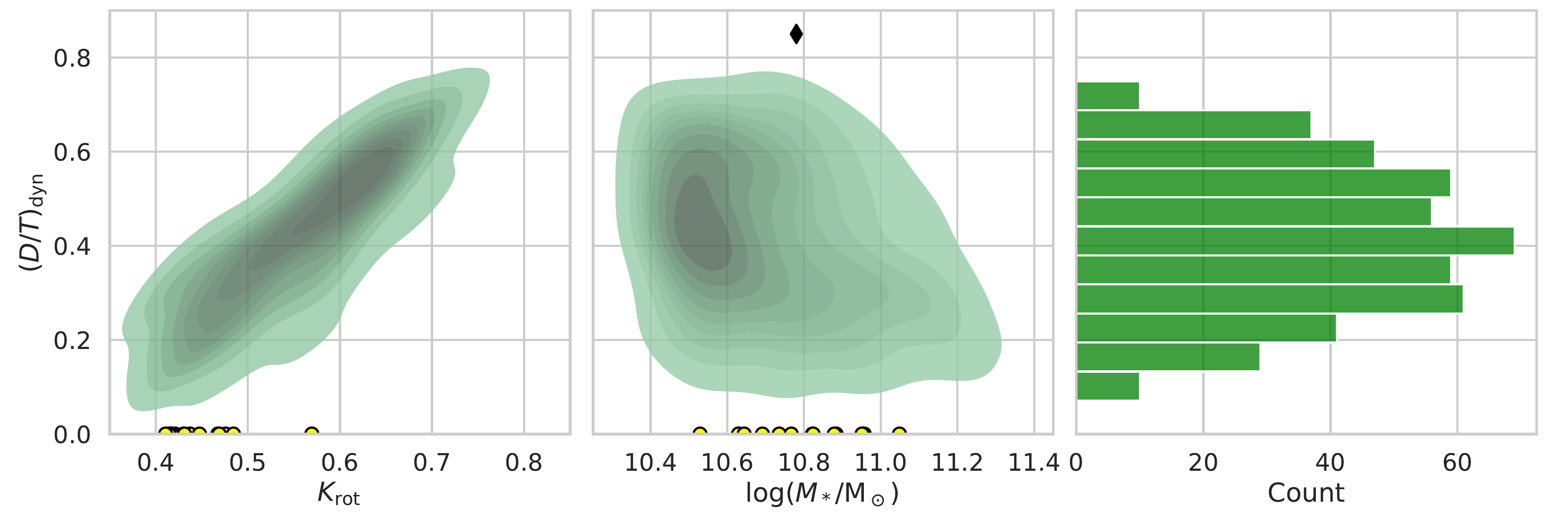}
\caption{Dynamical disk-to-total mass ratio (calculated for each galaxy as $(M_{\rm thin-disk} + M_{\rm thick-disk})/M_{*}$) vs the rotational support as quantified by $K_{\rm rot}$ (left), and vs stellar mass ($M_{*}$, centre). The levels correspond to iso-proportions of the kernel density estimate. The right most panel shows the histogram of $(D/T)_{\rm dyn}$. The yellow points in the two panels on the left show the galaxies which have no component classified as a disk, while the black diamond symbol in the middle panel gives the MW estimate of \citet{Cautun:2020}.}
\label{fig:D2T}
\end{figure*}

\subsection{Density profiles}
\label{sec:profiles}

The classical, photometry based method to classify galaxies involves fitting broad band images or radial surface density profiles constructed from said images. These fits usually can combine a few types of functions, among which the most common ones are the S{\`e}rsic function \citep{Sersic:1963}, typically expected of bulges/spheroids, and the declining exponential (or the S{\`e}rsic function with index $n=1$, typically expected of disks).

Using the more general S{\`e}rsic function in radial profiles fits comes with its own problems. Two of its three parameters are significantly correlated, namely the half-light/mass radius also known as effective radius and the S{\`e}rsic index \citep[e.g.][]{Graham:1997,Graham:2001}. Also, the best fit values can vary widely when changing the radial range for the fit \citep[e.g.][]{Saiz2001, Cairos:2003}. Notwithstanding these challenges, the S{\`e}rsic function is still widely used to decompose galaxies, and the conventional wisdom associated with it is that disks have large effective radii and small S{\`e}rsic indices, while bulges have small effective radii and $n$ varying from $<2$ for pseudo-bulges to $4$ and higher for classical bulges \citep[e.g.][]{Kormendy:2004}.              
By defining `generic' types of galaxy components, we can actually circumvent at least some of the problems associated with fitting S{\`e}rsic function to radial surface mass density profiles.  Also doing such fits is a useful exercise to further clarify the connection of these components to the photometrically defined substructures in galaxies.

We proceed as follows. We construct the surface mass density profile $\Sigma(R)$ from the face-on mass density of each component of each galaxy in the sample. We normalise the individual $\Sigma(R)$ profiles to their corresponding effective (half mass) radius $R_{\rm 50}$ and effective surface mass density $\Sigma_{\rm 50}=\Sigma(R_{\rm 50})$. We fit the resulting profiles with \citep[][]{Caon1993, Balcells2001}:
\begin{equation}
    f(x) = \exp\left[-k(n)(x^{1/n}-1) \right], \quad k(n)=1.9992n-0.3271,
\label{eq:Sersic}
\end{equation}
therefore reducing the problem to a fit with only one unknown parameter, the S{\`e}rsic index. 

For the stacked profile of the thin disks we only do the comparison with the expected exponential:   
\begin{equation}
    f(x) = \exp[1.678(1-x)],
    \label{eq:exponential}
\end{equation}
where the factor 1.678 in the exponent comes from transforming from effective radius to scalelength  
$R_{\rm 50} = 1.678R_d$.

For each type of component we stack together all the $\Sigma/\Sigma_{\rm 50}(R/R_{\rm 50})$, and plot the resulting surface mass density profiles in Figure~\ref{fig:densityprofile}. The dashed black curve in the right most top panel is the parameter free function in Equation~\ref{eq:exponential}. Using the same criteria as for the other features considered throughout the study, we plot the S{\`e}rsic profiles for each type of component taking the $50^{\rm th}$ (dashed black curve), $16^{\rm th}$ and $84^{\rm th}$ (dashed grey curves) percentiles of their distributions. The distributions of $n$ for the four types of components fitted are available in Appendix~\ref{appendix:sersic}. Even if they seem low compared to typical observational values, they are consistent with the expectations for EAGLE galaxies \citep[][]{EAGLESersic}. In Appendix~\ref{appendix:sersic}, we also give the results of fitting the stacked distribution, which was discarded because small changes in the fit range can result in big variations in $n$ (see Figure~\ref{fig:n_with_limit}).

The four fits result in a reasonable sequence of indices $n_{\rm spheroid }>n_{\rm bulge}>n_{\rm halo}>n_{\rm thick \, disk}$. The thin disk is well described by an exponential everywhere except in the innermost region, where it deviates to lower $\Sigma$. This kind of central dip below an exponential is in line with the analysis of observed galaxies by \citet{Breda2020}.

The components we classify as `bulges' have a wide distribution in S{\`e}rsic index (see Figure~\ref{fig:n_pdfs}), with a median value of $n=1.82$ very close to the limit sometimes used to separate classical from pseudo-bulges in observations \citep[$n=2.0$, e.g.][]{Kormendy:2004}. Based on 
our previous work with \texttt{gsf} \citep[][]{Obreja2016, Obreja2018}, and also on observed galaxies decomposed into dispersion and rotation dominated components \citep[e.g.][]{Zhu2018b}, we argue that using the S{\`e}rsic index to separate bulges in these two classes is not a good option given that rotating, central components can sometimes have $n<2$. Instead, the distribution of rotational velocities for bulges shown in the bottom right panel of Figure~\ref{fig:shapessizes} provides a more physics-based classification, as it seems to blend together two components, one centred on $v_{\rm\phi,50}=0$ that could be associated with classical, non-rotating bulges, and another one centred on slightly positive values of mass-weighted median rotational velocities, which could be associated with rotating, pseudo-bulges. The $c/a$ axes ratio of bulges in the same figure also seems to blend two components, so we speculate that a robust classification within the bulge class can be derived using some clustering method in the ($v_{\rm\phi,50}$,$c/a$) space. We do not do such a further split of bulges in this work, leaving it for a future study. In any case, it is important to keep in mind that we excluded barred objects galaxies from our sample, and these are precisely the type of objects which are thought to be often hosting rotating bulges or inner disks structures.

The bottom panels of Figure~\ref{fig:densityprofile} give the 2D half mass radii as a function of mass for the five types of components. These panels show that even though the surface mass density profiles look misleadingly similar for all but the thin disk, their normalisation both in terms of radius and in terms of effective surface mass density are significantly different. The bottom panels also give the 50$^{\rm th}$ percentiles for $R_{\rm 50,i}$ together with inter-percentile ranges. The black diamond symbols in the two right most panels are the MW estimates of \citet{Cautun:2020} for their Navarro-Frenk-White dark-matter model \citep{NFW:1997} constrained on observations. The MW thick disk estimate falls within the blob of simulated thick disks, but the MW thin disks is an outlier when compared to the simulations, being more massive and more compact than these.  

In the bottom panels of Figure~\ref{fig:densityprofile} we also give the total number of components in the sample associated with each `generic' substructure ($N$). While virtually all galaxies (464) have a bulge (461) and a stellar halo (464), only 68\% host also a spheroid. Regarding the disks, almost all galaxies (458) have a thick disk, while only 60\% have also a thin disk. This means that 60\% galaxies in the sample host a double (thin+thick) disk, while the rest have only one large scale single disk, which resembles more thick rather then thin disks.

\subsection{Disk-to-total ratios}
As a sanity check, we compare the mass fraction in disks $(D/T)_{\rm dyn}$ 
with the rotational support measure $K_{\rm rot}$. The left panel of Figure~\ref{fig:D2T} shows that indeed $(D/T)_{\rm dyn}$ correlates strongly with $K_{\rm rot}$, but not along a 1:1 relation.   
Actually, it is not expected to obtain an exact match between $(D/T)_{\rm dyn}$ and $K_{\rm rot}$. Even if both fractions quantify the amount of ``rotation" in a galaxy, the dispersion dominated components can contribute to $K_{\rm rot}$ if they have some net rotation, but this effect will not be reflected on the disks mass fraction.

In the central panel of Figure~\ref{fig:D2T} we can see that the dynamical disk fraction is not a function of stellar mass, and that the simulated galaxies in this sample do not reach $(D/T)_{\rm dyn}$ as high as those estimated for the MW  \citep[the black diamond show the MW position from][]{Cautun:2020}. For completeness, the right panels gives the histogram of $(D/T)_{\rm dyn}$, which shows that this sample peaks at $(D/T)_{\rm dyn}\approx0.4$. The fact that this subsample of EAGLE galaxies does not exceed $(D/T)_{\rm dyn}\approx0.8$ is a well-known issue in simulated galaxies, at least in some cosmological hydrodynamical simulations. Using different decomposition methods, similar upper limits are found for the disk fraction in IllustrisTNG (TNG50: \citealt{Zana:2022}) and EAGLE \citep{Trayford:2019, Tissera:2019, Thob2019}. In the particular case of EAGLE, this lack of strongly disk-dominated galaxies can be explained, at least partially, by the relatively low resolution. Higher resolution simulations (in this case, zoom-in simulations) mitigate this problem to some extent, \citep[e.g.][]{Garrison-Kimmel:2018,Gargiulo:2019,Buck:2020}, yielding to distributions which are closer to the observed values \citep{Peebles:2020}. However, simulating MW-mass galaxies lacking completely a classical bulge, as it is thought to be the case for MW \citep[see][and references therein]{BlandHawthorn:2016}, is still a work in progress.

%%%%%%%%%%%%%%%%%%%%%%%%%%%%%%%%%%

\begin{figure}
\includegraphics[width=0.45\textwidth]{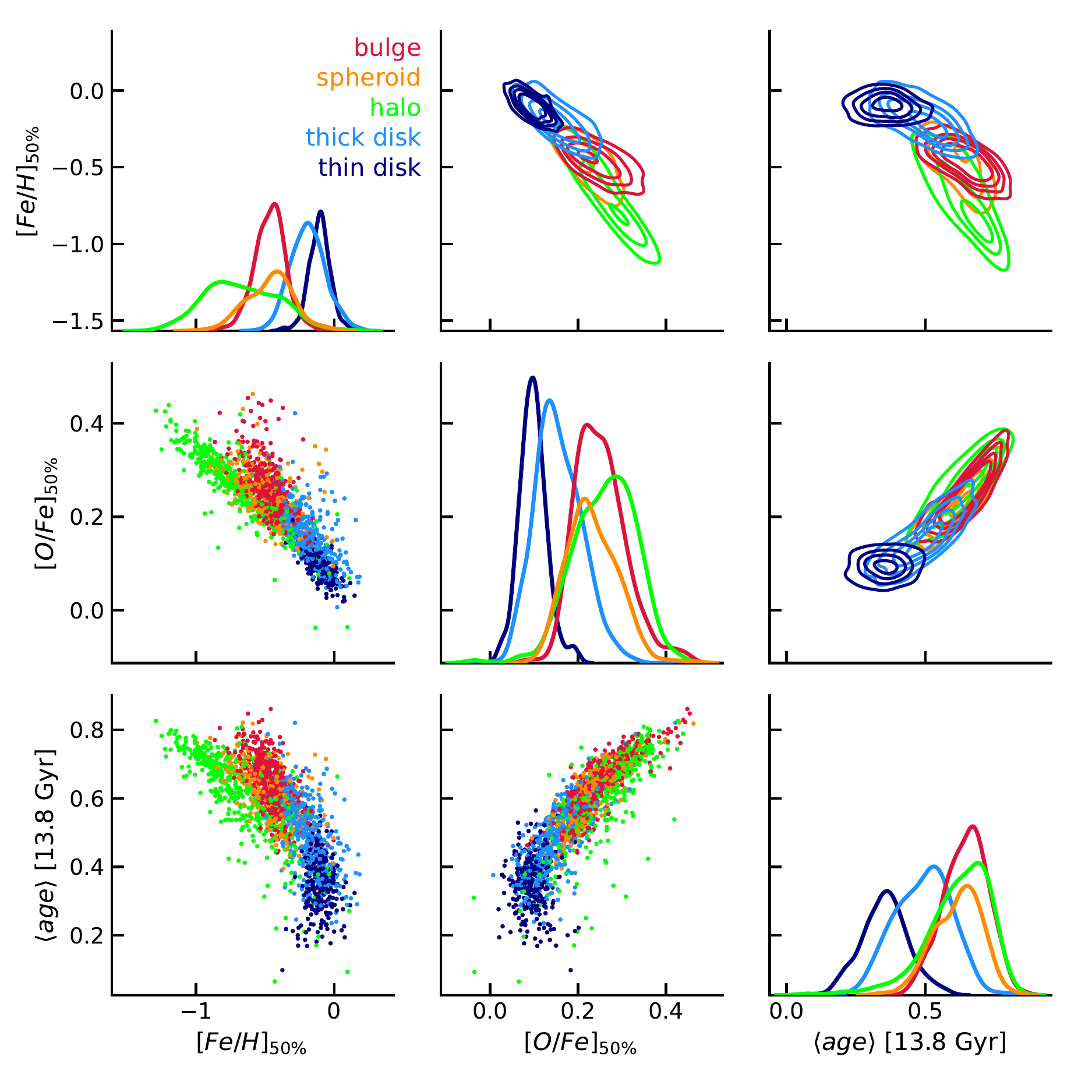}
\caption{Stellar population properties for the five types of dynamical components (different colours) defined in Figure~\ref{corner-SP-galaxy-fit}. For $[\rm Fe/H]$ and $[\rm O/Fe]$ we use the mass-weighted medians, while for the stellar age we used the mass-weighted means. The ages are in units of 13.8 Gyr.}
\label{fig:population_properties}
\end{figure}

\begin{figure}
\centering
\includegraphics[trim=0 0 475 494,clip, width=0.48\textwidth]{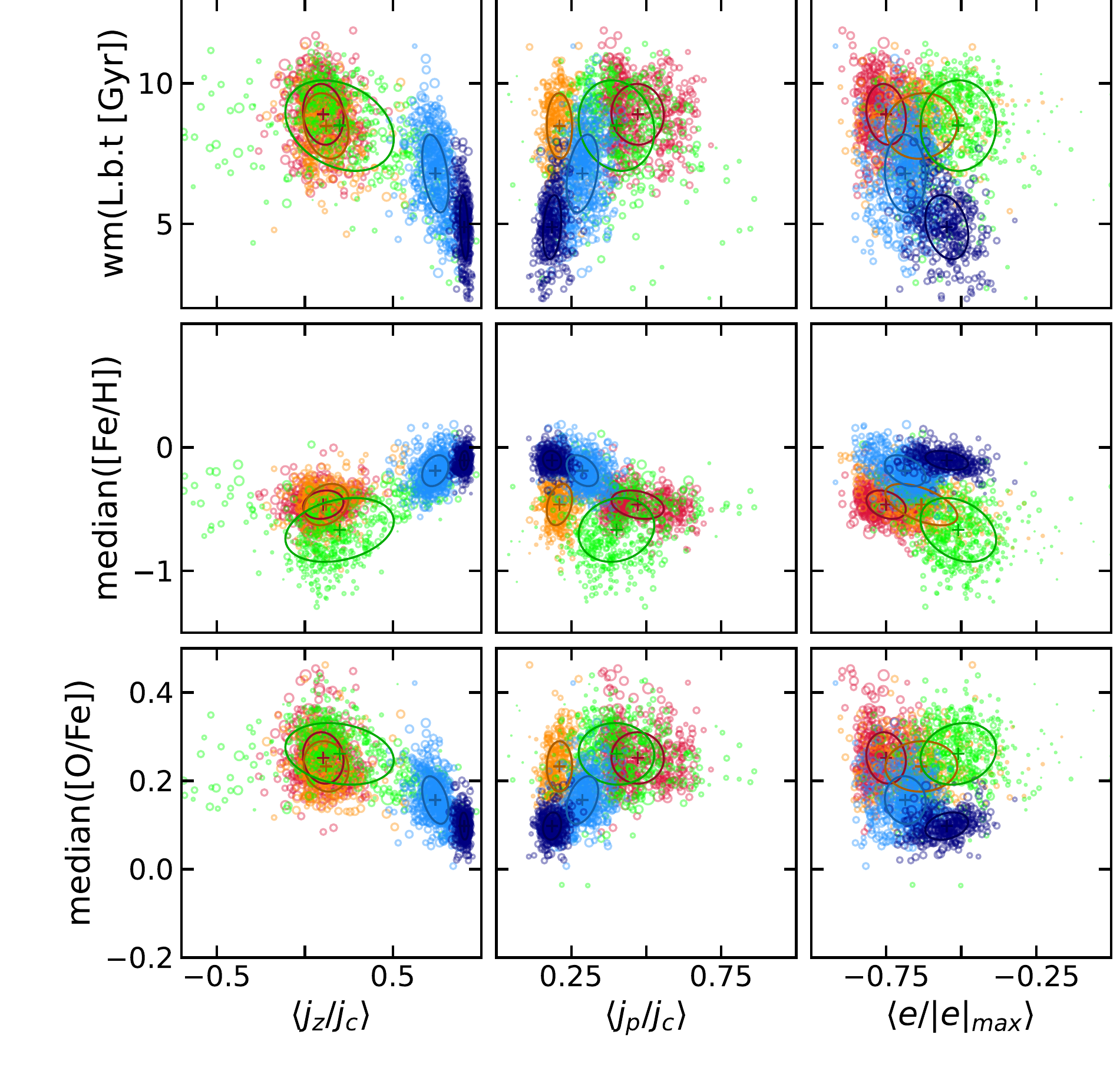}
\caption{The SP properties mean lookback time (top), median [Fe/H] (centre), and median [$\alpha/\rm Fe$] (bottom) as functions of the dynamical variables $\langle j_z/j_c\rangle$ (left), $\langle j_p/j_c\rangle$ (centre), and $\langle e/|e|_{\text{max}}\rangle$ (right). The colour code is the same as in Figure~\ref{corner-SP-galaxy-fit}. Ellipses represent the 1$\sigma$ confidence regions (calculated using the 2D covariance matrix).}
\label{corner-SP-galaxy}
\end{figure}

\begin{figure*}
\includegraphics[width=0.33\textwidth]{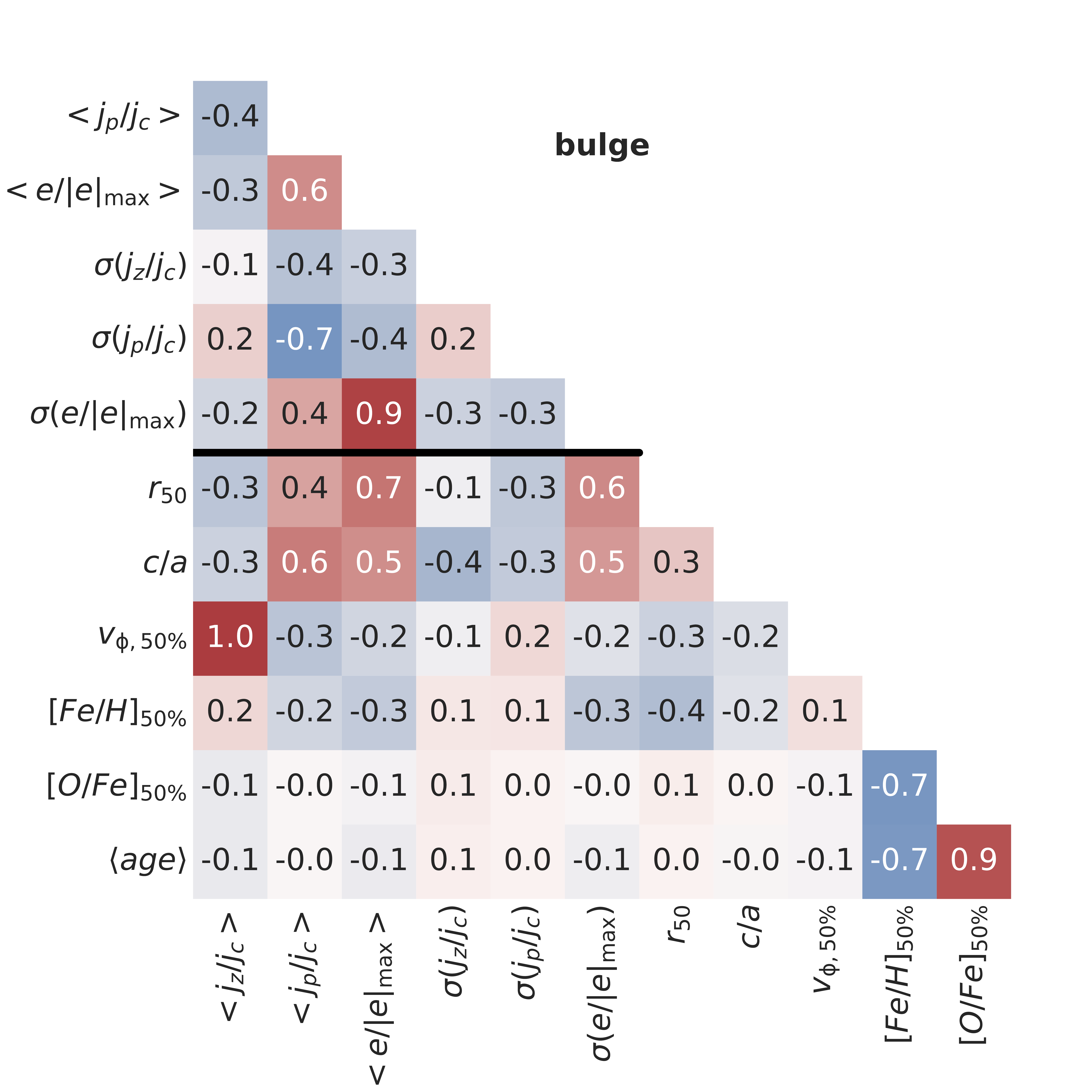}
\includegraphics[width=0.33\textwidth]{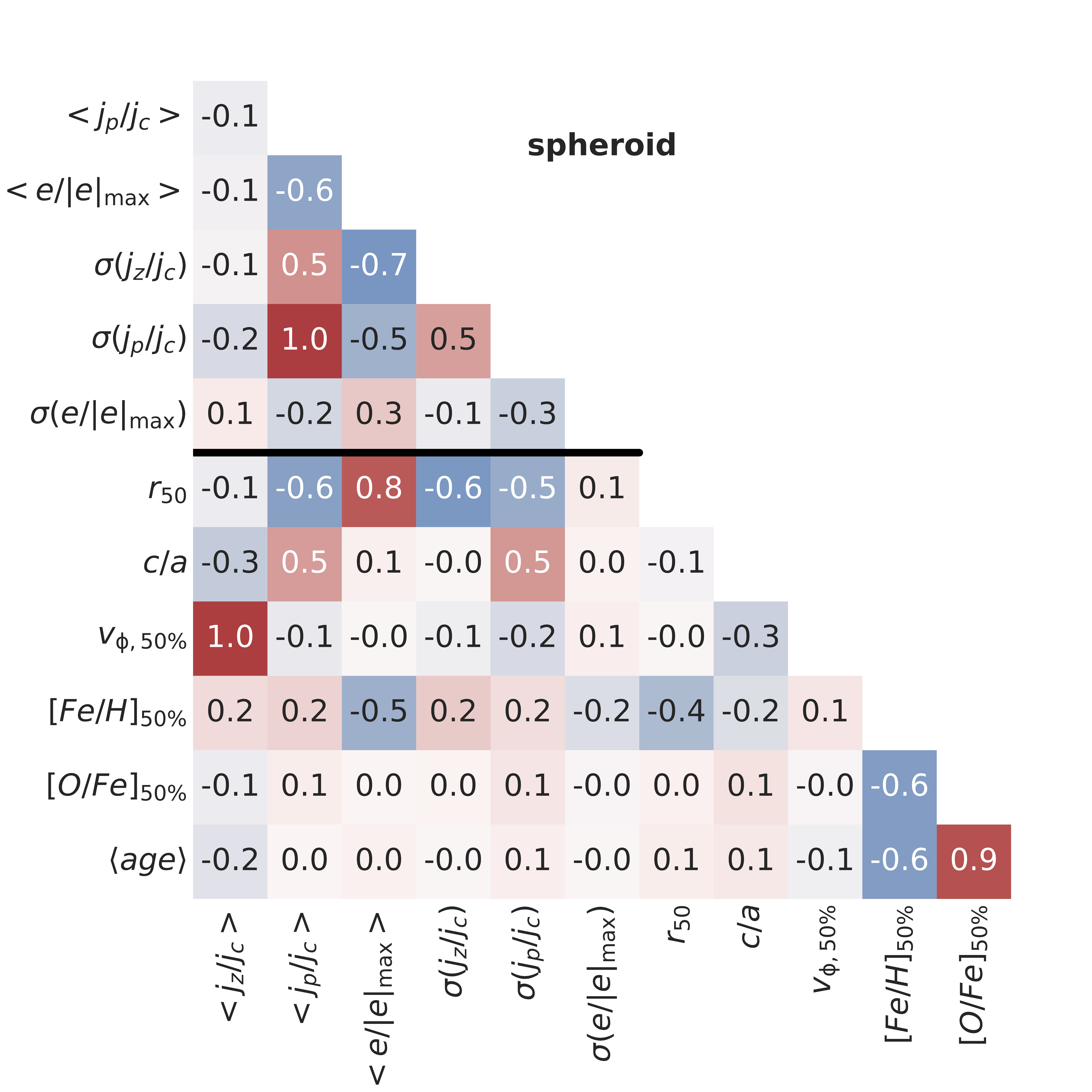}
\includegraphics[width=0.33\textwidth]{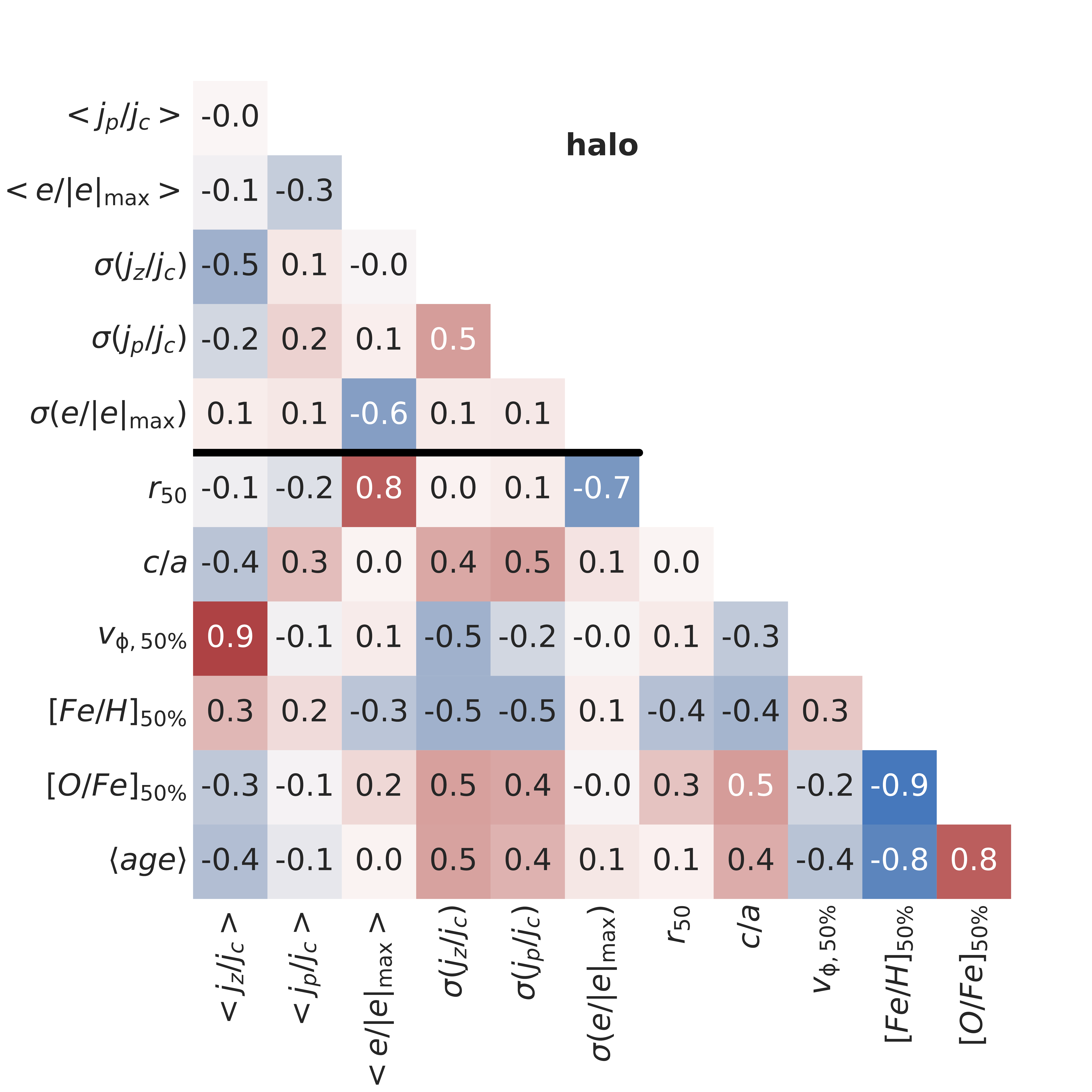}\\
\includegraphics[width=0.33\textwidth]{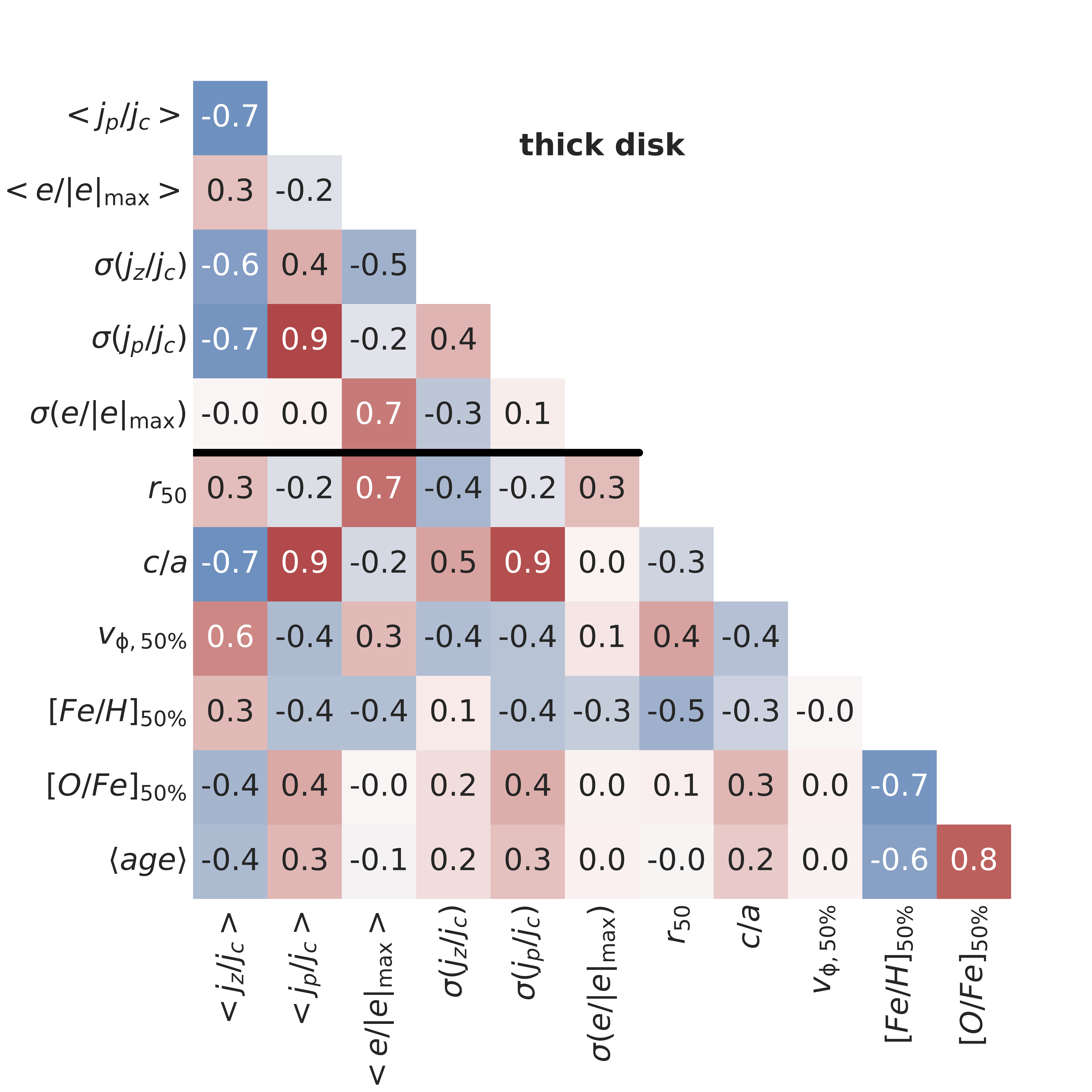}
\includegraphics[width=0.33\textwidth]{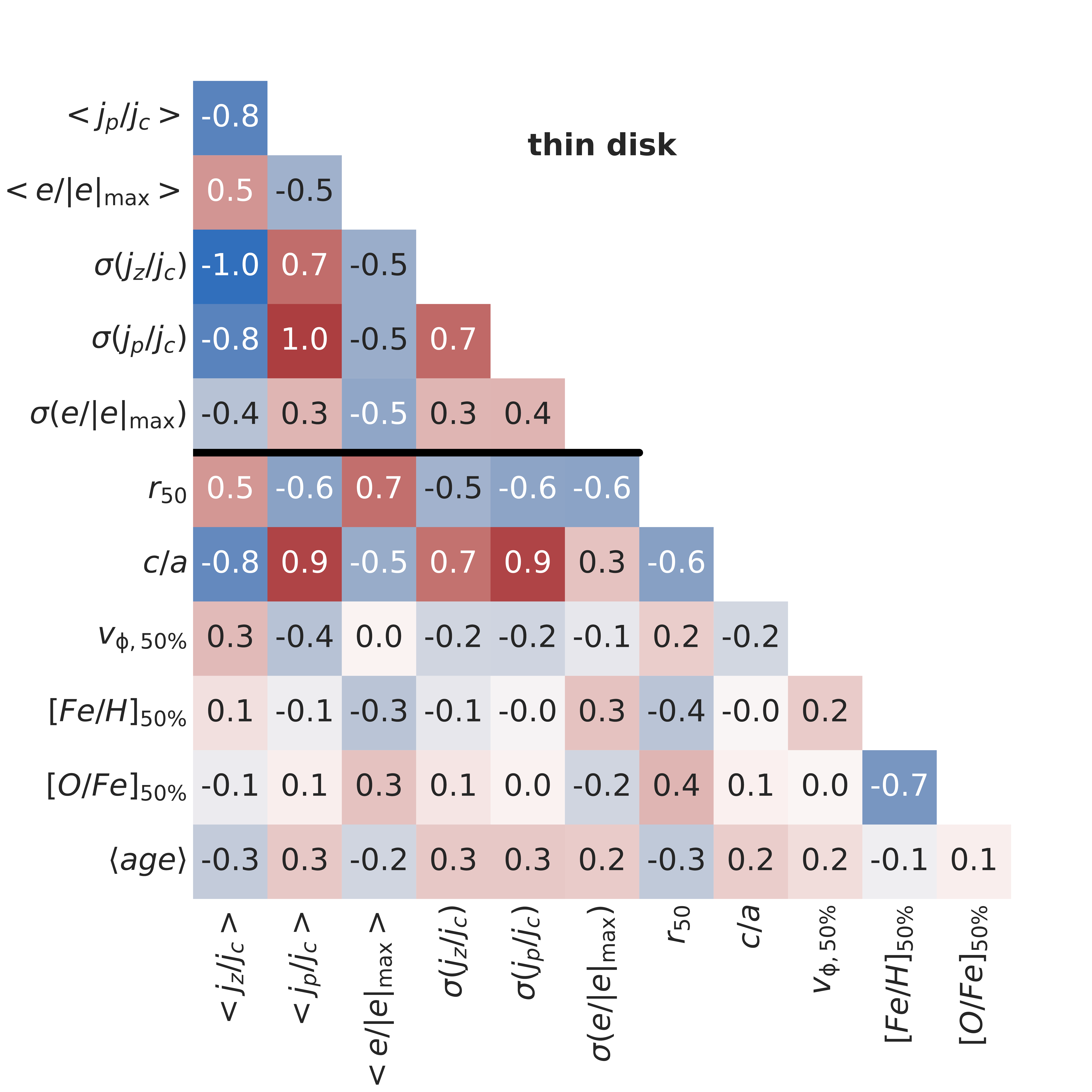}
\caption{The Spearman correlation coefficients for each type of generic galaxy components found in the galaxy sample.}
\label{fig:appendix_spearman}
\end{figure*}

\subsection{Stellar population properties}
\label{pop_properties}

Once the shapes, sizes, density profiles and rotation properties of the different dynamical components have been analysed, we address the stellar populations (SPs) properties. 

The MW in particular, has been extensively studied through the lens of stellar abundances.
E.g. most thin--thick disk classifications of solar neighbourhood stars have been made using the $[\alpha/Fe]$ vs $[Fe/H]$ plane. Therefore, it is interesting to quantify to what extent our dynamically defined components translate into separable structures in abundance space. In this work, we focus on stellar age (calculated as the look-back time), [Fe/H], and [$\alpha$/Fe] (where we use oxygen as a proxy for $\alpha$ elements).

To address this point, corner diagrams for the SPs parameter space are shown in
Figure~\ref{fig:population_properties}, while the distributions of mean/median SP properties for each type of component are quantified in Table~\ref{tab:bigtable}. 

Figure~\ref{fig:population_properties} and \autoref{tab:bigtable} show that disks are well separated from the three dispersion-dominated components (bulge, spheroid and halo) in SP properties. Thick disks are in-between, so that a smooth transition in the SP properties from rotation-dominated (indigo and light blue) to dispersion-dominated components (orange, red and lime) is clearly visible in the kde contours of the upper right panels, where halo stars (lime) deviate at low [Fe/H] values. Between the two types of disks, we can see the expected behaviour: thin disks are younger (mean half mass age of 4.90$^{\rm +0.99}_{-1.08}$ Gyr vs 6.88$^{\rm +1.28}_{-1.60}$ Gyr), more metal rich (median half mass metallicity of -0.10$^{\rm +0.07}_{-0.08}$ vs -0.19$^{\rm +0.12}_{-0.12}$), and less $\alpha$-enhanced than thick disks (median half mass [$\alpha$/Fe] of 0.10$^{\rm +0.03}_{-0.03}$ vs 0.15$^{\rm +0.06}_{-0.04}$). Thick disks (light blue) show mild correlations between any two of the three SP properties, while thin disks only display a correlation between [$\alpha$/Fe] and [Fe/H].
Compared to figure 12 of \cite{Amarsi2019}, the dynamical thin disk component follows the expected properties of observational disks.
We quantify how much each property correlates with another within a single `generic' component in the next section. 

The three dispersion dominated components overlap almost completely in ages and $[\alpha/Fe]$. In total metallicity, the lime component stands out as the [Fe/H]-poorest (median half mass metallicity of -0.68$^{\rm +0.30}_{-0.28}$ vs $\sim$-0.45 of spheroids/bulges, clearly visible in the 1D distribution of the top-left panel), in agreement with its interpretation as a stellar halo. This component displays very clear correlations between SP properties: younger stellar halos having higher [Fe/H] and lower [$\alpha$/Fe] than older halos. Interestingly, there is no clear difference in any of the SP properties considered between bulges (red) and spheroids (orange), which are, nevertheless, clearly separated in shape space (Figure~\ref{fig:formas}).

As expected, thick disks appear in between thin disks and spheroids/bulges in all three SP properties. Therefore, spheroids/bulges are older and formed on shorter times-scales than thick disks, while thick disks are older and formed on shorter times-scales than thin disks.

Figure~\ref{corner-SP-galaxy} shows how the SP variables mean half mass lookback time, median half mass [Fe/H], and median half mass [$\alpha/Fe$] vary as functions of the dynamical variables $\langle j_z/j_c\rangle$, $\langle j_p/j_c\rangle$, and $\langle e/|e|_{\text{max}}\rangle$.
The SP parameters of the two types of disks mildly correlate with $\langle j_z/j_c\rangle$ (left column): the higher $\langle j_z/j_c\rangle$, and the younger the average SPs, the [Fe/H] richer, and the less $\alpha$ enriched they are.
Conversely, the three dispersion-dominated components show no correlations. 
A similar behaviour appears in the SP vs planarity $\langle j_p/j_c\rangle$ plots (central column), where the lower $\langle j_p/j_c\rangle$ and the younger the average SPs, the [Fe/H] richer, and the less [$\alpha$/Fe] enriched they are. This is a consequence of the $\langle j_p/j_c\rangle$ vs $\langle j_z/j_c\rangle$ correlation shown in Figure~\ref{corner-SP-galaxy-fit}.
A stratified correlation appears for disks and bulges in the SP properties vs energy: the less bound a component is, the older, [Fe/H] poorer, and more $\alpha$ enriched it is. SPs of stellar halos (lime), on the other hand, do not follow these trends. These results are in agreement with MW observations \citep[APOGEE:][]{Bovy:2014, Hayden:2015, Bovy:2015}, reinforcing our interpretation that dynamically defined substructures do have observational counterparts.

%%%%%%%%%%%%%%%%%%%%%%%%%%%%%%%%%%%%%%%%%%%%%%%%%%%%%%%%%%%%%%%%%%%%%%%%%%%%%%%%%%%%%
% CORRELATIONS
%%%%%%%%%%%%%%%%%%%%%%%%%%%%%%%%%%%%%%%%%%%%%%%%%%%%%%%%%%%%%%%%%%%%%%%%%%%%%%%%%%%%%%%%%%

\subsection{Correlations}
\label{sec:correlations}

\begin{figure}
    \centering
    \includegraphics[width=0.235\textwidth]{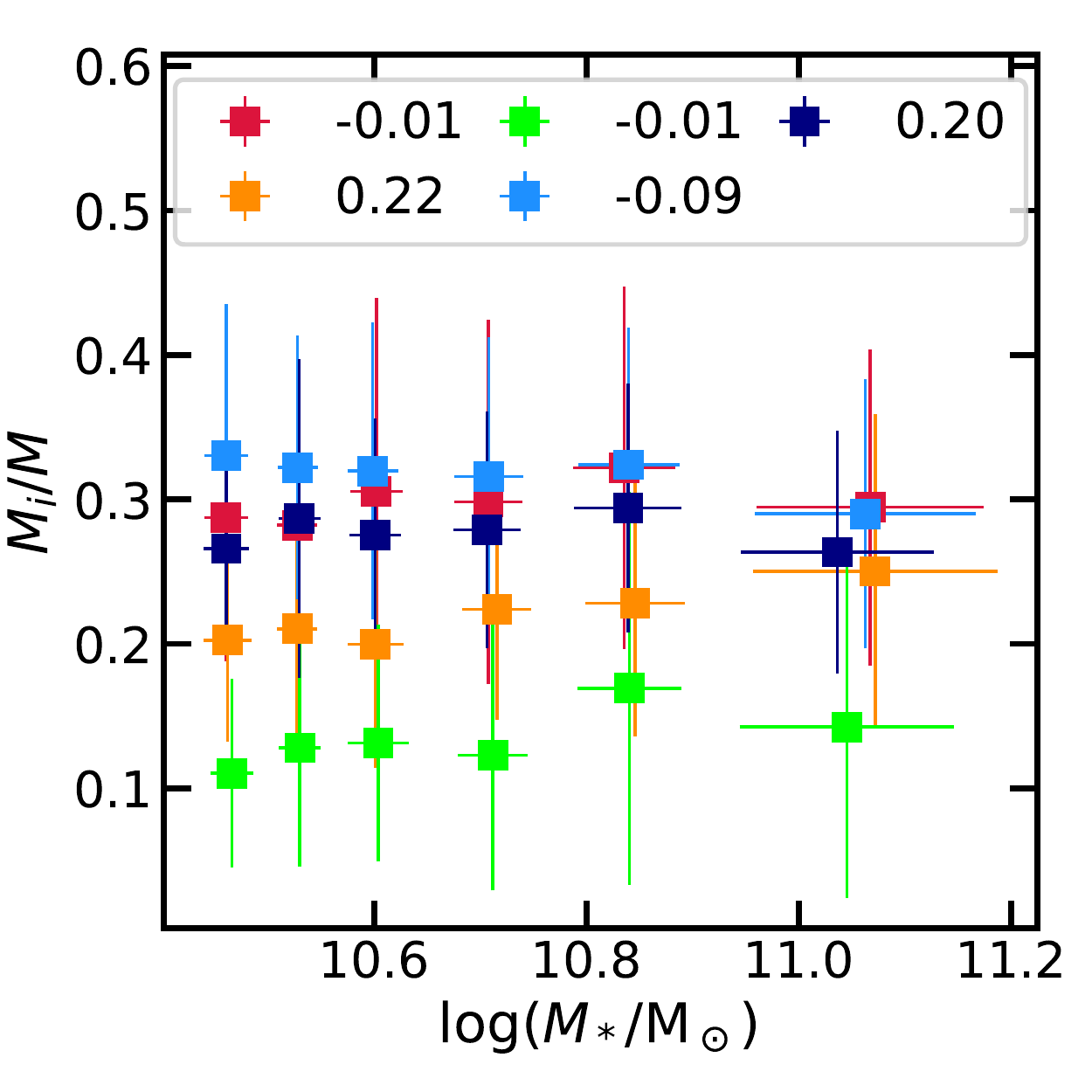}
    \includegraphics[width=0.235\textwidth]{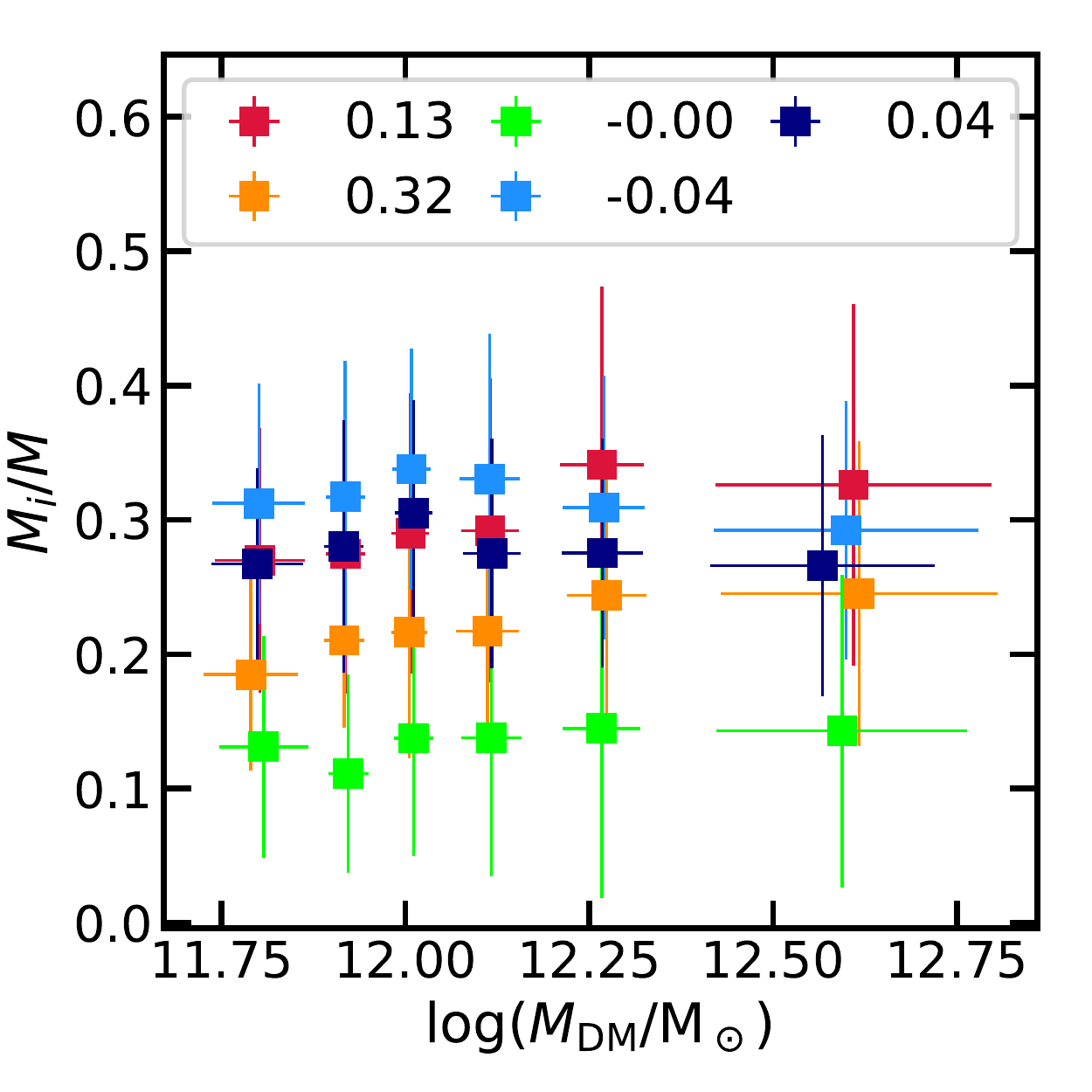}
    \caption{Mass fractions of components (mean and standard deviation) as function of total stellar mass of host galaxy (left), and of dark matter halo mass (right). The colour-code is the same as in Figure~\ref{corner-SP-galaxy-fit}. The legends in each panel give the MCD-derived correlation coefficients $\rho_{\rm MCD}$.}
    \label{fig:massdep_massfrac}
\end{figure}

\begin{figure*}
    \centering
    \includegraphics[width=0.99\textwidth]{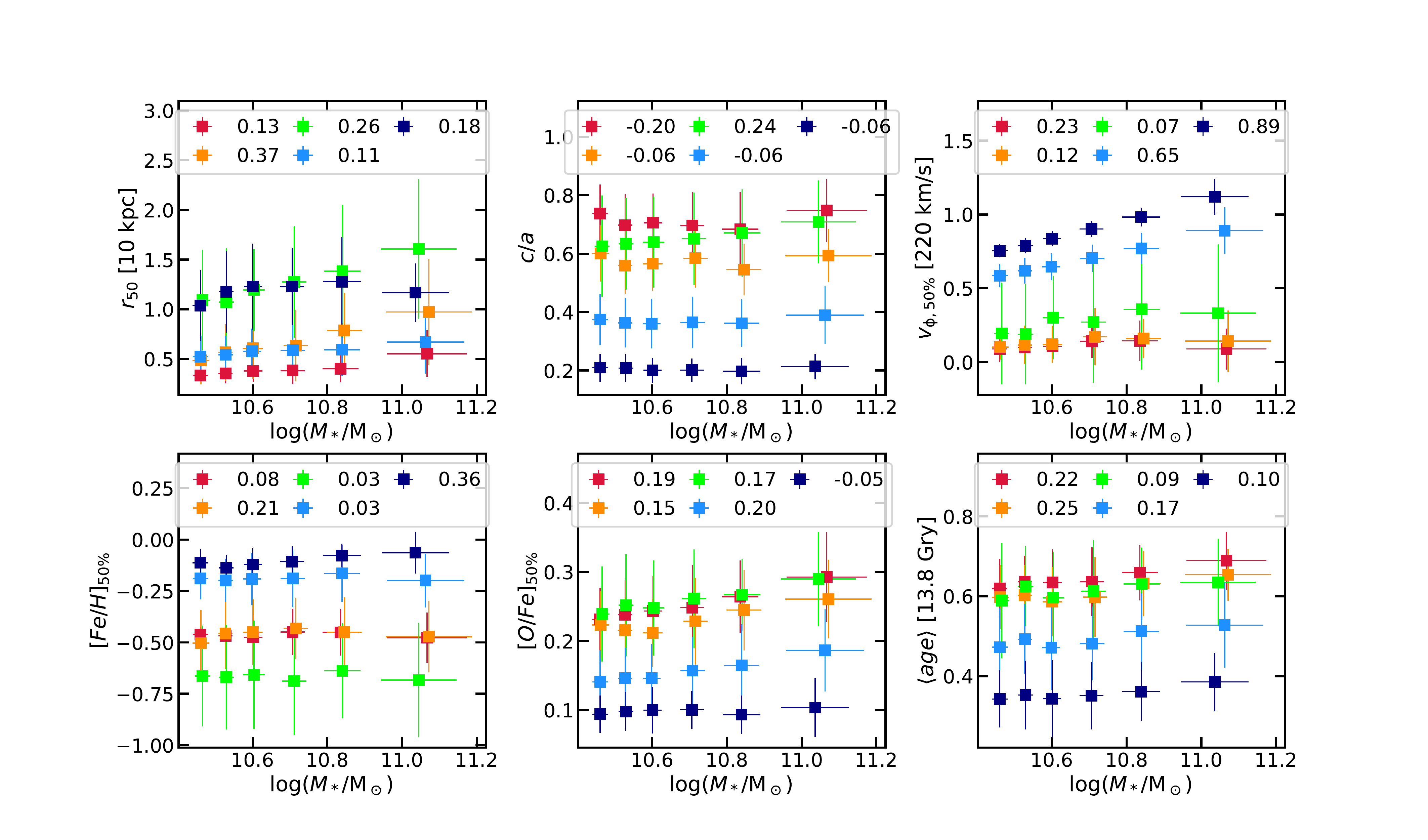}
    \caption{Same as Figure~\ref{fig:massdep_massfrac}, for the variation of observable properties discussed in Sections~\ref{sec:formas} and \ref{pop_properties} with the total stellar mass of the host galaxy.}
    \label{fig:massdep_observables}
\end{figure*}

We quantified the correlations between all the parameters discussed throughout Section~\ref{sec:results} using the Spearman's rank correlation coefficient $C_S$. Figure~\ref{fig:appendix_spearman} gives the $C_S$ for all combinations of two properties, for each of the five generic galaxy components separately. The horizontal black lines separate the part of the table containing only the six variables used to define the components (above) from the rest of properties (below).

Using Figure~\ref{fig:appendix_spearman}, it becomes clear what we were discussing in Section~\ref{pop_properties} for example. The strongest correlations between ages and $\alpha$-enhancement are found in bulges and spheroids ($C_S=0.9$), while for thin disks these two properties do not correlate with each other at all ($C_S=0.1$). Conversely, for all types of components [$\alpha$/Fe] anti-correlates significantly with [Fe/H], with the stellar halos standing out as previously discussed at $C_S=-0.9$.

\subsection{Mass dependencies}
\label{sec:massdep}

Our galaxy sample was chosen to match an extended mass range compatible with MW, which covers almost one order of magnitude in $M_{\rm *}$ (see Section~\ref{sec:sample}). Therefore, it is possible to check for mass dependencies if we restrict to using only a few bins. For this purpose we split the sample in 6 equally populated mass bins (82 galaxies per bin), and for each bin and each type of component we compute the mean and standard deviation of the properties discussed in the current section (see Table~\ref{tab:bigtable} for a reference). We also checked for global correlations over the entire stellar mass range using the Minimum Covariance Determinant (MCD), which is a robust estimator of covariance for normally distributed data \citep{Rousseeuw:1984,Rousseeuw:1985}. 

The mass fractions of the five types of components as function of stellar mass are given in the left panel of Figure~\ref{fig:massdep_massfrac}. The stellar halos, bulges and thick disks are consistent with no correlation between $M_i/M$ and $M_{\rm *}$, as quantified by the MCD-derived correlation coefficient in the legend ($\rho_{\rm MCD}$)\footnote{For all MCDs we require a 0.75 support fraction. Using a support fraction closer to the minimum of $\sim$0.5 can make weaker already weak correlations, but has little effect on strong ones.}. The only components that show a very weak correlations are the spheroids 
($\rho_{\rm MCD}=0.22$) and the thin disks ($\rho_{\rm MCD}=0.20$). The reason for the weak correlation for thin disks, as opposed to the no-correlation the binned data suggests, is the fact that MCD considers a large fraction of the points in the last mass bin as outliers. The right panel of Figure~\ref{fig:massdep_massfrac} shows the variation of the mass fractions with the mass of the dark matter host halo $M_{\rm DM}$. Same as in the left panel, none of the components shows a strong trend with $M_{\rm DM}$ as quantified by the $\rho_{\rm MCD}$, but interestingly, the mass fraction of both thin and thick disks seems to peak at $M_{\rm DM}\approx10^{\rm 12}M_{\rm\odot}$. This $M_{\rm DM}$ value also corresponds to the peak of the $M_{\rm *}$--$M_{\rm DM}$ relation \citep[e.g.][]{Moster:2018,Behroozi:2019}, and provides a good reason for looking also for correlations between properties of stellar components and $M_{\rm DM}$ since in this regime, galaxies with roughly equal $M_{\rm *}$ can be hosted by DM halos with masses covering a range larger than 1~dex \citep[e.g.][]{Schaller:2015}.

The dynamical properties $\langle j_z/j_c\rangle$, $\langle j_p/j_c\rangle$,$\langle e/|e|_{\rm max}\rangle$, $\sigma(j_z/j_c)$, $\sigma(j_p/j_c)$, $\sigma(e/|e|_{\rm max})$, used for the clustering defining the generic components also do not show any significant mass dependence. Figure~\ref{fig:massdep_input} shows only the expected separation among the five types of components. 

However, some of the observable properties show notable trends with $M_{\rm *}$.
Figure~\ref{fig:massdep_observables} shows the variation with $M_{\rm *}$ for the observable properties discussed in Sections~\ref{sec:formas} and \ref{pop_properties}. The only strong correlations appear between $v_{\rm\phi,50}$ and $M_{\rm *}$ for the two types of disks (top right panel). The correlation for thin disks in particular is so strong ($\rho_{\rm MCD}=0.89$) that it can be used, for example, to estimate the MW stellar mass from the median rotational velocity of thin disk stars. This variation evokes the Tully-Fisher relation \citep[TF,][]{Tully1977}, but it is important to take it into account that we are analysing the median mass-weighted velocity, and not the asymptotic rotational velocity as done in TF analysis. This two velocities should be similar in magnitude, but obtaining the exact correspondence between the two is beyond the scope of this paper. 
A strong correlations also appears between the median mass-weighted rotational velocity of thin disks and the mass of the dark matter host halo (see Figure~\ref{fig:virmassdep_observables}). In Appendix~\ref{appendix:massdep}, we quantified the $v_{\rm\phi,50}$--$M_{\rm *}$, and $v_{\rm\phi,50}$--$M_{\rm DM}$ relations in the disk components using linear regressions. 

The highest stellar mass bin with $\log (M_{\rm *}/ M_\odot)>10.7 $ shows a kind of decoupling from trends with increasing $M_*$ in the lower mass bins, of which the most important are: i) bulges become less planar (increase in $\langle j_p/j_c\rangle$), ii) the planarity dispersion $\sigma(j_p/j_c)$ of bulges and halos decreases, iii) both bulges and spheroids become more spherical ($c/a$ increases),
iv) the sizes $r_{50}$ of thin disks decrease, while those of bulges increase, v) the mass fractions of both disk types decrease, as well as that of stellar halos. We note that, quantitatively, these changes are lower than the corresponding error bars. However, they always occur so as to endow these SPs with a slightly  more marked spherical-like characteristics, in consistency  with  observations.  
Indeed, early analyses of the SDSS catalogue \citep[][]{York:2000,Stoughton:2002}  have shown that the properties of  stellar populations of galaxies with  $M_{\rm *}$ in the high mass end tend to be spherical-like \citep{Bernardi:2003}. Analyses of $\sim$ 600.000 galaxies in SSDS DR7 by  \citet{Thanjavur2016} have shown that at  $\log (M_{\rm *}/ M_\odot)>10.7 $ galaxies SPs are  dominated by spherical-like components (
 see, for example, their Figure 5). 
 
Apart from this decoupling of the high mass end trends from those in the lower mass bins, and the correlation between $v_{\rm\phi,50}$ and $M_{\rm *}$ for the two types of disks, no relevant mass effects show up in Figures \ref{fig:massdep_massfrac}, 
\ref{fig:massdep_observables}, \ref{fig:virmassdep_observables} and \ref{fig:massdep_input}. In general, mass trends are more visible for the rotation dominated components than for dispersion dominated ones.

\section{Summary and discussion}
\label{sec:summary}

We used a sub-sample of EAGLE galaxies from the 100 Mpc sized box to analyse in detail the stellar components present in MW-mass objects, 10.4 $<$ log($M_\text{star}$(in 30kpc) [$M_\odot$]) $<$ 11.2. The sub-sample was further reduced to contain only galaxies dominated by rotation according to the criteria of \citet{Sales2010}, $K_{\text{rot}} > 0.4$. We have also excluded galaxies with significant bars. 

To separate the stellar components in the simulated galaxies we used the new version of \texttt{galactic structure finder} (\texttt{gsf2}, Obreja et al. in prep.), which is based on a data-driven approach (multi-variate Gaussian Mixture models as unsupervised clustering) to identify galaxy substructure. Once each galaxy had its unique components defined, we quantified a large panoply of properties for them (mass fractions, shapes, sizes, rotational velocities, velocity dispersions, circularities, planarities, binding energies, metallicities, stellar ages, and $\alpha$-enhancements). 
To define the `generic' type of components present in the sample, we applied Gaussian Mixtures to a 6-dimensional space of dynamical properties, 
($\langle j_z/j_c\rangle$, $\langle j_p/j_c\rangle$,$\langle e/|e|_{\text{max}}\rangle$, $\sigma(j_z/j_c)$, $\sigma(j_p/j_c)$, $\sigma(e/|e|_{\text{max}})$), where $\langle$ and $\rangle$ denote mass-weighted distribution means, and $\sigma(x)$ quantify the spread of these distributions, and found five distinct substructures, all of which have observational counterparts. The five types of components and their features are:     
\begin{itemize}
    \item Thin disks (strongly rotation dominated, $v/\sigma$\footnote{We define the rotation-to-dispersion as $v/\sigma\equiv v_{\rm\phi,50}/\sigma(v_{\rm z}) $}$\sim$2.85): cold and planar components, radial extended, with high median rotational velocities and low velocity dispersions. Their surface density profiles follow the pure exponential, and they are composed of young, metal rich stellar populations.
    \item Thick disks (rotation dominated, $v/\sigma\sim$1.43): mostly cold and planar components, thicker, less radial extended and more $\alpha$-enhanced than thin disks, with a wider distribution of metallicities (less metal rich) and stellar ages (older). Their surface density profiles follow well pure exponentials. 
    \item Halos (dispersion dominated, $v/\sigma\sim$0.24): hot, loosely bound components, radial extensions comparable to disks and higher, populated by old stars with low metalliticy. Their surface density profiles also follow pure exponentials closely.
    \item Bulges (dispersion dominated, $v/\sigma\sim$0.15): hot, tightly bound, concentrated ($r_{\rm 50}$=3.54$^{+1.76}_{-0.84}$~kpc), spherical components whose surface density profiles can be fitted by S{\`e}rsic functions with $n=\rm1.82^{\rm+2.00}_{\rm-0.64}$. Stars belonging to this type of component are old and with intermediate metallicities on average.
    \item Spheroids (dispersion dominated, $v/\sigma\sim$0.17): hot, planar components with the same stellar population features as bulges, but more extended ($r_{\rm 50}$=6.07$^{+3.85}_{-2.78}$~kpc) and with higher S{\`e}rsic indices ($n=\rm2.35^{\rm+0.86}_{\rm-1.01}$).
\end{itemize}

The quantifiable differences between these types of components are summarised in Table~\ref{tab:bigtable}, and in Figures~\ref{corner-SP-galaxy-fit}, \ref{fig:shapessizes} \ref{fig:densityprofile}, and \ref{fig:population_properties}.
Virtually all galaxies in the sample contain a bulge, a stellar halo and a disk. 60\% of objects host two disks, a thin and a thick one, and 68\% host also a spheroid. If a galaxy hosts only one disk, this tends to be more similar to thick rather than thin disks. By comparison, observations tend to find a larger fraction of double disks, and a smaller fraction of bulges \citep[e.g.][]{Comeron:2014}, but in observations the components are typically separated photometrically, and not dynamically as we do here. Actually, EAGLE galaxies are known to have rather small S{\`e}rsic indices \citep{EAGLESersic}, and it has been shown in various works that the disk-to-total ratios based on photometry can be significantly larger than those based on dynamics \citep[e.g.][]{Scannapieco:2010,Obreja2016}. Our dynamically derived disk-to-total ratios are tightly related to the $K_{\rm rot}$ parameter, but the two do not follow a 1:1 relation (Figure~\ref{fig:D2T}). 
We also find that $(D/T)_{\rm dyn}$ does not depend on stellar mass (Figure~\ref{fig:D2T}).

With the definition of these five types of `generic' stellar components, we are able to look for correlations between intrinsic and observable properties on a component by component base.

The small mass range ($\sim$1~dex) probed by this galaxy sample, both in $M_{\rm *}$ and in $M_{\rm DM}$, 
makes it hard to unveil trends with mass. Thus, the only strong correlations appeared between the median rotational velocity $v_{\rm\phi,50}$ and $M_{\rm *}$ for the two types of disks (Figure~\ref{fig:massdep_observables}). Since $M_{\rm *}$ depends on $M_{\rm DM}$, we also find correlations of the disks $v_{\rm\phi,50}$ with the dark matter halo mass, but weaker than with $M_{\rm *}$. 
In particular, the relations of $v_{\rm\phi,50}$ with $M_{\rm *}$, and with $M_{\rm DM}$ for thin disks are so tight that they could be use to estimate the stellar and dark matter halo MW-masses (linear regressions in Figure~\ref{fig:mass_vphi_fit} and Table~\ref{tab:v_vs_mass}), if observational effects like selection functions can be properly accounted for. 

Two of the main aims of this work were to uncover to what extent the different dynamical components separate in the phase space of stellar population properties (Figure~\ref{fig:population_properties}), and to what extent the dynamical properties dictate 
stellar population properties (Figure~\ref{corner-SP-galaxy}). Thin disks SP properties do not overlap with those of dispersion dominated ones (bulges, spheroids and stellar halos). Thick disks have ages, [Fe/H], and [O/Fe] values bridging both groups, so that, we find a fair degree of overlap in stellar ages, [Fe/H], and [O/Fe] among the various components, with a smooth transition from bulges to disks. The stellar halos do not always follow these smooth transitions. With the exception of thin disks, all components show correlations among their stellar population properties: older ages mean lower metallicities and larger $\alpha$-enhancements (quantified in Figure~\ref{fig:appendix_spearman}). Not unexpected, we found that the positioning of a galaxy component in the dynamical space only loosely determines its stellar population properties, when looking component by component (Figure~\ref{corner-SP-galaxy}, with within component type correlations quantified in Figure~\ref{fig:appendix_spearman}). However, the information in Table ~\ref{tab:bigtable} and Figure~\ref{fig:appendix_spearman} can be straightforwardly used in semi-analytic models of galaxy formation, at least in the estimated MW-mass range. For a complete model of stellar components in galaxies, we would need to expand the galaxy mass range probed and relax the rotation-dominated criteria.  

As it was pointed in Section~\ref{pop_properties}, there is no clear separation between bulges and spheroids in terms of SP properties, even if they are well differentiated by their shapes and dynamical properties. In fact, even if we consider only four types of substructures instead of five, the bulges and the spheroids would still be considered different components (e.g., the thin and thick disks would just be lumped together as `disk'). 
The spheroids found using \texttt{gsf} on a sub-sample of zoom-in galaxies from the NIHAO project \citep{Obreja2018, Obreja2019} are different from (classical and pseudo) bulges in their formation histories (the progenitor material of spheroids is the first to be incorporated to the galaxies), so we can speculate here that these two types of dispersion dominated components might have different formation histories. %, a scenario that can be tested in future works. 
Components with the same properties as the spheroids in the current galaxy sample are also found in similar decompositions performed on other simulations. In their analysis of the Illustris-TNG simulation, \cite{Du2019} also find components with low average $\langle j_p/j_c\rangle$ values that have average circularities either close to 1 or in the [0, 0.25] range, with a rather  unpopulated area in-between. 
However, they classify this components as bulges or halos based on their $\langle j_z/j_c\rangle$ and $\langle e/|e|_{\text{max}}\rangle$ values, since that is the plane these authors use to label the different components.

A detailed analysis of the stellar tangential velocity, age  and chemical composition of MW stars has been recently presented by \cite{Belokurov2022}. They use data coming from GAIA EDR3 and APOGEE (DR17) \citep{Abdurro:2022}, and find a set of stars with low tangential velocity, old ages and low metallicities, located in the disk. These authors compare their observational analysis to results based on Auriga \citep{Grand:2017} and FIRE \citep{Hopkins:2018} simulations of MW-sized galaxies, to conclude that these stars have formed in an early, fast, disordered phase of galaxy assembly. 
Other possible scenarios for the origin of the spheroid component include debris of in-plane accreted satellites (i.e., with orbital poles parallel to the disk spin) or boxy orbits in a perturbed axis-symmetric potential \citep[][]{Libro}. 
Another possibility is that the spheroids in this work are linked to the presence of weak bars. In any case, this component emerges out of our analysis even if we assume a more restrictive criteria to exclude barred galaxies. However, the fact that the galaxies in our sample do not have a bar now does not necessarily imply they never had one, as it is still an open question whether bars are transient features \citep{Bournaud:2002, Peschken:2018, Cavanagh:2022} or not \citep{Athanassoula:2005, Athanassoula:2013, Kim:2016}. To clarify the nature of the spheroids in these EAGLE galaxies, we have to constrain how these components formed, an analysis which we leave for future work.

Finally, it is worth mentioning that the analysis method we use in this paper can lead to new insights into the properties of dark matter halos beyond their masses. \citet{Obreja:2022} showed that the angular momentum of dynamical components like disks and stellar halos is a good proxy for the angular momentum of DM halos, and used the relations derived from simulations to infer the spin of MW's dark matter halo. We plan to explore these avenues in future works using statistical significant sample of simulated galaxies.

\section{Acknowledgements}
We would like to thank the anonymous reviewer for their comments and suggestions.
This work was supported through MICIIN/FEDER (Spain) PGC2018-094975-C21 grants. This project has received funding from the European Union’s Horizon 2020 Research and Innovation Programme under the Marie Skłodowska-Curie grant agreement Nº 734374- LACEGAL.
%%%%%
SOM thanks the Ayuda para el fomento de la investigaci{\'o}n en estudios de master from Univ. Aut{\'o}noma de Madrid, the funding by the Spanish Ministry of Science and Innovation under grant number PRE2020-095788, and the warm welcome of the BACCO-project group. 
%%%%%%
AO is funded by the Deutsche Forschungsgemeinschaft (DFG, German Research Foundation) – 443044596.
%%%%%%
SP acknowledges funding from the same Horizon 2020 grant for a secondment at the Astrophysics group of the Univ. Aut{\'o}noma de Madrid (Spain), and partial support through PIP CONICET 11220170100638CO. 
YRG acknowledges the support of the
“Juan de la Cierva Incorporation” fellowship (ĲC2019-041131-I).
PBT acknowledges partial funding by Fondecyt 1200703/2020 (ANID),  ANID Basal projects ACE210002 and FB210003.
%%%%%%
This work used the DiRAC Data Centric system at Durham University, operated by the Institute for Computational Cosmology on behalf of the STFC DiRAC HPC Facility (www.dirac.ac.uk). This equipment was funded by BIS National E-infrastructure capital grant ST/K00042X/1, STFC capital grants ST/H008519/1 and ST/K00087X/1, STFC DiRAC Operations grant ST/K003267/1 and Durham University. DiRAC is part of the National E-Infrastructure.
For the analysis, we used the following libraries: \texttt{matplotlib} \citep{Hunter:2007}, \texttt{numpy}
\citep{Walt:2011}, \texttt{scipy} \citep{Jones:2001}, \texttt{scikit-learn} \citep{Pedregosa:2011}, \texttt{pandas} \citep{mckinney-proc-scipy-2010}, \texttt{seaborn} \citep{Waskom2021} and \texttt{F2PY} \citep{Peterson:2009}. We are grateful to all the people that have contributed to the development of these open source tools.

\section{Data availability}
The data shown in the figures of this article, as well as the new version of \texttt{gsf}, will be shared on reasonable request to the corresponding author.



\bibliographystyle{mnras}
\bibliography{example} % if your bibtex file is called example.bib

\begin{thebibliography}{}
\makeatletter
\relax
\def\mn@urlcharsother{\let\do\@makeother \do\$\do\&\do\#\do\^\do\_\do\%\do\~}
\def\mn@doi{\begingroup\mn@urlcharsother \@ifnextchar [ {\mn@doi@}
  {\mn@doi@[]}}
\def\mn@doi@[#1]#2{\def\@tempa{#1}\ifx\@tempa\@empty \href
  {http://dx.doi.org/#2} {doi:#2}\else \href {http://dx.doi.org/#2} {#1}\fi
  \endgroup}
\def\mn@eprint#1#2{\mn@eprint@#1:#2::\@nil}
\def\mn@eprint@arXiv#1{\href {http://arxiv.org/abs/#1} {{\tt arXiv:#1}}}
\def\mn@eprint@dblp#1{\href {http://dblp.uni-trier.de/rec/bibtex/#1.xml}
  {dblp:#1}}
\def\mn@eprint@#1:#2:#3:#4\@nil{\def\@tempa {#1}\def\@tempb {#2}\def\@tempc
  {#3}\ifx \@tempc \@empty \let \@tempc \@tempb \let \@tempb \@tempa \fi \ifx
  \@tempb \@empty \def\@tempb {arXiv}\fi \@ifundefined
  {mn@eprint@\@tempb}{\@tempb:\@tempc}{\expandafter \expandafter \csname
  mn@eprint@\@tempb\endcsname \expandafter{\@tempc}}}

\bibitem[\protect\citeauthoryear{{Abadi}, {Navarro}, {Steinmetz}  \&
  {Eke}}{{Abadi} et~al.}{2003}]{Abadi2003}
{Abadi} M.~G.,  {Navarro} J.~F.,  {Steinmetz} M.,   {Eke} V.~R.,  2003, \mn@doi
  [Astrophysical Journal] {10.1086/378316}, \href
  {https://ui.adsabs.harvard.edu/abs/2003ApJ...597...21A} {597, 21}

\bibitem[\protect\citeauthoryear{{Abdurro'uf} et~al.,}{{Abdurro'uf}
  et~al.}{2022}]{Abdurro:2022}
{Abdurro'uf} et~al., 2022, \mn@doi [\apjs] {10.3847/1538-4365/ac4414}, \href
  {https://ui.adsabs.harvard.edu/abs/2022ApJS..259...35A} {259, 35}

\bibitem[\protect\citeauthoryear{{Agertz}, {Teyssier}  \& {Moore}}{{Agertz}
  et~al.}{2011}]{Agertz2011}
{Agertz} O.,  {Teyssier} R.,   {Moore} B.,  2011, \mn@doi [\mnras]
  {10.1111/j.1365-2966.2010.17530.x}, \href
  {https://ui.adsabs.harvard.edu/abs/2011MNRAS.410.1391A} {410, 1391}

\bibitem[\protect\citeauthoryear{{Akaike}}{{Akaike}}{1974}]{Akaike:1974}
{Akaike} H.,  1974, IEEE Transactions on Automatic Control, \href
  {https://ui.adsabs.harvard.edu/abs/1974ITAC...19..716A} {19, 716}

\bibitem[\protect\citeauthoryear{{Amarsi}, {Nissen}  \&
  {Sk{\'u}lad{\'o}ttir}}{{Amarsi} et~al.}{2019}]{Amarsi2019}
{Amarsi} A.~M.,  {Nissen} P.~E.,   {Sk{\'u}lad{\'o}ttir} {\'A}.,  2019, \mn@doi
  [\aap] {10.1051/0004-6361/201936265}, \href
  {https://ui.adsabs.harvard.edu/abs/2019A&A...630A.104A} {630, A104}

\bibitem[\protect\citeauthoryear{{Angl{\'e}s-Alc{\'a}zar}, {Dav{\'e}},
  {Faucher-Gigu{\`e}re}, {{\"O}zel}  \& {Hopkins}}{{Angl{\'e}s-Alc{\'a}zar}
  et~al.}{2017}]{AnglesAlcazar:2017}
{Angl{\'e}s-Alc{\'a}zar} D.,  {Dav{\'e}} R.,  {Faucher-Gigu{\`e}re} C.-A.,
  {{\"O}zel} F.,   {Hopkins} P.~F.,  2017, \mn@doi [\mnras]
  {10.1093/mnras/stw2565}, \href
  {https://ui.adsabs.harvard.edu/abs/2017MNRAS.464.2840A} {464, 2840}

\bibitem[\protect\citeauthoryear{{Athanassoula}}{{Athanassoula}}{2005}]{Athanassoula:2005}
{Athanassoula} E.,  2005, \mn@doi [\mnras] {10.1111/j.1365-2966.2005.08872.x},
  \href {https://ui.adsabs.harvard.edu/abs/2005MNRAS.358.1477A} {358, 1477}

\bibitem[\protect\citeauthoryear{{Athanassoula} \& {Misiriotis}}{{Athanassoula}
  \& {Misiriotis}}{2002}]{Athanassoula2002}
{Athanassoula} E.,  {Misiriotis} A.,  2002, \mn@doi [\mnras]
  {10.1046/j.1365-8711.2002.05028.x}, \href
  {https://ui.adsabs.harvard.edu/abs/2002MNRAS.330...35A} {330, 35}

\bibitem[\protect\citeauthoryear{{Athanassoula}, {Lambert}  \&
  {Dehnen}}{{Athanassoula} et~al.}{2005}]{Athanassoula:2005b}
{Athanassoula} E.,  {Lambert} J.~C.,   {Dehnen} W.,  2005, \mn@doi [\mnras]
  {10.1111/j.1365-2966.2005.09445.x}, \href
  {https://ui.adsabs.harvard.edu/abs/2005MNRAS.363..496A} {363, 496}

\bibitem[\protect\citeauthoryear{{Athanassoula}, {Machado}  \&
  {Rodionov}}{{Athanassoula} et~al.}{2013}]{Athanassoula:2013}
{Athanassoula} E.,  {Machado} R. E.~G.,   {Rodionov} S.~A.,  2013, \mn@doi
  [\mnras] {10.1093/mnras/sts452}, \href
  {https://ui.adsabs.harvard.edu/abs/2013MNRAS.429.1949A} {429, 1949}

\bibitem[\protect\citeauthoryear{{Balcells}, {Dom{\'\i}nguez-Palmero}, {Graham}
   \& {Peletier}}{{Balcells} et~al.}{2001}]{Balcells2001}
{Balcells} M.,  {Dom{\'\i}nguez-Palmero} L.,  {Graham} A.,   {Peletier} R.~F.,
  2001, in {Knapen} J.~H.,  {Beckman} J.~E.,  {Shlosman} I.,   {Mahoney} T.~J.,
   eds,  Astronomical Society of the Pacific Conference Series Vol. 249, The
  Central Kiloparsec of Starbursts and AGN: The La Palma Connection. p.~167
  (\mn@eprint {arXiv} {astro-ph/0106413})

\bibitem[\protect\citeauthoryear{{Behroozi}, {Wechsler}, {Hearin}  \&
  {Conroy}}{{Behroozi} et~al.}{2019}]{Behroozi:2019}
{Behroozi} P.,  {Wechsler} R.~H.,  {Hearin} A.~P.,   {Conroy} C.,  2019,
  \mn@doi [\mnras] {10.1093/mnras/stz1182}, \href
  {https://ui.adsabs.harvard.edu/abs/2019MNRAS.488.3143B} {488, 3143}

\bibitem[\protect\citeauthoryear{{Belokurov} \& {Kravtsov}}{{Belokurov} \&
  {Kravtsov}}{2022}]{Belokurov2022}
{Belokurov} V.,  {Kravtsov} A.,  2022, arXiv e-prints, \href
  {https://ui.adsabs.harvard.edu/abs/2022arXiv220304980B} {p. arXiv:2203.04980}

\bibitem[\protect\citeauthoryear{{Bensby}, {Feltzing}  \&
  {Lundstr{\"o}m}}{{Bensby} et~al.}{2003}]{Bensby:2003}
{Bensby} T.,  {Feltzing} S.,   {Lundstr{\"o}m} I.,  2003, \mn@doi [\aap]
  {10.1051/0004-6361:20031213}, \href
  {https://ui.adsabs.harvard.edu/abs/2003A&A...410..527B} {410, 527}

\bibitem[\protect\citeauthoryear{{Bernardi} et~al.,}{{Bernardi}
  et~al.}{2003}]{Bernardi:2003}
{Bernardi} M.,  et~al., 2003, \mn@doi [\aj] {10.1086/367776}, \href
  {https://ui.adsabs.harvard.edu/abs/2003AJ....125.1817B} {125, 1817}

\bibitem[\protect\citeauthoryear{{Bignone}, {Helmi}  \& {Tissera}}{{Bignone}
  et~al.}{2019}]{Bignone2019}
{Bignone} L.~A.,  {Helmi} A.,   {Tissera} P.~B.,  2019, \mn@doi [\apjl]
  {10.3847/2041-8213/ab3e0e}, \href
  {https://ui.adsabs.harvard.edu/abs/2019ApJ...883L...5B} {883, L5}

\bibitem[\protect\citeauthoryear{{Binney} \& {Tremaine}}{{Binney} \&
  {Tremaine}}{2008}]{Libro}
{Binney} J.,  {Tremaine} S.,  2008, {Galactic Dynamics: Second Edition}

\bibitem[\protect\citeauthoryear{{Bird}, {Kazantzidis}, {Weinberg}, {Guedes},
  {Callegari}, {Mayer}  \& {Madau}}{{Bird} et~al.}{2013}]{Bird2013}
{Bird} J.~C.,  {Kazantzidis} S.,  {Weinberg} D.~H.,  {Guedes} J.,  {Callegari}
  S.,  {Mayer} L.,   {Madau} P.,  2013, \mn@doi [\apj]
  {10.1088/0004-637X/773/1/43}, \href
  {https://ui.adsabs.harvard.edu/abs/2013ApJ...773...43B} {773, 43}

\bibitem[\protect\citeauthoryear{{Bland-Hawthorn} \&
  {Gerhard}}{{Bland-Hawthorn} \& {Gerhard}}{2016}]{BlandHawthorn:2016}
{Bland-Hawthorn} J.,  {Gerhard} O.,  2016, \mn@doi [\araa]
  {10.1146/annurev-astro-081915-023441}, \href
  {https://ui.adsabs.harvard.edu/abs/2016ARA&A..54..529B} {54, 529}

\bibitem[\protect\citeauthoryear{{Bournaud} \& {Combes}}{{Bournaud} \&
  {Combes}}{2002}]{Bournaud:2002}
{Bournaud} F.,  {Combes} F.,  2002, \mn@doi [\aap]
  {10.1051/0004-6361:20020920}, \href
  {https://ui.adsabs.harvard.edu/abs/2002A&A...392...83B} {392, 83}

\bibitem[\protect\citeauthoryear{{Bovy} et~al.,}{{Bovy}
  et~al.}{2014}]{Bovy:2014}
{Bovy} J.,  et~al., 2014, \mn@doi [\apj] {10.1088/0004-637X/790/2/127}, \href
  {http://cdsads.u-strasbg.fr/abs/2014ApJ...790..127B} {790, 127}

\bibitem[\protect\citeauthoryear{{Bovy}, {Rix}, {Schlafly}, {Nidever},
  {Holtzman}, {Shetrone}  \& {Beers}}{{Bovy} et~al.}{2016}]{Bovy:2015}
{Bovy} J.,  {Rix} H.-W.,  {Schlafly} E.~F.,  {Nidever} D.~L.,  {Holtzman}
  J.~A.,  {Shetrone} M.,   {Beers} T.~C.,  2016, \mn@doi [\apj]
  {10.3847/0004-637X/823/1/30}, \href
  {http://cdsads.u-strasbg.fr/abs/2016ApJ...823...30B} {823, 30}

\bibitem[\protect\citeauthoryear{{Breda}, {Papaderos}  \& {Gomes}}{{Breda}
  et~al.}{2020}]{Breda2020}
{Breda} I.,  {Papaderos} P.,   {Gomes} J.-M.,  2020, \mn@doi [\aap]
  {10.1051/0004-6361/202037889}, \href
  {https://ui.adsabs.harvard.edu/abs/2020A&A...640A..20B} {640, A20}

\bibitem[\protect\citeauthoryear{{Brook}, {Kawata}, {Gibson}  \&
  {Freeman}}{{Brook} et~al.}{2004}]{Brook2004}
{Brook} C.~B.,  {Kawata} D.,  {Gibson} B.~K.,   {Freeman} K.~C.,  2004, \mn@doi
  [\apj] {10.1086/422709}, \href
  {https://ui.adsabs.harvard.edu/abs/2004ApJ...612..894B} {612, 894}

\bibitem[\protect\citeauthoryear{Buck, Obreja, Macciò, Minchev, Dutton  \&
  Ostriker}{Buck et~al.}{2019}]{Buck_2019}
Buck T.,  Obreja A.,  Macciò A.~V.,  Minchev I.,  Dutton A.~A.,   Ostriker
  J.~P.,  2019, \mn@doi [Monthly Notices of the Royal Astronomical Society]
  {10.1093/mnras/stz3241}

\bibitem[\protect\citeauthoryear{{Buck}, {Obreja}, {Macci{\`o}}, {Minchev},
  {Dutton}  \& {Ostriker}}{{Buck} et~al.}{2020}]{Buck:2020}
{Buck} T.,  {Obreja} A.,  {Macci{\`o}} A.~V.,  {Minchev} I.,  {Dutton} A.~A.,
  {Ostriker} J.~P.,  2020, \mn@doi [\mnras] {10.1093/mnras/stz3241}, \href
  {https://ui.adsabs.harvard.edu/abs/2020MNRAS.491.3461B} {491, 3461}

\bibitem[\protect\citeauthoryear{{Bulteel}, {Wilderjans}, {Tuerlinckx}  \&
  {Ceulemans}}{{Bulteel} et~al.}{2013}]{Bulteel2013}
{Bulteel} K.,  {Wilderjans} T.~F.,  {Tuerlinckx} F.,   {Ceulemans} E.,  2013,
  \mn@doi [Behavior Research Methods] {10.3758/s13428-012-0293-y}, \href
  {https://doi.org/10.3758/s13428-012-0293-y} {45, 782}

\bibitem[\protect\citeauthoryear{{Cair{\'o}s}, {Caon}, {Papaderos}, {Noeske},
  {V{\'\i}lchez}, {Garc{\'\i}a Lorenzo}  \&
  {Mu{\~n}oz-Tu{\~n}{\'o}n}}{{Cair{\'o}s} et~al.}{2003}]{Cairos:2003}
{Cair{\'o}s} L.~M.,  {Caon} N.,  {Papaderos} P.,  {Noeske} K.,  {V{\'\i}lchez}
  J.~M.,  {Garc{\'\i}a Lorenzo} B.,   {Mu{\~n}oz-Tu{\~n}{\'o}n} C.,  2003,
  \mn@doi [\apj] {10.1086/376516}, \href
  {https://ui.adsabs.harvard.edu/abs/2003ApJ...593..312C} {593, 312}

\bibitem[\protect\citeauthoryear{{Caon}, {Capaccioli}  \& {D'Onofrio}}{{Caon}
  et~al.}{1993}]{Caon1993}
{Caon} N.,  {Capaccioli} M.,   {D'Onofrio} M.,  1993, \mn@doi [\mnras]
  {10.1093/mnras/265.4.1013}, \href
  {https://ui.adsabs.harvard.edu/abs/1993MNRAS.265.1013C} {265, 1013}

\bibitem[\protect\citeauthoryear{{Cattell}}{{Cattell}}{1966}]{Cattell1966}
{Cattell} R.~B.,  1966, \mn@doi [Multivariate Behavioral Research]
  {10.1207/s15327906mbr0102_10}, \href
  {https://doi.org/10.1207/s15327906mbr0102_10} {1, 245}

\bibitem[\protect\citeauthoryear{{Cautun} et~al.,}{{Cautun}
  et~al.}{2020}]{Cautun:2020}
{Cautun} M.,  et~al., 2020, \mn@doi [\mnras] {10.1093/mnras/staa1017}, \href
  {https://ui.adsabs.harvard.edu/abs/2020MNRAS.494.4291C} {494, 4291}

\bibitem[\protect\citeauthoryear{{Cavanagh}, {Bekki}, {Groves}  \&
  {Pfeffer}}{{Cavanagh} et~al.}{2022}]{Cavanagh:2022}
{Cavanagh} M.~K.,  {Bekki} K.,  {Groves} B.~A.,   {Pfeffer} J.,  2022, \mn@doi
  [\mnras] {10.1093/mnras/stab3786}, \href
  {https://ui.adsabs.harvard.edu/abs/2022MNRAS.510.5164C} {510, 5164}

\bibitem[\protect\citeauthoryear{{Ceulemans} \& {Kiers}}{{Ceulemans} \&
  {Kiers}}{2006}]{Ceulemans2006}
{Ceulemans} E.,  {Kiers} H. A.~L.,  2006, \mn@doi [British Journal of
  Mathematical and Statistical Psychology] {10.1348/000711005X64817}, \href
  {https://doi.org/10.1348/000711005X64817} {59, 133}

\bibitem[\protect\citeauthoryear{{Chabrier}}{{Chabrier}}{2003}]{Chabrier:2003}
{Chabrier} G.,  2003, \mn@doi [\pasp] {10.1086/376392}, \href
  {https://ui.adsabs.harvard.edu/abs/2003PASP..115..763C} {115, 763}

\bibitem[\protect\citeauthoryear{{Comer{\'o}n}, {Elmegreen}, {Knapen}, {Salo}
  \& {et al.}}{{Comer{\'o}n} et~al.}{2011}]{Comeron:2011}
{Comer{\'o}n} S.,  {Elmegreen} B.~G.,  {Knapen} J.~H.,  {Salo} H.,   {et al.}
  2011, \mn@doi [\apj] {10.1088/0004-637X/741/1/28}, \href
  {http://cdsads.u-strasbg.fr/abs/2011ApJ...741...28C} {741, 28}

\bibitem[\protect\citeauthoryear{{Comer{\'o}n}, {Elmegreen}, {Salo},
  {Laurikainen}, {Holwerda}  \& {Knapen}}{{Comer{\'o}n}
  et~al.}{2014}]{Comeron:2014}
{Comer{\'o}n} S.,  {Elmegreen} B.~G.,  {Salo} H.,  {Laurikainen} E.,
  {Holwerda} B.~W.,   {Knapen} J.~H.,  2014, \mn@doi [\aap]
  {10.1051/0004-6361/201424412}, \href
  {https://ui.adsabs.harvard.edu/abs/2014A&A...571A..58C} {571, A58}

\bibitem[\protect\citeauthoryear{{Correa}, {Schaye}, {Clauwens}, {Bower},
  {Crain}, {Schaller}, {Theuns}  \& {Thob}}{{Correa} et~al.}{2017}]{Correa2017}
{Correa} C.~A.,  {Schaye} J.,  {Clauwens} B.,  {Bower} R.~G.,  {Crain} R.~A.,
  {Schaller} M.,  {Theuns} T.,   {Thob} A. C.~R.,  2017, \mn@doi [\mnras]
  {10.1093/mnrasl/slx133}, \href
  {https://ui.adsabs.harvard.edu/abs/2017MNRAS.472L..45C} {472, L45}

\bibitem[\protect\citeauthoryear{Crain et~al.,}{Crain et~al.}{2015}]{Crain}
Crain R.~A.,  et~al., 2015, Monthly Notices of the Royal Astronomical Society,
  450, 1937

\bibitem[\protect\citeauthoryear{{Dalcanton} \& {Bernstein}}{{Dalcanton} \&
  {Bernstein}}{2002}]{Dalcanton}
{Dalcanton} J.~J.,  {Bernstein} R.~A.,  2002, \mn@doi [\aj] {10.1086/342286},
  \href {https://ui.adsabs.harvard.edu/abs/2002AJ....124.1328D} {124, 1328}

\bibitem[\protect\citeauthoryear{{Dalla Vecchia} \& {Schaye}}{{Dalla Vecchia}
  \& {Schaye}}{2008}]{DallaVecchia:2008}
{Dalla Vecchia} C.,  {Schaye} J.,  2008, \mn@doi [\mnras]
  {10.1111/j.1365-2966.2008.13322.x}, \href
  {https://ui.adsabs.harvard.edu/abs/2008MNRAS.387.1431D} {387, 1431}

\bibitem[\protect\citeauthoryear{{Dalla Vecchia} \& {Schaye}}{{Dalla Vecchia}
  \& {Schaye}}{2012}]{DallaVecchia2012}
{Dalla Vecchia} C.,  {Schaye} J.,  2012, \mn@doi [\mnras]
  {10.1111/j.1365-2966.2012.21704.x}, \href
  {https://ui.adsabs.harvard.edu/abs/2012MNRAS.426..140D} {426, 140}

\bibitem[\protect\citeauthoryear{{Dehnen}}{{Dehnen}}{2000}]{Dehnen:2000}
{Dehnen} W.,  2000, \mn@doi [\aj] {10.1086/301226}, \href
  {https://ui.adsabs.harvard.edu/abs/2000AJ....119..800D} {119, 800}

\bibitem[\protect\citeauthoryear{Dhillon, Guan  \& Kulis}{Dhillon
  et~al.}{2004}]{Dhillon}
Dhillon I.~S.,  Guan Y.,   Kulis B.~J.,  2004, in ACM SIGKDD International
  Conference on Knowledge Discovery and Data Mining (KDD).

\bibitem[\protect\citeauthoryear{{Dom{\'e}nech-Moral}, {Mart{\'\i}nez-Serrano},
  {Dom{\'\i}nguez-Tenreiro}  \& {Serna}}{{Dom{\'e}nech-Moral}
  et~al.}{2012}]{Domenech2012}
{Dom{\'e}nech-Moral} M.,  {Mart{\'\i}nez-Serrano} F.~J.,
  {Dom{\'\i}nguez-Tenreiro} R.,   {Serna} A.,  2012, \mn@doi [\mnras]
  {10.1111/j.1365-2966.2012.20534.x}, \href
  {https://ui.adsabs.harvard.edu/abs/2012MNRAS.421.2510D} {421, 2510}

\bibitem[\protect\citeauthoryear{{Dom{\'\i}nguez-Tenreiro}, {Tissera}  \&
  {S{\'a}iz}}{{Dom{\'\i}nguez-Tenreiro} et~al.}{1998}]{Rosa1998}
{Dom{\'\i}nguez-Tenreiro} R.,  {Tissera} P.~B.,   {S{\'a}iz} A.,  1998, \mn@doi
  [\apjl] {10.1086/311733}, \href
  {https://ui.adsabs.harvard.edu/abs/1998ApJ...508L.123D} {508, L123}

\bibitem[\protect\citeauthoryear{{Du}, {Ho}, {Zhao}, {Shi}, {Debattista},
  {Hernquist}  \& {Nelson}}{{Du} et~al.}{2019}]{Du2019}
{Du} M.,  {Ho} L.~C.,  {Zhao} D.,  {Shi} J.,  {Debattista} V.~P.,  {Hernquist}
  L.,   {Nelson} D.,  2019, \mn@doi [Astrophysical Journal]
  {10.3847/1538-4357/ab43cc}, \href
  {https://ui.adsabs.harvard.edu/abs/2019ApJ...884..129D} {884, 129}

\bibitem[\protect\citeauthoryear{{Edvardsson}, {Andersen}, {Gustafsson},
  {Lambert}, {Nissen}  \& {Tomkin}}{{Edvardsson} et~al.}{1993}]{Edvardsson1993}
{Edvardsson} B.,  {Andersen} J.,  {Gustafsson} B.,  {Lambert} D.~L.,  {Nissen}
  P.~E.,   {Tomkin} J.,  1993, \aap, \href
  {https://ui.adsabs.harvard.edu/abs/1993A&A...275..101E} {275, 101}

\bibitem[\protect\citeauthoryear{Eric~Jones et~al.}{Eric~Jones
  et~al.}{2001}]{Jones:2001}
Eric~Jones Travis~Oliphant P.~P.,  et~al., 2001, {SciPy: Open source scientific
  tools for Python}, \url {http://www.scipy.org/}

\bibitem[\protect\citeauthoryear{{Erwin}, {Pohlen}  \& {Beckman}}{{Erwin}
  et~al.}{2008}]{Erwin:2008}
{Erwin} P.,  {Pohlen} M.,   {Beckman} J.~E.,  2008, \mn@doi [\aj]
  {10.1088/0004-6256/135/1/20}, \href
  {http://adsabs.harvard.edu/abs/2008AJ....135...20E} {135, 20}

\bibitem[\protect\citeauthoryear{{Erwin} et~al.,}{{Erwin}
  et~al.}{2015}]{Erwin:2015}
{Erwin} P.,  et~al., 2015, \mn@doi [\mnras] {10.1093/mnras/stu2376}, \href
  {https://ui.adsabs.harvard.edu/abs/2015MNRAS.446.4039E} {446, 4039}

\bibitem[\protect\citeauthoryear{{Fisher} \& {Drory}}{{Fisher} \&
  {Drory}}{2011}]{Fisher:2011}
{Fisher} D.~B.,  {Drory} N.,  2011, \mn@doi [\apjl]
  {10.1088/2041-8205/733/2/L47}, \href
  {https://ui.adsabs.harvard.edu/abs/2011ApJ...733L..47F} {733, L47}

\bibitem[\protect\citeauthoryear{{Font}, {McCarthy}, {Crain}, {Theuns},
  {Schaye}, {Wiersma}  \& {Dalla Vecchia}}{{Font} et~al.}{2011}]{Font2012}
{Font} A.~S.,  {McCarthy} I.~G.,  {Crain} R.~A.,  {Theuns} T.,  {Schaye} J.,
  {Wiersma} R.~P.~C.,   {Dalla Vecchia} C.,  2011, \mn@doi [\mnras]
  {10.1111/j.1365-2966.2011.19227.x}, \href
  {https://ui.adsabs.harvard.edu/abs/2011MNRAS.416.2802F} {416, 2802}

\bibitem[\protect\citeauthoryear{{F{\"o}rster Schreiber} et~al.,}{{F{\"o}rster
  Schreiber} et~al.}{2006}]{ForsterSchreiber:2006}
{F{\"o}rster Schreiber} N.~M.,  et~al., 2006, \mn@doi [\apj] {10.1086/504403},
  \href {https://ui.adsabs.harvard.edu/abs/2006ApJ...645.1062F} {645, 1062}

\bibitem[\protect\citeauthoryear{{Fragkoudi}, {Di Matteo}, {Haywood},
  {G{\'o}mez}, {Combes}, {Katz}  \& {Semelin}}{{Fragkoudi}
  et~al.}{2017}]{Fragkoudi:2017}
{Fragkoudi} F.,  {Di Matteo} P.,  {Haywood} M.,  {G{\'o}mez} A.,  {Combes} F.,
  {Katz} D.,   {Semelin} B.,  2017, \mn@doi [\aap]
  {10.1051/0004-6361/201630244}, \href
  {https://ui.adsabs.harvard.edu/abs/2017A&A...606A..47F} {606, A47}

\bibitem[\protect\citeauthoryear{{Fraternali}, {Karim}, {Magnelli},
  {G{\'o}mez-Guijarro}, {Jim{\'e}nez-Andrade}  \& {Posses}}{{Fraternali}
  et~al.}{2021}]{Fraternali2021}
{Fraternali} F.,  {Karim} A.,  {Magnelli} B.,  {G{\'o}mez-Guijarro} C.,
  {Jim{\'e}nez-Andrade} E.~F.,   {Posses} A.~C.,  2021, \mn@doi [\aap]
  {10.1051/0004-6361/202039807}, \href
  {https://ui.adsabs.harvard.edu/abs/2021A&A...647A.194F} {647, A194}

\bibitem[\protect\citeauthoryear{{Freeman}}{{Freeman}}{1970}]{Freeman1970}
{Freeman} K.~C.,  1970, \mn@doi [\apj] {10.1086/150474}, \href
  {https://ui.adsabs.harvard.edu/abs/1970ApJ...160..811F} {160, 811}

\bibitem[\protect\citeauthoryear{{Fuhrmann}}{{Fuhrmann}}{1998}]{Fuhrmann:1998}
{Fuhrmann} K.,  1998, \aap, \href
  {https://ui.adsabs.harvard.edu/abs/1998A&A...338..161F} {338, 161}

\bibitem[\protect\citeauthoryear{{Gadotti}}{{Gadotti}}{2009}]{Gadotti:2009}
{Gadotti} D.~A.,  2009, \mn@doi [\mnras] {10.1111/j.1365-2966.2008.14257.x},
  \href {https://ui.adsabs.harvard.edu/abs/2009MNRAS.393.1531G} {393, 1531}

\bibitem[\protect\citeauthoryear{{Gargiulo} et~al.,}{{Gargiulo}
  et~al.}{2019}]{Gargiulo:2019}
{Gargiulo} I.~D.,  et~al., 2019, \mn@doi [\mnras] {10.1093/mnras/stz2536},
  \href {https://ui.adsabs.harvard.edu/abs/2019MNRAS.489.5742G} {489, 5742}

\bibitem[\protect\citeauthoryear{{Garrison-Kimmel} et~al.,}{{Garrison-Kimmel}
  et~al.}{2018}]{Garrison-Kimmel:2018}
{Garrison-Kimmel} S.,  et~al., 2018, \mn@doi [\mnras] {10.1093/mnras/sty2513},
  \href {https://ui.adsabs.harvard.edu/abs/2018MNRAS.481.4133G} {481, 4133}

\bibitem[\protect\citeauthoryear{{Gilmore} \& {Reid}}{{Gilmore} \&
  {Reid}}{1983}]{Gilmore}
{Gilmore} G.,  {Reid} N.,  1983, \mn@doi [\mnras] {10.1093/mnras/202.4.1025},
  \href {https://ui.adsabs.harvard.edu/abs/1983MNRAS.202.1025G} {202, 1025}

\bibitem[\protect\citeauthoryear{{Gilmore}, {Wyse}  \& {Kuijken}}{{Gilmore}
  et~al.}{1989}]{Gilmore:1989}
{Gilmore} G.,  {Wyse} R. F.~G.,   {Kuijken} K.,  1989, \mn@doi [\araa]
  {10.1146/annurev.aa.27.090189.003011}, \href
  {https://ui.adsabs.harvard.edu/abs/1989ARA&A..27..555G} {27, 555}

\bibitem[\protect\citeauthoryear{{G{\'o}mez} et~al.,}{{G{\'o}mez}
  et~al.}{2017}]{Gomez2017}
{G{\'o}mez} F.~A.,  et~al., 2017, \mn@doi [\mnras] {10.1093/mnras/stx2149},
  \href {https://ui.adsabs.harvard.edu/abs/2017MNRAS.472.3722G} {472, 3722}

\bibitem[\protect\citeauthoryear{{G{\'o}rski}, {Hivon}, {Banday}, {Wandelt},
  {Hansen}, {Reinecke}  \& {Bartelmann}}{{G{\'o}rski}
  et~al.}{2005}]{Gorski:2005}
{G{\'o}rski} K.~M.,  {Hivon} E.,  {Banday} A.~J.,  {Wandelt} B.~D.,  {Hansen}
  F.~K.,  {Reinecke} M.,   {Bartelmann} M.,  2005, \mn@doi [\apj]
  {10.1086/427976}, \href
  {https://ui.adsabs.harvard.edu/abs/2005ApJ...622..759G} {622, 759}

\bibitem[\protect\citeauthoryear{{Graham} \& {Colless}}{{Graham} \&
  {Colless}}{1997}]{Graham:1997}
{Graham} A.,  {Colless} M.,  1997, \mn@doi [\mnras] {10.1093/mnras/287.1.221},
  \href {https://ui.adsabs.harvard.edu/abs/1997MNRAS.287..221G} {287, 221}

\bibitem[\protect\citeauthoryear{{Graham}, {Trujillo}  \& {Caon}}{{Graham}
  et~al.}{2001}]{Graham:2001}
{Graham} A.~W.,  {Trujillo} I.,   {Caon} N.,  2001, \mn@doi [\aj]
  {10.1086/323090}, \href
  {https://ui.adsabs.harvard.edu/abs/2001AJ....122.1707G} {122, 1707}

\bibitem[\protect\citeauthoryear{{Grand} et~al.,}{{Grand}
  et~al.}{2017}]{Grand:2017}
{Grand} R. J.~J.,  et~al., 2017, \mn@doi [\mnras] {10.1093/mnras/stx071}, \href
  {https://ui.adsabs.harvard.edu/abs/2017MNRAS.467..179G} {467, 179}

\bibitem[\protect\citeauthoryear{{Grand} et~al.,}{{Grand}
  et~al.}{2020}]{Grand2020}
{Grand} R. J.~J.,  et~al., 2020, \mn@doi [\mnras] {10.1093/mnras/staa2057},
  \href {https://ui.adsabs.harvard.edu/abs/2020MNRAS.497.1603G} {497, 1603}

\bibitem[\protect\citeauthoryear{{Hayden}, {Bovy}, {Holtzman}, {Nidever}  \&
  {et al.}}{{Hayden} et~al.}{2015}]{Hayden:2015}
{Hayden} M.~R.,  {Bovy} J.,  {Holtzman} J.~A.,  {Nidever} D.~L.,   {et al.}
  2015, \mn@doi [\apj] {10.1088/0004-637X/808/2/132}, \href
  {http://cdsads.u-strasbg.fr/abs/2015ApJ...808..132H} {808, 132}

\bibitem[\protect\citeauthoryear{{Hopkins} et~al.,}{{Hopkins}
  et~al.}{2018}]{Hopkins:2018}
{Hopkins} P.~F.,  et~al., 2018, \mn@doi [\mnras] {10.1093/mnras/sty1690}, \href
  {https://ui.adsabs.harvard.edu/abs/2018MNRAS.480..800H} {480, 800}

\bibitem[\protect\citeauthoryear{{Hunter}}{{Hunter}}{2007}]{Hunter:2007}
{Hunter} J.~D.,  2007, \mn@doi [Computing in Science and Engineering]
  {10.1109/MCSE.2007.55}, \href
  {http://adsabs.harvard.edu/abs/2007CSE.....9...90H} {9, 90}

\bibitem[\protect\citeauthoryear{{Iodice} et~al.,}{{Iodice}
  et~al.}{2016}]{Iodice:2016}
{Iodice} E.,  et~al., 2016, \mn@doi [\apj] {10.3847/0004-637X/820/1/42}, \href
  {https://ui.adsabs.harvard.edu/abs/2016ApJ...820...42I} {820, 42}

\bibitem[\protect\citeauthoryear{{Irodotou} \& {Thomas}}{{Irodotou} \&
  {Thomas}}{2021}]{Irodotou2021}
{Irodotou} D.,  {Thomas} P.~A.,  2021, \mn@doi [\mnras]
  {10.1093/mnras/staa3804}, \href
  {https://ui.adsabs.harvard.edu/abs/2021MNRAS.501.2182I} {501, 2182}

\bibitem[\protect\citeauthoryear{{Ivezi{\'c}}, {Beers}  \&
  {Juri{\'c}}}{{Ivezi{\'c}} et~al.}{2012}]{Ivezic:2012}
{Ivezi{\'c}} {\v Z}.,  {Beers} T.~C.,   {Juri{\'c}} M.,  2012, \mn@doi [\araa]
  {10.1146/annurev-astro-081811-125504}, \href
  {http://cdsads.u-strasbg.fr/abs/2012ARA%26A..50..251I} {50, 251}

\bibitem[\protect\citeauthoryear{Karatzoglou, Smola, Hornik  \&
  Zeileis}{Karatzoglou et~al.}{2004}]{Karat}
Karatzoglou A.,  Smola A.,  Hornik K.,   Zeileis A.,  2004, \mn@doi [Journal of
  Statistical Software, Articles] {10.18637/jss.v011.i09}, 11, 1

\bibitem[\protect\citeauthoryear{{Katz}}{{Katz}}{1992}]{Katz:1992}
{Katz} N.,  1992, \mn@doi [\apj] {10.1086/171366}, \href
  {https://ui.adsabs.harvard.edu/abs/1992ApJ...391..502K} {391, 502}

\bibitem[\protect\citeauthoryear{{Kazantzidis}, {Bullock}, {Zentner},
  {Kravtsov}  \& {Moustakas}}{{Kazantzidis} et~al.}{2008}]{Kazantzidis:2008}
{Kazantzidis} S.,  {Bullock} J.~S.,  {Zentner} A.~R.,  {Kravtsov} A.~V.,
  {Moustakas} L.~A.,  2008, \mn@doi [\apj] {10.1086/591958}, \href
  {https://ui.adsabs.harvard.edu/abs/2008ApJ...688..254K} {688, 254}

\bibitem[\protect\citeauthoryear{{Kim}, {Gadotti}, {Athanassoula}, {Bosma},
  {Sheth}  \& {Lee}}{{Kim} et~al.}{2016}]{Kim:2016}
{Kim} T.,  {Gadotti} D.~A.,  {Athanassoula} E.,  {Bosma} A.,  {Sheth} K.,
  {Lee} M.~G.,  2016, \mn@doi [\mnras] {10.1093/mnras/stw1899}, \href
  {https://ui.adsabs.harvard.edu/abs/2016MNRAS.462.3430K} {462, 3430}

\bibitem[\protect\citeauthoryear{{Kormendy} \& {Kennicutt}}{{Kormendy} \&
  {Kennicutt}}{2004}]{Kormendy:2004}
{Kormendy} J.,  {Kennicutt} Robert~C. J.,  2004, \mn@doi [\araa]
  {10.1146/annurev.astro.42.053102.134024}, \href
  {https://ui.adsabs.harvard.edu/abs/2004ARA&A..42..603K} {42, 603}

\bibitem[\protect\citeauthoryear{{Lagos}, {Schaye}, {Bah{\'e}}, {van de Sande},
  {Kay}, {Barnes}, {Davis}  \& {Dalla Vecchia}}{{Lagos}
  et~al.}{2018}]{Lagos2018}
{Lagos} C. d.~P.,  {Schaye} J.,  {Bah{\'e}} Y.,  {van de Sande} J.,  {Kay}
  S.~T.,  {Barnes} D.,  {Davis} T.~A.,   {Dalla Vecchia} C.,  2018, \mn@doi
  [\mnras] {10.1093/mnras/sty489}, \href
  {https://ui.adsabs.harvard.edu/abs/2018MNRAS.476.4327L} {476, 4327}

\bibitem[\protect\citeauthoryear{{Laurikainen}, {Salo}, {Athanassoula}, {Bosma}
   \& {Herrera-Endoqui}}{{Laurikainen} et~al.}{2014}]{Laurikainen:2014}
{Laurikainen} E.,  {Salo} H.,  {Athanassoula} E.,  {Bosma} A.,
  {Herrera-Endoqui} M.,  2014, \mn@doi [\mnras] {10.1093/mnrasl/slu118}, \href
  {https://ui.adsabs.harvard.edu/abs/2014MNRAS.444L..80L} {444, L80}

\bibitem[\protect\citeauthoryear{{Lelli}, {Di Teodoro}, {Fraternali}, {Man},
  {Zhang}, {De Breuck}, {Davis}  \& {Maiolino}}{{Lelli}
  et~al.}{2021}]{Lelli2021}
{Lelli} F.,  {Di Teodoro} E.~M.,  {Fraternali} F.,  {Man} A. W.~S.,  {Zhang}
  Z.-Y.,  {De Breuck} C.,  {Davis} T.~A.,   {Maiolino} R.,  2021, \mn@doi
  [Science] {10.1126/science.abc1893}, \href
  {https://ui.adsabs.harvard.edu/abs/2021Sci...371..713L} {371, 713}

\bibitem[\protect\citeauthoryear{{Marinacci}, {Pakmor}  \&
  {Springel}}{{Marinacci} et~al.}{2014}]{Marinacci2014}
{Marinacci} F.,  {Pakmor} R.,   {Springel} V.,  2014, \mn@doi [\mnras]
  {10.1093/mnras/stt2003}, \href
  {https://ui.adsabs.harvard.edu/abs/2014MNRAS.437.1750M} {437, 1750}

\bibitem[\protect\citeauthoryear{{Marinacci} et~al.,}{{Marinacci}
  et~al.}{2018}]{Marinacci2018}
{Marinacci} F.,  et~al., 2018, \mn@doi [\mnras] {10.1093/mnras/sty2206}, \href
  {https://ui.adsabs.harvard.edu/abs/2018MNRAS.480.5113M} {480, 5113}

\bibitem[\protect\citeauthoryear{{Mart{\'\i}nez-Delgado}
  et~al.,}{{Mart{\'\i}nez-Delgado} et~al.}{2010}]{MartinezDelgado:2010}
{Mart{\'\i}nez-Delgado} D.,  et~al., 2010, \mn@doi [\aj]
  {10.1088/0004-6256/140/4/962}, \href
  {https://ui.adsabs.harvard.edu/abs/2010AJ....140..962M} {140, 962}

\bibitem[\protect\citeauthoryear{{M{\'e}ndez-Abreu}, {Ruiz-Lara},
  {S{\'a}nchez-Menguiano}  \& {et al.}}{{M{\'e}ndez-Abreu}
  et~al.}{2017}]{MendezAbreu:2017}
{M{\'e}ndez-Abreu} J.,  {Ruiz-Lara} T.,  {S{\'a}nchez-Menguiano} L.,   {et al.}
  2017, \mn@doi [\aap] {10.1051/0004-6361/201629525}, \href
  {http://adsabs.harvard.edu/abs/2017A%26A...598A..32M} {598, A32}

\bibitem[\protect\citeauthoryear{{Monachesi} et~al.,}{{Monachesi}
  et~al.}{2019}]{Monachesi2019}
{Monachesi} A.,  et~al., 2019, \mn@doi [\mnras] {10.1093/mnras/stz538}, \href
  {https://ui.adsabs.harvard.edu/abs/2019MNRAS.485.2589M} {485, 2589}

\bibitem[\protect\citeauthoryear{{Moster}, {Naab}  \& {White}}{{Moster}
  et~al.}{2018}]{Moster:2018}
{Moster} B.~P.,  {Naab} T.,   {White} S. D.~M.,  2018, \mn@doi [\mnras]
  {10.1093/mnras/sty655}, \href
  {https://ui.adsabs.harvard.edu/abs/2018MNRAS.477.1822M} {477, 1822}

\bibitem[\protect\citeauthoryear{{Naiman} et~al.,}{{Naiman}
  et~al.}{2018}]{Naiman2018}
{Naiman} J.~P.,  et~al., 2018, \mn@doi [\mnras] {10.1093/mnras/sty618}, \href
  {https://ui.adsabs.harvard.edu/abs/2018MNRAS.477.1206N} {477, 1206}

\bibitem[\protect\citeauthoryear{{Navarro}, {Frenk}  \& {White}}{{Navarro}
  et~al.}{1997}]{NFW:1997}
{Navarro} J.~F.,  {Frenk} C.~S.,   {White} S. D.~M.,  1997, \mn@doi [\apj]
  {10.1086/304888}, \href
  {https://ui.adsabs.harvard.edu/abs/1997ApJ...490..493N} {490, 493}

\bibitem[\protect\citeauthoryear{{Neeleman}, {Prochaska}, {Kanekar}  \&
  {Rafelski}}{{Neeleman} et~al.}{2020}]{Neeleman2020}
{Neeleman} M.,  {Prochaska} J.~X.,  {Kanekar} N.,   {Rafelski} M.,  2020,
  \mn@doi [\nat] {10.1038/s41586-020-2276-y}, \href
  {https://ui.adsabs.harvard.edu/abs/2020Natur.581..269N} {581, 269}

\bibitem[\protect\citeauthoryear{{Nelson} et~al.,}{{Nelson}
  et~al.}{2018}]{Nelson2018}
{Nelson} D.,  et~al., 2018, \mn@doi [\mnras] {10.1093/mnras/stx3040}, \href
  {https://ui.adsabs.harvard.edu/abs/2018MNRAS.475..624N} {475, 624}

\bibitem[\protect\citeauthoryear{{Obreja}, {Dom{\'\i}nguez-Tenreiro}, {Brook},
  {Mart{\'\i}nez-Serrano}, {Dom{\'e}nech-Moral}, {Serna}, {Moll{\'a}}  \&
  {Stinson}}{{Obreja} et~al.}{2013}]{Obreja:2013}
{Obreja} A.,  {Dom{\'\i}nguez-Tenreiro} R.,  {Brook} C.,
  {Mart{\'\i}nez-Serrano} F.~J.,  {Dom{\'e}nech-Moral} M.,  {Serna} A.,
  {Moll{\'a}} M.,   {Stinson} G.,  2013, \mn@doi [\apj]
  {10.1088/0004-637X/763/1/26}, \href
  {https://ui.adsabs.harvard.edu/abs/2013ApJ...763...26O} {763, 26}

\bibitem[\protect\citeauthoryear{{Obreja}, {Stinson}, {Dutton}, {Macci{\`o}},
  {Wang}  \& {Kang}}{{Obreja} et~al.}{2016}]{Obreja2016}
{Obreja} A.,  {Stinson} G.~S.,  {Dutton} A.~A.,  {Macci{\`o}} A.~V.,  {Wang}
  L.,   {Kang} X.,  2016, \mn@doi [\mnras] {10.1093/mnras/stw690}, \href
  {https://ui.adsabs.harvard.edu/abs/2016MNRAS.459..467O} {459, 467}

\bibitem[\protect\citeauthoryear{{Obreja}, {Macci{\`o}}, {Moster}, {Dutton},
  {Buck}, {Stinson}  \& {Wang}}{{Obreja} et~al.}{2018}]{Obreja2018}
{Obreja} A.,  {Macci{\`o}} A.~V.,  {Moster} B.,  {Dutton} A.~A.,  {Buck} T.,
  {Stinson} G.~S.,   {Wang} L.,  2018, \mn@doi [\mnras]
  {10.1093/mnras/sty1022}, \href
  {https://ui.adsabs.harvard.edu/abs/2018MNRAS.477.4915O} {477, 4915}

\bibitem[\protect\citeauthoryear{{Obreja} et~al.,}{{Obreja}
  et~al.}{2019}]{Obreja2019}
{Obreja} A.,  et~al., 2019, \mn@doi [\mnras] {10.1093/mnras/stz1563}, \href
  {https://ui.adsabs.harvard.edu/abs/2019MNRAS.487.4424O} {487, 4424}

\bibitem[\protect\citeauthoryear{{Obreja}, {Buck}  \& {Macci{\`o}}}{{Obreja}
  et~al.}{2022}]{Obreja:2022}
{Obreja} A.,  {Buck} T.,   {Macci{\`o}} A.~V.,  2022, \mn@doi [\aap]
  {10.1051/0004-6361/202140983}, \href
  {https://ui.adsabs.harvard.edu/abs/2022A&A...657A..15O} {657, A15}

\bibitem[\protect\citeauthoryear{{Okamoto}}{{Okamoto}}{2013}]{Okamoto:2013}
{Okamoto} T.,  2013, \mn@doi [\mnras] {10.1093/mnras/sts067}, \href
  {https://ui.adsabs.harvard.edu/abs/2013MNRAS.428..718O} {428, 718}

\bibitem[\protect\citeauthoryear{Pedregosa, Varoquaux, Gramfort  \& {et
  al.}}{Pedregosa et~al.}{2011}]{Pedregosa:2011}
Pedregosa F.,  Varoquaux G.,  Gramfort A.,   {et al.} 2011, Journal of Machine
  Learning Research, 12, 2825

\bibitem[\protect\citeauthoryear{{Pedregosa} et~al.,}{{Pedregosa}
  et~al.}{2012}]{scikit}
{Pedregosa} F.,  et~al., 2012, arXiv e-prints, \href
  {https://ui.adsabs.harvard.edu/abs/2012arXiv1201.0490P} {p. arXiv:1201.0490}

\bibitem[\protect\citeauthoryear{{Pedrosa} \& {Tissera}}{{Pedrosa} \&
  {Tissera}}{2015}]{Pedrosa2015}
{Pedrosa} S.~E.,  {Tissera} P.~B.,  2015, \mn@doi [\aap]
  {10.1051/0004-6361/201526440}, \href
  {https://ui.adsabs.harvard.edu/abs/2015A&A...584A..43P} {584, A43}

\bibitem[\protect\citeauthoryear{{Peebles}}{{Peebles}}{2020}]{Peebles:2020}
{Peebles} P.~J.~E.,  2020, \mn@doi [\mnras] {10.1093/mnras/staa2649}, \href
  {https://ui.adsabs.harvard.edu/abs/2020MNRAS.498.4386P} {498, 4386}

\bibitem[\protect\citeauthoryear{Peschken \& {\L}okas}{Peschken \&
  {\L}okas}{2018}]{Peschken:2018}
Peschken N.,  {\L}okas E.~L.,  2018, \mn@doi [Monthly Notices of the Royal
  Astronomical Society] {10.1093/mnras/sty3277}, 483, 2721

\bibitem[\protect\citeauthoryear{Peterson}{Peterson}{2009}]{Peterson:2009}
Peterson P.,  2009, International Journal of Computational Science and
  Engineering, 4, 296

\bibitem[\protect\citeauthoryear{{Pillepich} et~al.,}{{Pillepich}
  et~al.}{2018}]{Pillepich2018}
{Pillepich} A.,  et~al., 2018, \mn@doi [\mnras] {10.1093/mnras/stx3112}, \href
  {https://ui.adsabs.harvard.edu/abs/2018MNRAS.475..648P} {475, 648}

\bibitem[\protect\citeauthoryear{{Planck Collaboration} et~al.,}{{Planck
  Collaboration} et~al.}{2016}]{Planck2016}
{Planck Collaboration} et~al., 2016, \mn@doi [\aap]
  {10.1051/0004-6361/201525830}, \href
  {https://ui.adsabs.harvard.edu/abs/2016A&A...594A..13P} {594, A13}

\bibitem[\protect\citeauthoryear{{Prochaska}, {Naumov}, {Carney}, {McWilliam}
  \& {Wolfe}}{{Prochaska} et~al.}{2000}]{Prochaska2000}
{Prochaska} J.~X.,  {Naumov} S.~O.,  {Carney} B.~W.,  {McWilliam} A.,   {Wolfe}
  A.~M.,  2000, \mn@doi [\aj] {10.1086/316818}, \href
  {https://ui.adsabs.harvard.edu/abs/2000AJ....120.2513P} {120, 2513}

\bibitem[\protect\citeauthoryear{{Reddy}, {Tomkin}, {Lambert}  \& {Allende
  Prieto}}{{Reddy} et~al.}{2003}]{Reddy2003}
{Reddy} B.~E.,  {Tomkin} J.,  {Lambert} D.~L.,   {Allende Prieto} C.,  2003,
  \mn@doi [\mnras] {10.1046/j.1365-8711.2003.06305.x}, \href
  {https://ui.adsabs.harvard.edu/abs/2003MNRAS.340..304R} {340, 304}

\bibitem[\protect\citeauthoryear{{Reddy}, {Lambert}  \& {Allende
  Prieto}}{{Reddy} et~al.}{2006a}]{Reddy2006}
{Reddy} B.~E.,  {Lambert} D.~L.,   {Allende Prieto} C.,  2006a, \mn@doi
  [\mnras] {10.1111/j.1365-2966.2006.10148.x}, \href
  {https://ui.adsabs.harvard.edu/abs/2006MNRAS.367.1329R} {367, 1329}

\bibitem[\protect\citeauthoryear{{Reddy}, {Steidel}, {Erb}, {Shapley}  \&
  {Pettini}}{{Reddy} et~al.}{2006b}]{Reddy:2006}
{Reddy} N.~A.,  {Steidel} C.~C.,  {Erb} D.~K.,  {Shapley} A.~E.,   {Pettini}
  M.,  2006b, \mn@doi [\apj] {10.1086/508851}, \href
  {https://ui.adsabs.harvard.edu/abs/2006ApJ...653.1004R} {653, 1004}

\bibitem[\protect\citeauthoryear{{Richings} \& {Schaye}}{{Richings} \&
  {Schaye}}{2016}]{Richings:2016}
{Richings} A.~J.,  {Schaye} J.,  2016, \mn@doi [\mnras] {10.1093/mnras/stw327},
  \href {https://ui.adsabs.harvard.edu/abs/2016MNRAS.458..270R} {458, 270}

\bibitem[\protect\citeauthoryear{{Rizzo}, {Vegetti}, {Powell}, {Fraternali},
  {McKean}, {Stacey}  \& {White}}{{Rizzo} et~al.}{2020}]{Rizzo2020}
{Rizzo} F.,  {Vegetti} S.,  {Powell} D.,  {Fraternali} F.,  {McKean} J.~P.,
  {Stacey} H.~R.,   {White} S.~D.~M.,  2020, \mn@doi [\nat]
  {10.1038/s41586-020-2572-6}, \href
  {https://ui.adsabs.harvard.edu/abs/2020Natur.584..201R} {584, 201}

\bibitem[\protect\citeauthoryear{{Rosas-Guevara} et~al.,}{{Rosas-Guevara}
  et~al.}{2015}]{Yetli2015}
{Rosas-Guevara} Y.~M.,  et~al., 2015, \mn@doi [\mnras] {10.1093/mnras/stv2056},
  \href {https://ui.adsabs.harvard.edu/abs/2015MNRAS.454.1038R} {454, 1038}

\bibitem[\protect\citeauthoryear{{Rosas-Guevara} et~al.,}{{Rosas-Guevara}
  et~al.}{2020}]{Yetli2020}
{Rosas-Guevara} Y.,  et~al., 2020, \mn@doi [\mnras] {10.1093/mnras/stz3180},
  \href {https://ui.adsabs.harvard.edu/abs/2020MNRAS.491.2547R} {491, 2547}

\bibitem[\protect\citeauthoryear{{Rosas-Guevara} et~al.,}{{Rosas-Guevara}
  et~al.}{2022}]{Yetli2022}
{Rosas-Guevara} Y.,  et~al., 2022, \mn@doi [\mnras] {10.1093/mnras/stac816},
  \href {https://ui.adsabs.harvard.edu/abs/2022MNRAS.512.5339R} {512, 5339}

\bibitem[\protect\citeauthoryear{{Rousseeuw}}{{Rousseeuw}}{1984}]{Rousseeuw:1984}
{Rousseeuw} P.,  1984, Journal of the American Statistical Association, 79,
  871–880

\bibitem[\protect\citeauthoryear{{Rousseeuw}}{{Rousseeuw}}{1985}]{Rousseeuw:1985}
{Rousseeuw} P.,  1985, Mathematical statistics and applications, B, 283–297

\bibitem[\protect\citeauthoryear{{Ro{\v{s}}kar}, {Debattista}, {Quinn},
  {Stinson}  \& {Wadsley}}{{Ro{\v{s}}kar} et~al.}{2008}]{Roskar:2008}
{Ro{\v{s}}kar} R.,  {Debattista} V.~P.,  {Quinn} T.~R.,  {Stinson} G.~S.,
  {Wadsley} J.,  2008, \mn@doi [\apjl] {10.1086/592231}, \href
  {https://ui.adsabs.harvard.edu/abs/2008ApJ...684L..79R} {684, L79}

\bibitem[\protect\citeauthoryear{{Ro{\v{s}}kar}, {Debattista}  \&
  {Loebman}}{{Ro{\v{s}}kar} et~al.}{2013}]{Roskar:2013}
{Ro{\v{s}}kar} R.,  {Debattista} V.~P.,   {Loebman} S.~R.,  2013, \mn@doi
  [\mnras] {10.1093/mnras/stt788}, \href
  {https://ui.adsabs.harvard.edu/abs/2013MNRAS.433..976R} {433, 976}

\bibitem[\protect\citeauthoryear{{S{\'a}iz}, {Dom{\'\i}nguez-Tenreiro},
  {Tissera}  \& {Courteau}}{{S{\'a}iz} et~al.}{2001}]{Saiz2001}
{S{\'a}iz} A.,  {Dom{\'\i}nguez-Tenreiro} R.,  {Tissera} P.~B.,   {Courteau}
  S.,  2001, \mn@doi [\mnras] {10.1046/j.1365-8711.2001.04391.x}, \href
  {https://ui.adsabs.harvard.edu/abs/2001MNRAS.325..119S} {325, 119}

\bibitem[\protect\citeauthoryear{{Sales}, {Navarro}, {Schaye}, {Dalla Vecchia},
  {Springel}  \& {Booth}}{{Sales} et~al.}{2010}]{Sales2010}
{Sales} L.~V.,  {Navarro} J.~F.,  {Schaye} J.,  {Dalla Vecchia} C.,  {Springel}
  V.,   {Booth} C.~M.,  2010, \mn@doi [\mnras]
  {10.1111/j.1365-2966.2010.17391.x}, \href
  {https://ui.adsabs.harvard.edu/abs/2010MNRAS.409.1541S} {409, 1541}

\bibitem[\protect\citeauthoryear{{Samland} \& {Gerhard}}{{Samland} \&
  {Gerhard}}{2003}]{Samland:2003}
{Samland} M.,  {Gerhard} O.~E.,  2003, \mn@doi [\aap]
  {10.1051/0004-6361:20021842}, \href
  {https://ui.adsabs.harvard.edu/abs/2003A&A...399..961S} {399, 961}

\bibitem[\protect\citeauthoryear{{S{\'a}nchez} et~al.,}{{S{\'a}nchez}
  et~al.}{2012}]{IFUCalifa}
{S{\'a}nchez} S.~F.,  et~al., 2012, \mn@doi [\aap]
  {10.1051/0004-6361/201117353}, \href
  {https://ui.adsabs.harvard.edu/abs/2012A&A...538A...8S} {538, A8}

\bibitem[\protect\citeauthoryear{{Scannapieco}, {Gadotti}, {Jonsson}  \&
  {White}}{{Scannapieco} et~al.}{2010}]{Scannapieco:2010}
{Scannapieco} C.,  {Gadotti} D.~A.,  {Jonsson} P.,   {White} S. D.~M.,  2010,
  \mn@doi [\mnras] {10.1111/j.1745-3933.2010.00900.x}, \href
  {https://ui.adsabs.harvard.edu/abs/2010MNRAS.407L..41S} {407, L41}

\bibitem[\protect\citeauthoryear{{Schaller}, {Dalla Vecchia}, {Schaye},
  {Bower}, {Theuns}, {Crain}, {Furlong}  \& {McCarthy}}{{Schaller}
  et~al.}{2015}]{Schaller:2015}
{Schaller} M.,  {Dalla Vecchia} C.,  {Schaye} J.,  {Bower} R.~G.,  {Theuns} T.,
   {Crain} R.~A.,  {Furlong} M.,   {McCarthy} I.~G.,  2015, \mn@doi [\mnras]
  {10.1093/mnras/stv2169}, \href
  {https://ui.adsabs.harvard.edu/abs/2015MNRAS.454.2277S} {454, 2277}

\bibitem[\protect\citeauthoryear{{Schaye} \& {Dalla Vecchia}}{{Schaye} \&
  {Dalla Vecchia}}{2008}]{Schaye2008}
{Schaye} J.,  {Dalla Vecchia} C.,  2008, \mn@doi [\mnras]
  {10.1111/j.1365-2966.2007.12639.x}, \href
  {https://ui.adsabs.harvard.edu/abs/2008MNRAS.383.1210S} {383, 1210}

\bibitem[\protect\citeauthoryear{Schaye et~al.,}{Schaye et~al.}{2015}]{Schaye}
Schaye J.,  et~al., 2015, \mn@doi [Monthly Notices of the Royal Astronomical
  Society] {10.1093/mnras/stu2058}, 446, 521

\bibitem[\protect\citeauthoryear{{Sch{\"o}nrich} \& {Binney}}{{Sch{\"o}nrich}
  \& {Binney}}{2012}]{Schonrich:2012}
{Sch{\"o}nrich} R.,  {Binney} J.,  2012, \mn@doi [\mnras]
  {10.1111/j.1365-2966.2011.19816.x}, \href
  {https://ui.adsabs.harvard.edu/abs/2012MNRAS.419.1546S} {419, 1546}

\bibitem[\protect\citeauthoryear{{Schwarz}}{{Schwarz}}{1978}]{Schwarz:1978}
{Schwarz} G.,  1978, Annals of Statistics, \href
  {https://ui.adsabs.harvard.edu/abs/1978AnSta...6..461S} {6, 461}

\bibitem[\protect\citeauthoryear{Schölkopf, Smola, Smola  \&
  Müller}{Schölkopf et~al.}{1998}]{Schol}
Schölkopf B.,  Smola A.,  Smola E.,   Müller K.-R.,  1998, Neural
  Computation, 10, 1299

\bibitem[\protect\citeauthoryear{{Sellwood} \& {Binney}}{{Sellwood} \&
  {Binney}}{2002}]{Sellwood:2002}
{Sellwood} J.~A.,  {Binney} J.~J.,  2002, \mn@doi [\mnras]
  {10.1046/j.1365-8711.2002.05806.x}, \href
  {https://ui.adsabs.harvard.edu/abs/2002MNRAS.336..785S} {336, 785}

\bibitem[\protect\citeauthoryear{{S{\`e}rsic}}{{S{\`e}rsic}}{1963}]{Sersic:1963}
{S{\`e}rsic} J.~L.,  1963, Boletin de la Asociacion Argentina de Astronomia La
  Plata Argentina, \href
  {https://ui.adsabs.harvard.edu/abs/1963BAAA....6...41S} {6, 41}

\bibitem[\protect\citeauthoryear{{S{\`e}rsic}}{{S{\`e}rsic}}{1968}]{Sersic1968}
{S{\`e}rsic} J.~L.,  1968, {Atlas de Galaxias Australes}

\bibitem[\protect\citeauthoryear{{Seth}, {Dalcanton}  \& {de Jong}}{{Seth}
  et~al.}{2005}]{Seth}
{Seth} A.~C.,  {Dalcanton} J.~J.,   {de Jong} R.~S.,  2005, \mn@doi [\aj]
  {10.1086/444620}, \href
  {https://ui.adsabs.harvard.edu/abs/2005AJ....130.1574S} {130, 1574}

\bibitem[\protect\citeauthoryear{{Soubiran}, {Bienaym{\'e}}  \&
  {Siebert}}{{Soubiran} et~al.}{2003}]{Soubiran:2003}
{Soubiran} C.,  {Bienaym{\'e}} O.,   {Siebert} A.,  2003, \mn@doi [\aap]
  {10.1051/0004-6361:20021615}, \href
  {https://ui.adsabs.harvard.edu/abs/2003A&A...398..141S} {398, 141}

\bibitem[\protect\citeauthoryear{{Springel}, {White}, {Tormen}  \&
  {Kauffmann}}{{Springel} et~al.}{2001}]{Springel2001}
{Springel} V.,  {White} S. D.~M.,  {Tormen} G.,   {Kauffmann} G.,  2001,
  \mn@doi [\mnras] {10.1046/j.1365-8711.2001.04912.x}, \href
  {https://ui.adsabs.harvard.edu/abs/2001MNRAS.328..726S} {328, 726}

\bibitem[\protect\citeauthoryear{{Springel} et~al.,}{{Springel}
  et~al.}{2018}]{Springel2018}
{Springel} V.,  et~al., 2018, \mn@doi [\mnras] {10.1093/mnras/stx3304}, \href
  {https://ui.adsabs.harvard.edu/abs/2018MNRAS.475..676S} {475, 676}

\bibitem[\protect\citeauthoryear{{Stinson}, {Seth}, {Katz}, {Wadsley},
  {Governato}  \& {Quinn}}{{Stinson} et~al.}{2006}]{Stinson:2006}
{Stinson} G.,  {Seth} A.,  {Katz} N.,  {Wadsley} J.,  {Governato} F.,   {Quinn}
  T.,  2006, \mn@doi [\mnras] {10.1111/j.1365-2966.2006.11097.x}, \href
  {https://ui.adsabs.harvard.edu/abs/2006MNRAS.373.1074S} {373, 1074}

\bibitem[\protect\citeauthoryear{{Stinson}, {Brook}, {Macci{\`o}}, {Wadsley},
  {Quinn}  \& {Couchman}}{{Stinson} et~al.}{2013a}]{Stinson:2013b}
{Stinson} G.~S.,  {Brook} C.,  {Macci{\`o}} A.~V.,  {Wadsley} J.,  {Quinn}
  T.~R.,   {Couchman} H.~M.~P.,  2013a, \mn@doi [\mnras]
  {10.1093/mnras/sts028}, \href
  {https://ui.adsabs.harvard.edu/abs/2013MNRAS.428..129S} {428, 129}

\bibitem[\protect\citeauthoryear{{Stinson} et~al.,}{{Stinson}
  et~al.}{2013b}]{Stinson2013}
{Stinson} G.~S.,  et~al., 2013b, \mn@doi [\mnras] {10.1093/mnras/stt1600},
  \href {https://ui.adsabs.harvard.edu/abs/2013MNRAS.436..625S} {436, 625}

\bibitem[\protect\citeauthoryear{{Stoughton} et~al.,}{{Stoughton}
  et~al.}{2002}]{Stoughton:2002}
{Stoughton} C.,  et~al., 2002, \mn@doi [\aj] {10.1086/324741}, \href
  {https://ui.adsabs.harvard.edu/abs/2002AJ....123..485S} {123, 485}

\bibitem[\protect\citeauthoryear{{Thanjavur}, {Simard}, {Bluck}  \&
  {Mendel}}{{Thanjavur} et~al.}{2016}]{Thanjavur2016}
{Thanjavur} K.,  {Simard} L.,  {Bluck} A. F.~L.,   {Mendel} T.,  2016, \mn@doi
  [\mnras] {10.1093/mnras/stw495}, \href
  {https://ui.adsabs.harvard.edu/abs/2016MNRAS.459...44T} {459, 44}

\bibitem[\protect\citeauthoryear{{Thob} et~al.,}{{Thob}
  et~al.}{2019}]{Thob2019}
{Thob} A. C.~R.,  et~al., 2019, \mn@doi [\mnras] {10.1093/mnras/stz448}, \href
  {https://ui.adsabs.harvard.edu/abs/2019MNRAS.485..972T} {485, 972}

\bibitem[\protect\citeauthoryear{{Tissera}, {White}  \&
  {Scannapieco}}{{Tissera} et~al.}{2012}]{Tissera2012}
{Tissera} P.~B.,  {White} S. D.~M.,   {Scannapieco} C.,  2012, \mn@doi [\mnras]
  {10.1111/j.1365-2966.2011.20028.x}, \href
  {https://ui.adsabs.harvard.edu/abs/2012MNRAS.420..255T} {420, 255}

\bibitem[\protect\citeauthoryear{{Tissera}, {Scannapieco}, {Beers}  \&
  {Carollo}}{{Tissera} et~al.}{2013}]{Tissera2013}
{Tissera} P.~B.,  {Scannapieco} C.,  {Beers} T.~C.,   {Carollo} D.,  2013,
  \mn@doi [\mnras] {10.1093/mnras/stt691}, \href
  {https://ui.adsabs.harvard.edu/abs/2013MNRAS.432.3391T} {432, 3391}

\bibitem[\protect\citeauthoryear{{Tissera}, {Beers}, {Carollo}  \&
  {Scannapieco}}{{Tissera} et~al.}{2014}]{Tissera2014}
{Tissera} P.~B.,  {Beers} T.~C.,  {Carollo} D.,   {Scannapieco} C.,  2014,
  \mn@doi [\mnras] {10.1093/mnras/stu181}, \href
  {https://ui.adsabs.harvard.edu/abs/2014MNRAS.439.3128T} {439, 3128}

\bibitem[\protect\citeauthoryear{{Tissera}, {Rosas-Guevara}, {Bower}, {Crain},
  {del P Lagos}, {Schaller}, {Schaye}  \& {Theuns}}{{Tissera}
  et~al.}{2019}]{Tissera:2019}
{Tissera} P.~B.,  {Rosas-Guevara} Y.,  {Bower} R.~G.,  {Crain} R.~A.,  {del P
  Lagos} C.,  {Schaller} M.,  {Schaye} J.,   {Theuns} T.,  2019, \mn@doi
  [\mnras] {10.1093/mnras/sty2817}, \href
  {https://ui.adsabs.harvard.edu/abs/2019MNRAS.482.2208T} {482, 2208}

\bibitem[\protect\citeauthoryear{{Trayford}, {Frenk}, {Theuns}, {Schaye}  \&
  {Correa}}{{Trayford} et~al.}{2019}]{Trayford:2019}
{Trayford} J.~W.,  {Frenk} C.~S.,  {Theuns} T.,  {Schaye} J.,   {Correa} C.,
  2019, \mn@doi [\mnras] {10.1093/mnras/sty2860}, \href
  {https://ui.adsabs.harvard.edu/abs/2019MNRAS.483..744T} {483, 744}

\bibitem[\protect\citeauthoryear{{Trujillo} \& {Fliri}}{{Trujillo} \&
  {Fliri}}{2016}]{Trujillo:2016}
{Trujillo} I.,  {Fliri} J.,  2016, \mn@doi [\apj]
  {10.3847/0004-637X/823/2/123}, \href
  {https://ui.adsabs.harvard.edu/abs/2016ApJ...823..123T} {823, 123}

\bibitem[\protect\citeauthoryear{{Tully} \& {Fisher}}{{Tully} \&
  {Fisher}}{1977}]{Tully1977}
{Tully} R.~B.,  {Fisher} J.~R.,  1977, \aap, \href
  {https://ui.adsabs.harvard.edu/abs/1977A&A....54..661T} {54, 661}

\bibitem[\protect\citeauthoryear{{Varela-Lavin}, {Tissera}, {G{\'o}mez},
  {Bignone}  \& {Lagos}}{{Varela-Lavin} et~al.}{2021}]{Varela2021}
{Varela-Lavin} S.,  {Tissera} P.~B.,  {G{\'o}mez} F.~A.,  {Bignone} L.~A.,
  {Lagos} C. d.~P.,  2021, arXiv e-prints, \href
  {https://ui.adsabs.harvard.edu/abs/2021arXiv211114126V} {p. arXiv:2111.14126}

\bibitem[\protect\citeauthoryear{{Walt}, {Colbert}  \& {Varoquaux}}{{Walt}
  et~al.}{2011}]{Walt:2011}
{Walt} S. v.~d.,  {Colbert} S.~C.,   {Varoquaux} G.,  2011, Computing in
  Science and Engineering, 13, 22

\bibitem[\protect\citeauthoryear{{Wang}, {Dutton}, {Stinson}, {Macci{\`o}},
  {Penzo}, {Kang}, {Keller}  \& {Wadsley}}{{Wang} et~al.}{2015}]{NIHAO}
{Wang} L.,  {Dutton} A.~A.,  {Stinson} G.~S.,  {Macci{\`o}} A.~V.,  {Penzo} C.,
   {Kang} X.,  {Keller} B.~W.,   {Wadsley} J.,  2015, \mn@doi [\mnras]
  {10.1093/mnras/stv1937}, \href
  {https://ui.adsabs.harvard.edu/abs/2015MNRAS.454...83W} {454, 83}

\bibitem[\protect\citeauthoryear{Waskom}{Waskom}{2021}]{Waskom2021}
Waskom M.~L.,  2021, \mn@doi [Journal of Open Source Software]
  {10.21105/joss.03021}, 6, 3021

\bibitem[\protect\citeauthoryear{{W}es {M}c{K}inney}{{W}es
  {M}c{K}inney}{2010}]{mckinney-proc-scipy-2010}
{W}es {M}c{K}inney 2010, in {S}t\'efan van~der {W}alt {J}arrod {M}illman eds,
  {P}roceedings of the 9th {P}ython in {S}cience {C}onference. pp 56 -- 61,
  \mn@doi{10.25080/Majora-92bf1922-00a}

\bibitem[\protect\citeauthoryear{{Wiersma}, {Schaye}, {Theuns}, {Dalla Vecchia}
   \& {Tornatore}}{{Wiersma} et~al.}{2009}]{Wiersma}
{Wiersma} R. P.~C.,  {Schaye} J.,  {Theuns} T.,  {Dalla Vecchia} C.,
  {Tornatore} L.,  2009, \mn@doi [\mnras] {10.1111/j.1365-2966.2009.15331.x},
  \href {https://ui.adsabs.harvard.edu/abs/2009MNRAS.399..574W} {399, 574}

\bibitem[\protect\citeauthoryear{{York} et~al.,}{{York}
  et~al.}{2000}]{York:2000}
{York} D.~G.,  et~al., 2000, \mn@doi [\aj] {10.1086/301513}, \href
  {https://ui.adsabs.harvard.edu/abs/2000AJ....120.1579Y} {120, 1579}

\bibitem[\protect\citeauthoryear{{Zana} et~al.,}{{Zana}
  et~al.}{2022}]{Zana:2022}
{Zana} T.,  et~al., 2022, \mn@doi [\mnras] {10.1093/mnras/stac1708}, \href
  {https://ui.adsabs.harvard.edu/abs/2022MNRAS.tmp.1668Z} {}

\bibitem[\protect\citeauthoryear{Zhu et~al.,}{Zhu et~al.}{2017}]{Zhu2017}
Zhu L.,  et~al., 2017, \mn@doi [Monthly Notices of the Royal Astronomical
  Society] {10.1093/mnras/stx2409}, 473, 3000

\bibitem[\protect\citeauthoryear{{Zhu} et~al.,}{{Zhu} et~al.}{2018a}]{Zhu2018a}
{Zhu} L.,  et~al., 2018a, \mn@doi [Nature Astronomy]
  {10.1038/s41550-017-0348-1}, \href
  {https://ui.adsabs.harvard.edu/abs/2018NatAs...2..233Z} {2, 233}

\bibitem[\protect\citeauthoryear{{Zhu}, {van de Ven}, {M{\'e}ndez-Abreu}  \&
  {Obreja}}{{Zhu} et~al.}{2018b}]{Zhu2018b}
{Zhu} L.,  {van de Ven} G.,  {M{\'e}ndez-Abreu} J.,   {Obreja} A.,  2018b,
  \mn@doi [Monthly Notices of the RAS] {10.1093/mnras/sty1521}, \href
  {https://ui.adsabs.harvard.edu/abs/2018MNRAS.479..945Z} {479, 945}

\bibitem[\protect\citeauthoryear{{Zhu} et~al.,}{{Zhu} et~al.}{2020}]{Zhu3}
{Zhu} L.,  et~al., 2020, \mn@doi [Monthly Notices of the RAS]
  {10.1093/mnras/staa1584}, \href
  {https://ui.adsabs.harvard.edu/abs/2020MNRAS.496.1579Z} {496, 1579}

\bibitem[\protect\citeauthoryear{de Graaff, Trayford, Franx, Schaller, Schaye
  \& van~der Wel}{de~Graaff et~al.}{2021}]{EAGLESersic}
de Graaff A.,  Trayford J.,  Franx M.,  Schaller M.,  Schaye J.,   van~der Wel
  A.,  2021, \mn@doi [Monthly Notices of the Royal Astronomical Society]
  {10.1093/mnras/stab3510}, 511, 2544–2564

\bibitem[\protect\citeauthoryear{{van de Sande} et~al.,}{{van de Sande}
  et~al.}{2017}]{vandesande2017}
{van de Sande} J.,  et~al., 2017, \mn@doi [\apj] {10.3847/1538-4357/835/1/104},
  \href {https://ui.adsabs.harvard.edu/abs/2017ApJ...835..104V} {835, 104}

\makeatother
\end{thebibliography}

%\begin{thebibliography}{99}
%\end{thebibliography}

%%%%%%%%%%%%%%%%%%%%%%%%%%%%%%%%%%%%%%%%%%%%%%%%%%

%%%%%%%%%%%%%%%%% APPENDICES %%%%%%%%%%%%%%%%%%%%%

\appendix

\section{A note on S{\`e}rsic indices}

\begin{figure}
\centering
\includegraphics[width=0.42\textwidth]{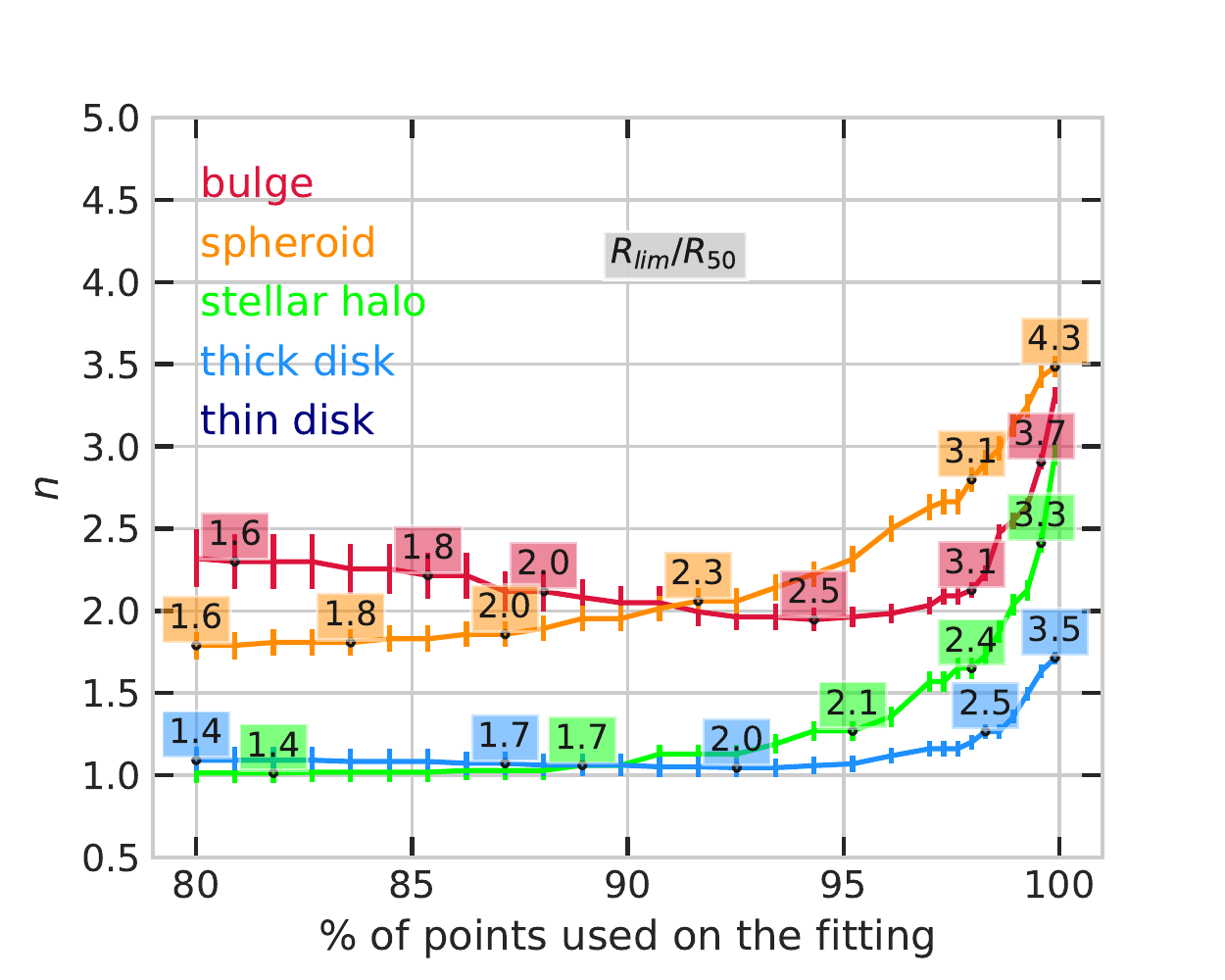}\\
\includegraphics[width=0.42\textwidth]{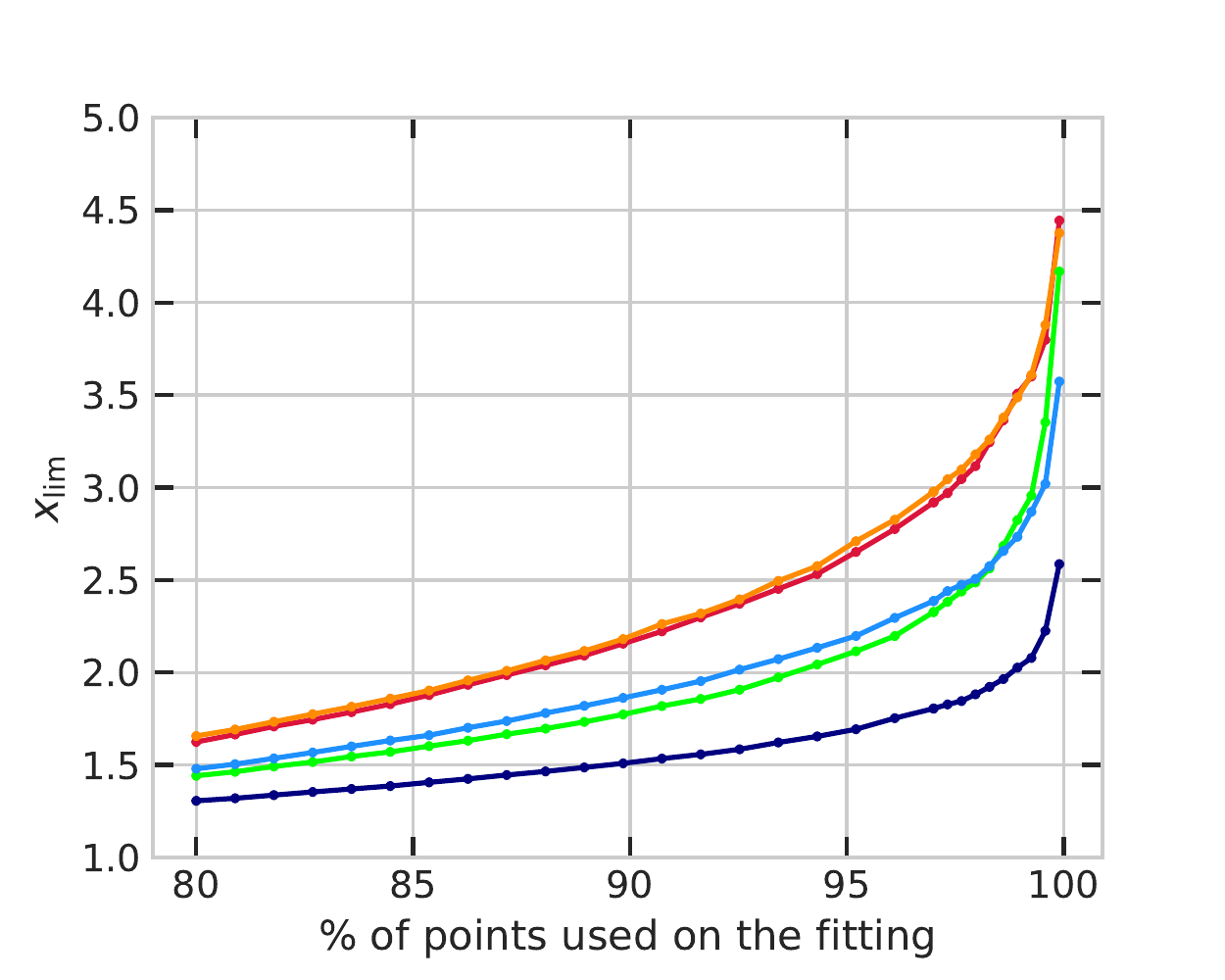}
\caption{\textbf{Top:} Variation of the cross-validated S{\`e}rsic index with the upper radial limit used for the fit of the stacked density distributions. \textbf{Bottom:} Distribution of the points of the density profiles of the components before the stacking for each type of component, where $x_{\rm lim} = R_{\rm lim}/R_{\rm 50}$. Colour coding is the same in both panels.}
\label{fig:n_with_limit}
\end{figure}

\label{appendix:sersic}
\begin{figure}
\centering
\includegraphics[width=0.42\textwidth]{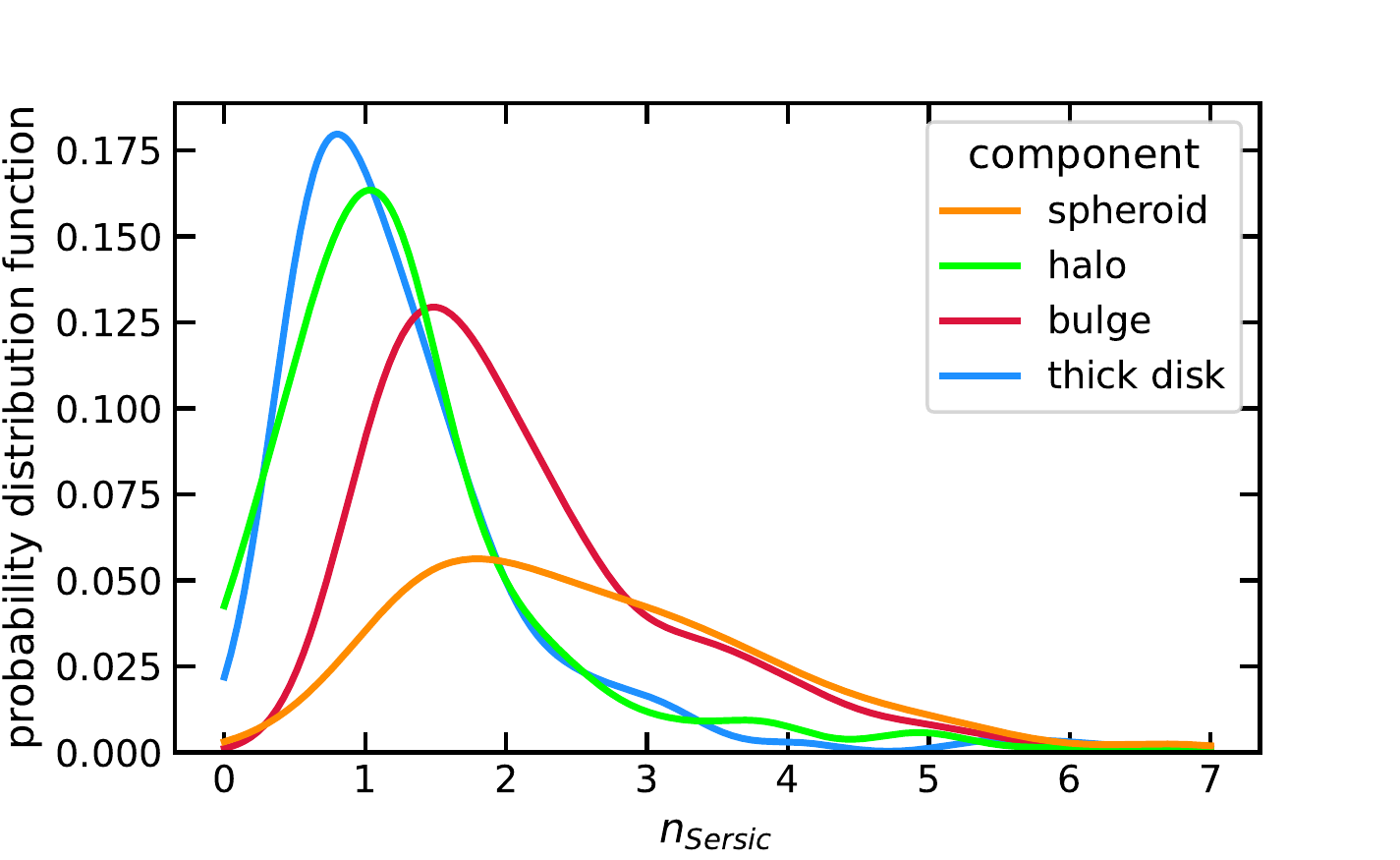}
\caption{The distribution of individual S{\`e}rsic indices $n$ for all the components in the sample when the upper radial limit is fixed to the radius enclosing 90 percent of the component mass.}
\label{fig:n_pdfs}
\end{figure}

The fit in Equation~\ref{eq:Sersic} is very sensitive to the radial range. Figure~\ref{fig:n_with_limit} shows how $n$ varies for each component when the upper radial limit for the fit is increased. The radial limit is defined in terms of the percentage of points considered, since not all the components show the same relations: the value of the normalisation radius ($R_{\rm 50}$) is larger for more radially extended components (such as the halo), such that the density profiles of individual galaxies mostly lie within smaller values of $R_{lim}$, as show by the numerical values of $R_{lim}$ written in the figure for some of the points. For the thick disk, bulge and spheroid, $n$ is relatively stable as long as the $R_{\rm lim}<2R_{\rm 50}$. For the stellar halo, $n$ increases rapidly with $R_{\rm lim}$ for $R_{\rm lim}>1.5R_{\rm 50}$. 

For completeness, we also show in Figure~\ref{fig:n_pdfs} the distribution of $n$ for bulges, spheroids, stellar halos and thick disks. The thick disks peak at $n$ slightly less than 1, while stellar halos peak at 1. As expected, bulges have larger $n$ than halos and thick disks, but there is no clear separation between what we call spheroids and bulges. 
For all the individual fits, we used twice the gravitational softening as lower radial limit, and as upper limit we used the radius enclosing 90 percent of the component's mass.

\section{Mass dependencies}
\label{appendix:massdep}

\begin{figure}
    \centering
    \includegraphics[width=0.23\textwidth]{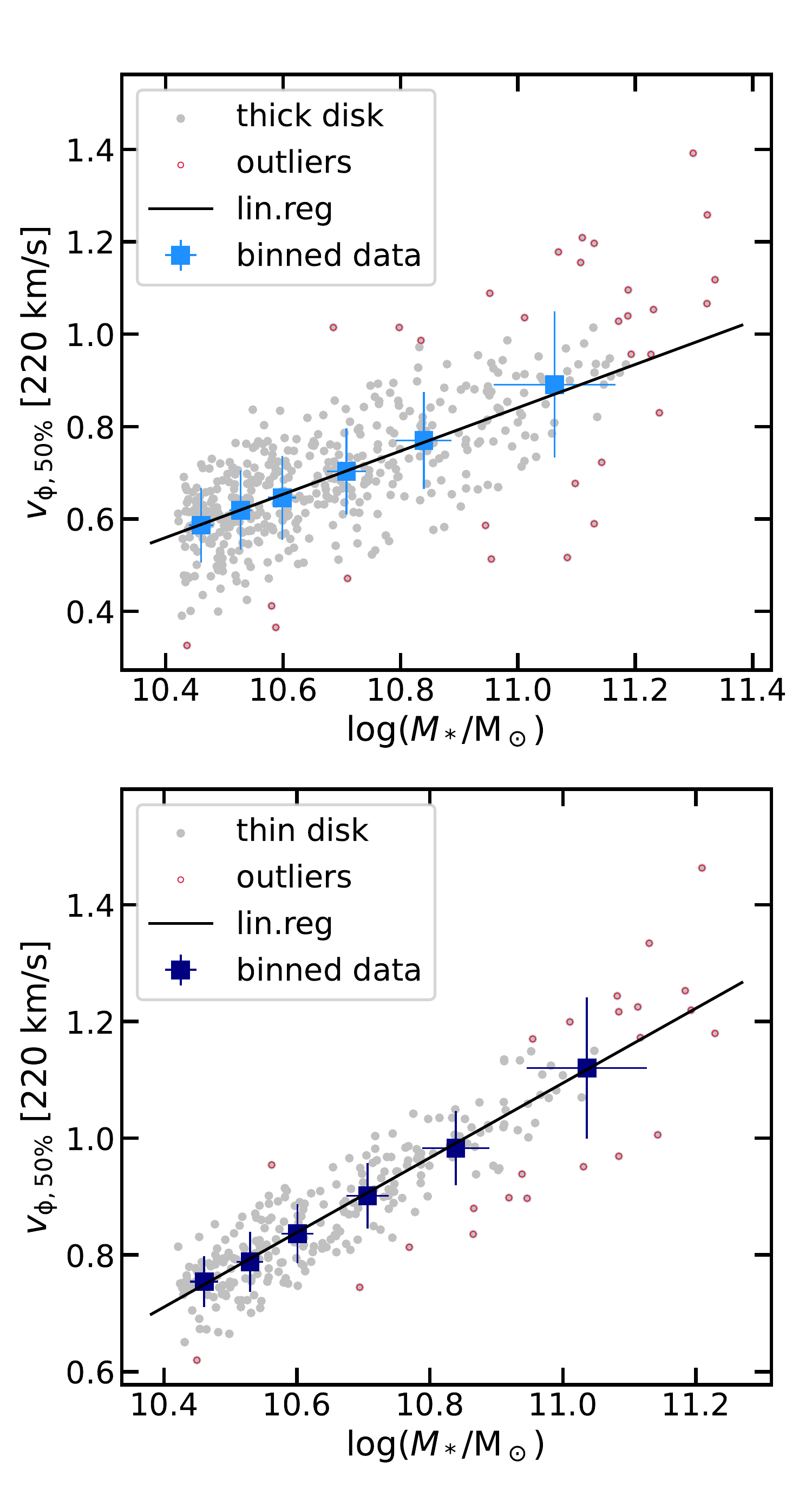}
    \includegraphics[width=0.23\textwidth]{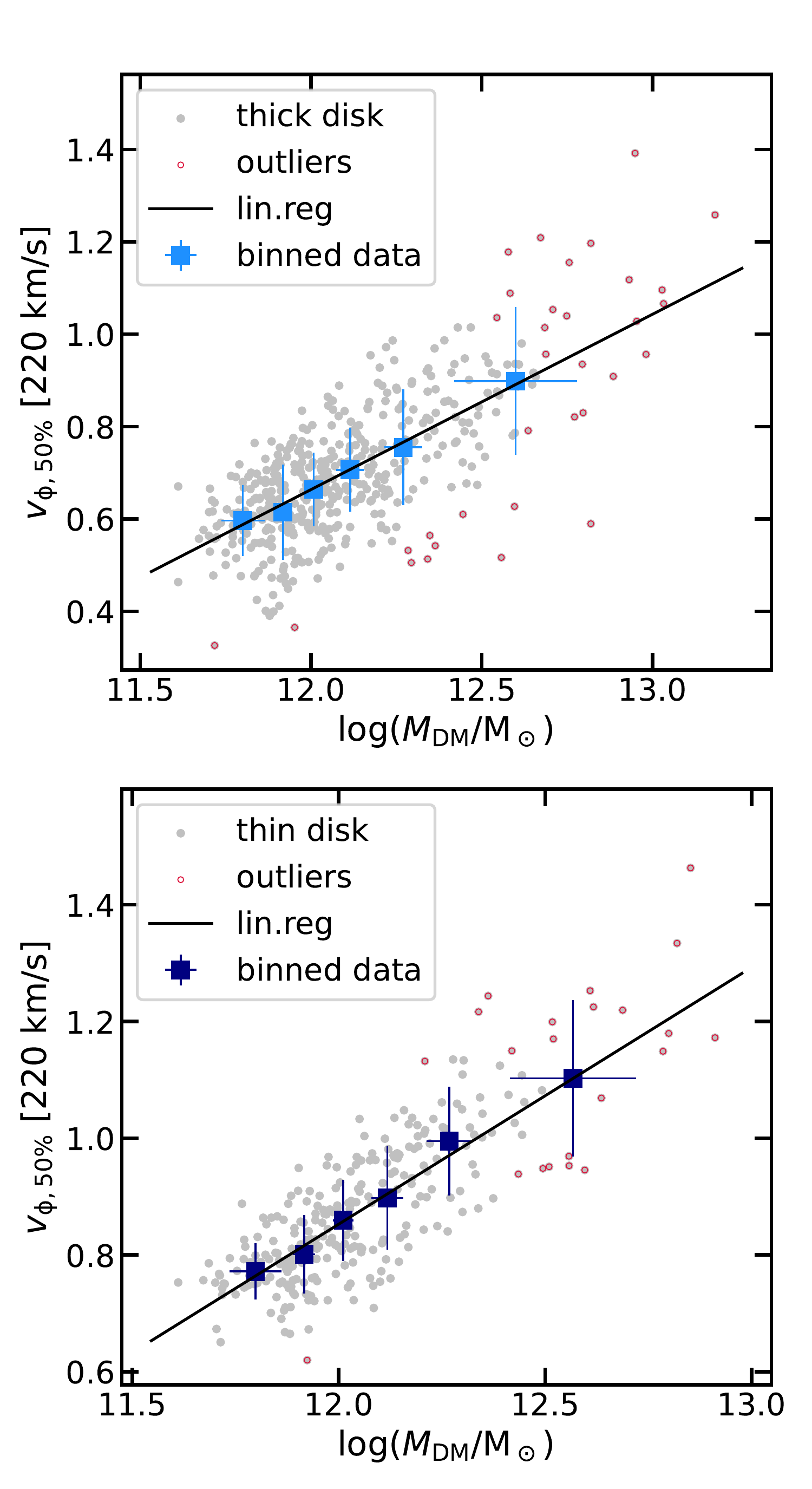}
    \caption{The rotational velocities of thick (top) and thin (bottom) disks as functions of stellar (left) and DM halo (right) mass. The red points are the data points classified as outliers by MCD. The linear regressions (Table~\ref{tab:v_vs_mass}) have been done only with the MCD inliers (grey). The binned data are the ones from Figure~\ref{fig:massdep_observables}. Velocities are normalised by 220~km~s$^{\rm -1}$, and the fitting parameters are given in Table~\ref{tab:v_vs_mass}.}
    \label{fig:mass_vphi_fit}
\end{figure}

\begin{figure}
    \centering
    \includegraphics[width=0.48\textwidth]{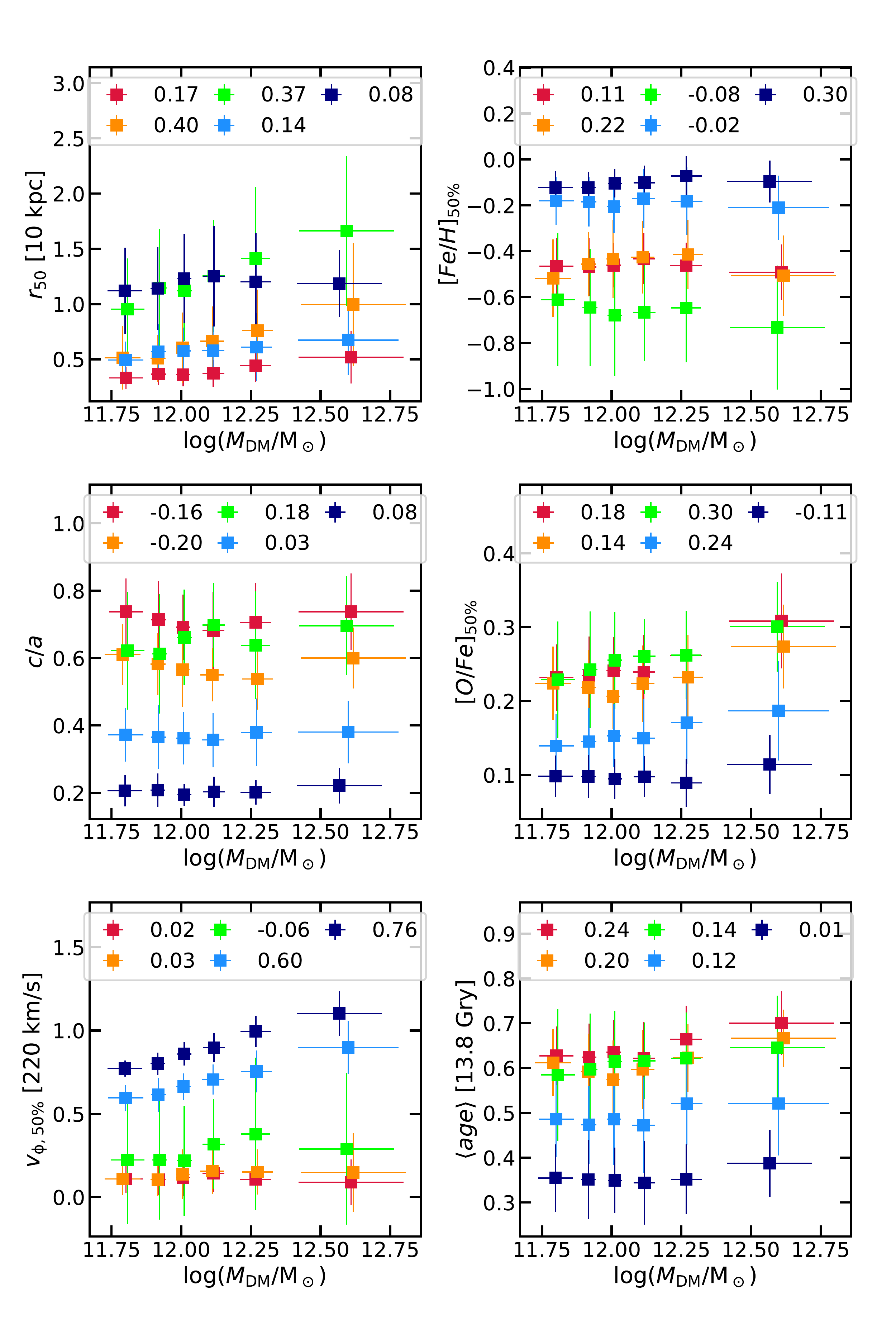}
    \caption{Variation of the observable properties discussed in Sections~\ref{sec:formas} and \ref{pop_properties} with the dark matter mass of the host halo.}
    \label{fig:virmassdep_observables}
\end{figure}

\begin{table}
\centering
\begin{tabular}{ccccc}
\hline
Component & $x$ & norm & slope & $\rho_{\rm MCD}$\\
\hline
\hline 
thick disk & log($M_{\rm *}$/M$_{\rm\odot}$) & -4.31$\pm$0.23 & 0.47$\pm$0.02 & 0.65\\
 & log($M_{\rm DM}$/M$_{\rm\odot}$) & -3.89$\pm$0.24 & 0.38$\pm$0.02 & 0.60\\
 \hline
thin disk & log($M_{\rm *}$/M$_{\rm\odot}$) & -5.94$\pm$0.20 & 0.64$\pm$0.02 & 0.89\\
 & log($M_{\rm DM}$/M$_{\rm\odot}$) & -4.43$\pm$0.29 & 0.44$\pm$0.024 & 0.76\\
 \hline
\end{tabular}
\caption{The linear regression fits though the $v_{\rm\phi,50}$ vs mass (total stellar and DM halo masses) for the thick and thin disks (shown in Figure~\ref{fig:mass_vphi_fit}). For the fits we only used the data points classified as inliers by MCD. The normalisations are in units of 220~km~s$^{\rm -1}$, while the slopes are in units of 220~km~s$^{\rm -1}$log(M$_{\rm\odot}$)$^{\rm -1}$.}
\label{tab:v_vs_mass}
\end{table}

\begin{figure*}
    \centering
    \includegraphics[width=0.99\textwidth]{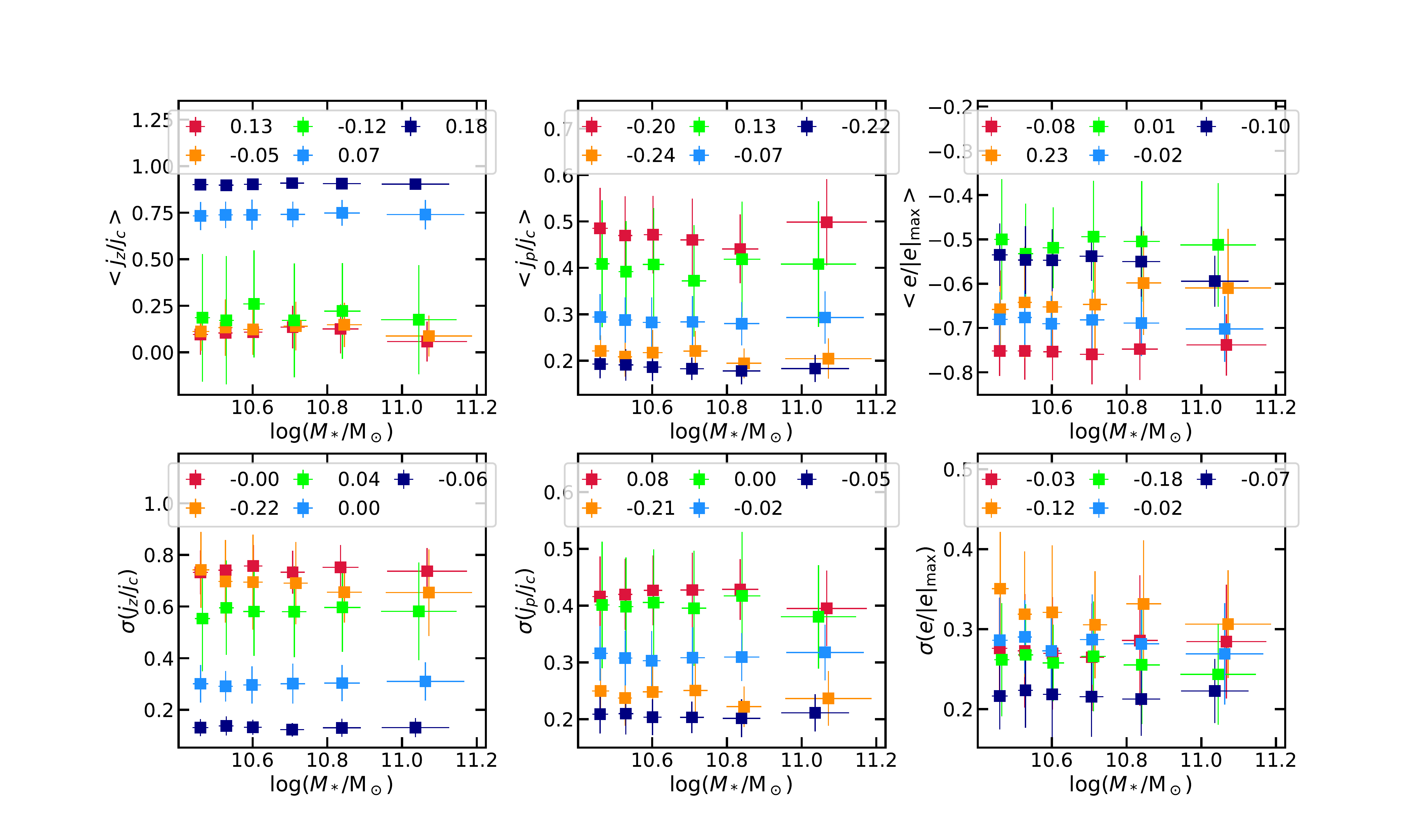}
    \rule{10cm}{1pt}
    \includegraphics[width=0.99\textwidth]{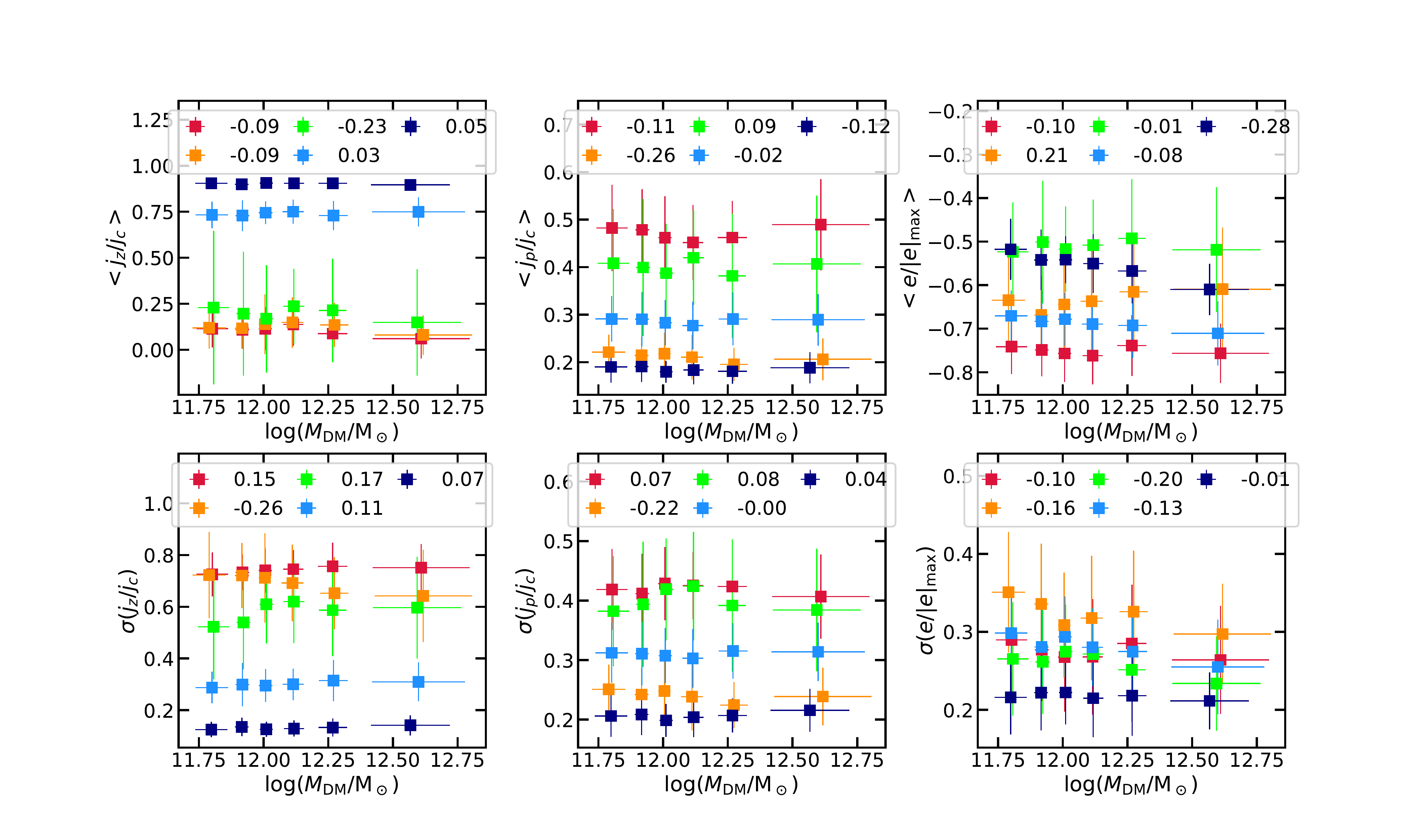}
    \caption{Variation of the dynamical properties used for clustering in Section~\ref{generic_components} with the total stellar mass of the host galaxy (top) and with the dark matter mass of the host halo (bottom).}
    \label{fig:massdep_input}
\end{figure*}

We quantified the $v_{\rm\phi,50}$--$M_{\rm *}$, and $v_{\rm\phi,50}$--$M_{\rm DM}$ relations in the disk components using linear regression in log(mass) (Figure~\ref{fig:mass_vphi_fit}), all fits parameters being reported in Table~\ref{tab:v_vs_mass}. Figure~\ref{fig:mass_vphi_fit} also nicely illustrates the decrease in thin disk frequency as the mass increases. 

Figure~\ref{fig:virmassdep_observables} gives the observable properties discussed in Sections~\ref{sec:formas} and \ref{pop_properties} as functions of the dark matter host halo, $M_{\rm DM}$. The only clean correlations that stand out are between the median rotational velocities of disks (both thin and thick) and $M_{\rm DM}$.

The dependencies of the dynamical variables used to define the generic components with the stellar mass are very weak or not at all. Same is true if we consider the mass of the dark matter host halos. Figure~\ref{fig:massdep_input} shows all six dynamical properties for all five components in the six equally populated mass bins, while the legends in each panel give the MCD-derived correlation coefficients $\rho_{\rm MCD}$. The top part of the figure shows the dependence with $M_{\rm *}$, while the bottom one the dependence with $M_{\rm DM}$.

% Don't change these lines
\bsp	% typesetting comment
\label{lastpage}
\end{document}